\newcommand\apjcls{1}
\newcommand\aastexcls{2}
\newcommand\othercls{3}

\newcommand\papercls{\apjcls}
\documentclass[iop,apj]{emulateapj}

\if\papercls \apjcls
\usepackage{apjfonts}
\else\if\papercls \othercls
\usepackage{epsfig}
\usepackage{margin}
\usepackage{times}
\fi\fi
\usepackage{ifthen}
\usepackage{natbib}
\usepackage{amssymb, amsmath}
\usepackage{appendix}
\usepackage{etoolbox}
\usepackage[T1]{fontenc}
\usepackage{paralist}

\if\papercls \apjcls
\newcommand\aas{\ref@jnl{AAS Meeting Abstracts}}% *** added by jh
\newcommand\dps{\ref@jnl{AAS/DPS Meeting Abstracts}}% *** added by jh
\newcommand\maps{\ref@jnl{MAPS}}% *** added by jh
\else\if\papercls \othercls
\usepackage{astjnlabbrev-jh}
\fi\fi

\if\papercls \aastexcls
\hypersetup{citecolor=blue, % color for \cite{...} links
            linkcolor=blue, % color for \ref{...} links
            menucolor=blue, % color for Acrobat menu buttons
            urlcolor=blue}  % color for \url{...} links
\else
\usepackage[%pdftex,      % hyper-references for pdflatex
bookmarks=true,           %%% generate bookmarks ...
bookmarksnumbered=true,   %%% ... with numbers
colorlinks=true,          % links are colored
citecolor=blue,           % color for \cite{...} links
linkcolor=blue,           % color for \ref{...} links
menucolor=blue,           % color for Acrobat menu buttons
urlcolor=blue,            % color for \url{...} links
linkbordercolor={0 0 1},  %%% blue frames around links
pdfborder={0 0 1},
frenchlinks=true]{hyperref}
\fi

\if\papercls \othercls

\else

\fi

\providecommand{\adsurl}[1]{\href{#1}{ADS}}

\makeatletter
\patchcmd{\NAT@citex}
  {\@citea\NAT@hyper@{%
     \NAT@nmfmt{\NAT@nm}%
     \hyper@natlinkbreak{\NAT@aysep\NAT@spacechar}{\@citeb\@extra@b@citeb}%
     \NAT@date}}
  {\@citea\NAT@nmfmt{\NAT@nm}%
   \NAT@aysep\NAT@spacechar\NAT@hyper@{\NAT@date}}{}{}

\patchcmd{\NAT@citex}
  {\@citea\NAT@hyper@{%
     \NAT@nmfmt{\NAT@nm}%
     \hyper@natlinkbreak{\NAT@spacechar\NAT@@open\if*#1*\else#1\NAT@spacechar\fi}%
       {\@citeb\@extra@b@citeb}%
     \NAT@date}}
  {\@citea\NAT@nmfmt{\NAT@nm}%
   \NAT@spacechar\NAT@@open\if*#1*\else#1\NAT@spacechar\fi\NAT@hyper@{\NAT@date}}
  {}{}
\makeatother

\makeatletter
\DeclareRobustCommand{\lowcase}[1]{\@lowcase#1\@nil}
\def\@lowcase#1\@nil{\if\relax#1\relax\else\MakeLowercase{#1}\fi}
\makeatother

\DeclareSymbolFont{UPM}{U}{eur}{m}{n}
\DeclareMathSymbol{\umu}{0}{UPM}{"16}
\let\oldumu=\umu
\renewcommand\umu{\ifmmode\oldumu\else\math{\oldumu}\fi}

\if\papercls \othercls

\else

\fi

\let\oldsim=\sim
\renewcommand\sim{\ifmmode\oldsim\else\math{\oldsim}\fi}
\let\oldpm=\pm
\renewcommand\pm{\ifmmode\oldpm\else\math{\oldpm}\fi}
\newcommand\by{\ifmmode\times\else\math{\times}\fi}
\newbox{\wdbox}
\renewcommand\c{\setbox\wdbox=\hbox{,}\hspace{\wd\wdbox}}
\renewcommand\i{\setbox\wdbox=\hbox{i}\hspace{\wd\wdbox}}

\newcount\timect
\newcount\hourct
\newcount\minct
\newcommand\now{\timect=\time \divide\timect by 60
         \hourct=\timect \multiply\hourct by 60
         \minct=\time \advance\minct by -\hourct
         \number\timect:\ifnum \minct < 10 0\fi\number\minct}

\catcode`@=11

\newcommand\comment[1]{}

\newcommand\commenton{\catcode`\%=14}

\renewcommand\math[1]{$#1$}
\newcommand\mathshifton{\catcode`\$=3}

\let\atab=&
\newcommand\atabon{\catcode`\&=4}

\let\oldmsp=\sp
\let\oldmsb=\sb
\def\sp#1{\ifmmode
           \oldmsp{#1}%
         \else\strut\raise.85ex\hbox{\scriptsize #1}\fi}
\def\sb#1{\ifmmode
           \oldmsb{#1}%
         \else\strut\raise-.54ex\hbox{\scriptsize #1}\fi}
\newbox\@sp
\newbox\@sb
\def\sbp#1#2{\ifmmode%
           \oldmsb{#1}\oldmsp{#2}%
         \else
           \setbox\@sb=\hbox{\sb{#1}}%
           \setbox\@sp=\hbox{\sp{#2}}%
           \rlap{\copy\@sb}\copy\@sp
           \ifdim \wd\@sb >\wd\@sp
             \hskip -\wd\@sp \hskip \wd\@sb
           \fi
        \fi}
\def\msp#1{\ifmmode
           \oldmsp{#1}
         \else \math{\oldmsp{#1}}\fi}
\def\msb#1{\ifmmode
           \oldmsb{#1}
         \else \math{\oldmsb{#1}}\fi}

\def\supon{\catcode`\^=7}

\def\subon{\catcode`\_=8}

\def\supsubon{\supon \subon}

\newcommand\actcharon{\catcode`\~=13}

\newcommand\paramon{\catcode`\#=6}

\newcommand\reservedcharson{ \commenton  \mathshifton  \atabon  \supsubon 
                             \actcharon  \paramon}

\catcode`@=12
\reservedcharson

\if\papercls \apjcls

\else

\fi

\newcommand\chisq{\ifmmode{\chi\sp{2}}\else\math{\chi\sp{2}}\fi}
\newcommand\redchisq{\ifmmode{ \chi\sp{2}\sb{\rm red}}
                    \else\math{\chi\sp{2}\sb{\rm red}}\fi}
\newcommand\Teq{\ifmmode{T\sb{\rm eq}}\else$T$\sb{eq}\fi}
\newcommand\mjup{\ifmmode{M\sb{\rm Jup}}\else$M$\sb{Jup}\fi}
\newcommand\rjup{\ifmmode{R\sb{\rm Jup}}\else$R$\sb{Jup}\fi}
\newcommand\mearth{\ifmmode{M\sb{\oplus}}\else$M\sb{\oplus}$\fi}
\newcommand\rearth{\ifmmode{R\sb{\oplus}}\else$R\sb{\oplus}$\fi}
\usepackage[utf8]{inputenc} % Required for inputting international characters
\usepackage[T1]{fontenc} % Output font encoding for international characters
\usepackage{stmaryrd}
\usepackage{soul}

\newcommand{\akari}{\textit{AKARI}}
\newcommand{\wise}{\textit{WISE}}
\newcommand{\iras}{\textit{IRAS}}
\newcommand{\herschel}{\textit{Herschel}}

\newcommand{\oiilong}{\hbox{[O\sc ii] 3726\AA\ 3729\AA}}

\newcommand{\oiiiblong}{\hbox{[O\sc iii] 5007\AA}}
\newcommand{\oiiilong}{\hbox{[O\sc iii] 4959\AA\ 5007\AA}}
\newcommand{\oiii}{\hbox{[O\sc iii]}}
\newcommand{\nii}{\hbox{[N\sc ii]}}
\newcommand{\niilong}{\hbox{[N\sc ii] 6548\AA\ 6583\AA}}

\newcommand{\siilong}{\hbox{[S\sc ii] 6716\AA\ 6731\AA}}
\newcommand{\feii}{\hbox{Fe\sc ii}}
\newcommand{\mgii}{\hbox{Mg\sc ii}}

\newcommand{\ha}{H$\alpha$}
\newcommand{\hb}{H$\beta$}
\newcommand{\hii}{\hbox{H\sc ii}}

\newcommand{\msun}{M$_{\odot}$}

\newcommand{\lsun}{L$_{\odot}$}
\newcommand{\kms}{km\,s$^{-1}$}

\newcommand{\sfrunit}{M$_{\odot}$\,yr$^{-1}$}
\newcommand{\ccm}{cm$^{-3}$}

\shorttitle{Evolution of \akari-selected ULIRGs}
\shortauthors{Chen {\em et al.}}

\begin{document}

\title{Tracing the co-evolution path of super massive black holes and spheroids with \akari-selected ultra-luminous IR galaxies at intermediate redshifts}

%% AUTHOR/INSTITUTIONS FOR AASTEX6.1:
%\author{Xiaoyang Chen}
%\affiliation{Astronomical Institute, Tohoku University, 6-3 Aramaki, Aoba-ku, Sendai, Miyagi 980-8578, Japan}
%\author{Masayuki Akiyama}
%\affiliation{Astronomical Institute, Tohoku University, 6-3 Aramaki, Aoba-ku, Sendai, Miyagi 980-8578, Japan}
%\author{Kohei Ichikawa}
%\affiliation{Astronomical Institute, Tohoku University, 6-3 Aramaki, Aoba-ku, Sendai, Miyagi 980-8578, Japan}

%% AUTHOR/INSTITUTIONS FOR EMULATE APJ:
 \author{Xiaoyang Chen\altaffilmark{1,2},
 Masayuki Akiyama\altaffilmark{1},
 Kohei Ichikawa\altaffilmark{1,2},
 Hirofumi Noda\altaffilmark{3},
 Yoshiki Toba\altaffilmark{4,5,6},
 Issei Yamamura\altaffilmark{7,8},
 Toshihiro Kawaguchi\altaffilmark{9},
 Abdurro'uf\altaffilmark{5},
 and
 Mitsuru Kokubo\altaffilmark{1}
 }
 \affil{\sp{1} Astronomical Institute, Tohoku University, 6-3 Aramaki, Aoba-ku, Sendai, Miyagi 980-8578, Japan\\
        \sp{2} Frontier Research Institute for Interdisciplinary Sciences, Tohoku University, Sendai 980-8578, Japan\\
        \sp{3} Department of Earth and Space Science, Graduate School of Science, Osaka University, 1-1 Machikaneyama-cho, Toyonaka-shi, Osaka 560-0043, Japan\\
        \sp{4} Department of Astronomy, Kyoto University, Kitashirakawa-Oiwake-cho, Sakyo-ku, Kyoto 606-8502, Japan\\
        \sp{5} Academia Sinica Institute of Astronomy and Astrophysics, 11F of Astronomy-Mathematics Building, AS/NTU, No.1, Section 4, Roosevelt Road, Taipei 10617, Taiwan\\
        \sp{6} Research Center for Space and Cosmic Evolution, Ehime University, 2-5 Bunkyo-cho, Matsuyama, Ehime 790-8577, Japan\\
        \sp{7} Institute of Space and Astronautical Science, JAXA, 3-1-1 Yoshinodai, Chuo-ku, Sagamihara, Kanagawa 252-5210, Japan\\
        \sp{8} Department of Space and Astronautical Science, SOKENDAI, 3-1-1 Yoshinodai, Chuo-ku, Sagamihara, Kanagawa 252-5210, Japan\\
        \sp{9} Department of Economics, Management and Information Science, Onomichi City University, Hisayamada 1600-2, Onomichi, Hiroshima 722-8506, Japan\\
        }

%National Astronomical Observatory of Japan, National Institutes of Natural Sciences, 2-21-1 Osawa, Mitaka, Tokyo 181-8588, Japan

\email{xy.chen@astr.tohoku.ac.jp}

% %% Extra info for aastex:
% \received{Yesterday}
% \revised{Today}
% \accepted{Tonight}
% \published{Tomorrow}
% \submitjournal{AASJournal}

\begin{abstract}
We present the stellar population and ionized-gas outflow properties of ultra-luminous IR galaxies (ULIRGs) at $z=$\,0.1--1.0, which are selected from \akari\ FIR all-sky survey. We construct a catalog of 1077 ULIRGs to examine feedback effect after major mergers. 202 out of the 1077 ULIRGs are spectroscopically identified by SDSS and Subaru/FOCAS observations. Thanks to deeper depth and higher resolution of \akari\ compared to the previous \iras\ survey, and reliable identification from \wise\ MIR pointing, the sample is unique in identifying optically-faint (i$\sim$20) IR-bright galaxies, which could be missed in previous surveys. 
A self-consistent spectrum-SED decomposition method, which constrains stellar population properties in SED modeling based on spectral fitting results, has been employed for 149 ULIRGs whose optical continua are dominated by host galaxies. They are massive galaxies ($M_{\rm star}\sim10^{11}$--$10^{12}$\,\msun), associated with intense star formation activities (SFR\,$\sim$\,200--2000\,\sfrunit). The sample covers a range of AGN bolometric luminosity of $10^{10}$--$10^{13}$\,\lsun, and the outflow velocity measured from \oiiiblong\ line shows a correlation with AGN luminosity. Eight galaxies show extremely fast outflows with velocity up to 1500-2000 \kms. 
However, the co-existence of vigorous starbursts and strong outflows suggests the star formation has not been quenched during the ULIRG phase. By deriving stellar mass and mass fraction of young stellar population, we find no significant discrepancies between stellar properties of ULIRGs with weak and powerful AGNs. The results are not consistent with the merger-induced evolutionary scenario, which predicts that SF-dominated ULIRGs show smaller stellar mass and younger stellar populations compared to AGN-dominated ULIRGs. 
\end{abstract}

% http://journals.aas.org/authors/keywords2013.html

\section{Introduction}
\label{introduction}

Major mergers (interacting galaxies with mass ratio smaller than three, e.g., \citealp{Toomre1977}; \citealp{Mihos1996}) of gas-rich disk galaxies are considered as the predominant mechanism to form the most massive super-massive black holes (SMBHs) and spheroidal populations in the universe from the normal spiral disk galaxies with moderately luminous active galactic nuclei (AGNs; e.g., 
\citealp{Bridge2007, Kartaltepe2010, Kauffmann2000, Hopkins2006}). 
%$M_{\rm B}\ge -23$, or $L_{\rm AGN}\le 10^{12}$\lsun, where $M_{\rm B}$ and $L_{\rm AGN}$ are the B band absolute magnitude and the bolometric luminosity of AGN). 
According to the merger-induced scenario, ultra-luminous IR galaxies (ULIRGs, $L_{\rm 8-1000 \mu m} > 10^{12}$\,\lsun, e.g., \citealp{Sanders1988}) appear during the final coalescence of the galaxies after the massive inflows of cool gas triggering intense starbursts in the nuclear regions 
\citep[e.g.,][]{Sanders1996, Rigopoulou1999, Dasyra2006, Hopkins2008}. 
The massive inflowing gas also feed the rapid growth of SMBHs, 
although they are initially small compared to the newly forming spheroid. 
The intense star formation (SF) activities enhance the formation of dust grains, which in turn, attenuate the UV\,/\,optical radiation of young stars and AGNs and re-emit in IR\,/\,sub-millimeter (submm) wavelength range. 
As the result of the rapid growth of SMBHs, the ULIRGs migrate from the previous SF-dominated phase to the AGN-dominated phase, 
with the young poststarburst stellar populations and the bright, but highly dust-reddened quasars (e.g., red quasars, \citealp{Glikman2007}, \citealp{Urrutia2008}; extremely red quasars, \citealp{Ross2015}, \citealp{Goulding2018}). 
The luminous quasars could ignite powerful outflowing wind and expel the remaining gas and dust out of the galaxies. 
The lifetime of ULIRGs is limited (e.g., $<100$\, Myr, \citealp{Inayoshi2018}) due to the quick consumption of gas by the intense star formation and the dispersion of gas by the supernovae- and\,/\,or AGN-driven outflow, and the galaxies eventually evolve into the elliptical galaxies with old stellar population which host the typical un-obscured quasars. 

%The merger-induced scenario suggests the evolution path from star formation dominated ULIRGs to AGN-dominated ULIRGs. 
%The AGN-dominated ULIRGs are associated with strong outflow, which is expected to blow out the gas and dust out of the galaxy and terminate the star formation as well as the growth of the SMBH. 
Recently, signatures of outflowing gas in multi-phase have been found in various ULIRGs 
(neutral gas, \citealp{Rupke2011}, \citealp{Perna2015}; %Na I D absorption line observation
ionized gas, \citealp{Soto2012}, \citealp{RodriguezZ2013}, \citealp{Toba2017}; 
and molecular gas, \citealp{Veilleux2013}, \citealp{Saito2017}). %CO and OH
Furthermore, spatially-resolved spectroscopic observations of ULIRGs show signatures that the outflows affect gas not only in their central nuclear regions ($<1$ kpc scale), but also in their outer regions (1--10 kpc scale). 
For example, \citet{Westmoquette2012} presented the integral field spectroscopic maps of 18 southern ULIRGs and found that 11 of them show evidence of spatially extended outflowing warm ionized gas. 
\citet{Harrison2012} presented kinematic measurements of eight ULIRGs at $1.4<z<3.4$ and reported that in four of them strong outflow signatures of \oiiiblong\ emission line with full width at half-maximum (FWHM) of 700--1400 km s$^{-1}$ extends to a large scale (4--15 kpc) from the nucleus with large velocity offsets from the systemic redshifts (up to 850 km s$^{-1}$). 
\citet{Chen2019} reported a discovery of a ULIRG which shows a fast outflow ($v_{\rm out} \sim 2000$\,\kms) extending to a radius of 4 kpc. 
Those observations support the idea that outflow plays an important role in the transition of a ULIRG from a vigorous starburst to a quiescent galaxy. 

The outflow in a ULIRG is expected to quench its vigorous star formation, as a result, it is expected that the early stage, SF-dominated ULIRGs have younger stellar population than in the late stage, AGN-dominated ULIRGs \citep{Netzer2007, Veilleux2009}. 
\citet{Hou2011} investigated the stellar population of 160 \iras-selected ULIRGs and found that the mean stellar age and stellar mass increase from SF- to AGN-dominated ULIRGs. 
However, the evolution sequence was questioned by other works on stellar populations of ULIRGs 
(e.g., \citealp{Rodriguez2010, Su2013}), making the evolutionary path of ULIRGs still not conclusive. 
A systematic research on the properties of a large sample of ULIRGs is still required for better understanding of the evolutionary path of ULIRGs, and their role in the co-evolution of SMBHs and massive galaxies. 

In this paper, in order to constrain the stellar population and outflow properties of ULIRGs, we apply a self-consistent spectrum-SED analysis for 149 \akari-selected ULIRGs whose optical spectra are dominated by their host galaxies.
The construction of this ULIRG sample is introduced in Section \ref{sec:chap2_sample}. 
The details of the self-consistent spectrum-SED analysis method are explained in Section \ref{sec:chap2_method}. 
In Section \ref{sec:chap2_results}, we report the results on the properties of AGN, host galaxy, and outflowing gas. 
The evolutionary scenario of the ULIRG population is discussed in Section \ref{sec:chap2_discussions} and the conclusion 
is summarized in \ref{sec:chap2_conclusions}. 
Throughout the paper we adopt the cosmological parameters, $H_0=$ 70 \kms\,Mpc$^{-1}$, $\Omega_m=0.3$ and $\Omega_{\Lambda}=0.7$. 

\section{Sample Construction}
\label{sec:chap2_sample}

\subsection{Selection of FIR sources from \akari\ catalog}
\label{subsec:akari_FIR_catalog}

%\akari\ (previously known as \textit{ASTRO-F}; \citet{Murakami2007}) is the second Japanese space mission for infrared astronomy, after the Infrared Telescope in Space (\textit{IRTS}; Murakami 1997) in 1995. The satellite was launched in 2006 February 22, and imaged the sky at the four FIR bands using Far-Infrared Surveyor (FIS) and additionally with the Infrared Camera (IRC) at 9 and 18 $\mu$m. 

The \akari\ all-sky survey possesses higher spatial resolution than the previous FIR all-sky survey by \iras\ (e.g., 0.5--0.8$^{\prime}$ compared to 6$^{\prime}$), and is also deeper in wavelength around 90--100 $\mu$m. 
The \akari\ Far-Infrared Surveyor (FIS) Bright Source Catalogue (Ver.2, hereafter FISBSCv2\footnote{https://www.ir.isas.jaxa.jp/AKARI/Observation/update/20160425\_\\ preliminary\_release.html}, Yamamura et al., 2016) 
contains total 918056 sources observed at 65, 90, 140, and 160 $\mu$m, in which 501444 sources are detected in the main catalog\footnote{The source in FISBSCv2 main catalog is confirmed in two or more bands, or detected in four or more scans in one band, while the source in FISBSCv2 supplemental catalog is detected only in one band and in three or two scans. See release note of FISBSCv2 catalog for details. 
}. 
FISBSCv2 covers 98\% of the sky at 65 and 90\,$\mu$m and 99\% of the sky at 140  and 160 $\mu$m. 
The detection limit is 2.4, 0.44, 3.4, and 7.5 Jy at 65, 90, 140, and 160 $\mu$m, respectively. 
In particular, the \akari\ FIR survey provides the deepest data in terms of the FIR all-sky survey and thus provides a unique dataset to construct a large sample of ULIRGs \citep[e.g.,][]{Goto2011, Kilerci2014, Toba2017a}. 

Since the Wide-S band (90\,$\mu$m) is deeper than the other three bands of \akari/FIS, 
we select FIR sources from the entire \akari\ catalog (main + supplemental) with signal-to-noise ratio (S\,/\,N) at least 3 and high quality (fqual90 = 3) in the Wide-S band. 
In order to avoid the contamination of 
Galactic Cirrus emission, 
%the star-forming clouds in the Milky Way, 
%and nearby galaxies, e.g., LMC and SMC, 
%Galactic objects in low galactic latitude regions, 
we limited the sample to sources 
%with Galactic extinction $E(B-V) < 0.10$, and 
at least 5 arcmin away from any dusty regions with Galactic extinction $E(B-V) > 0.10$ using the Galactic foreground dust map from \citet{Schlegel1998} and updated by \citet{Schlafly2011}. 
Additionally, in order to avoid the contamination from FIR flux of nearby bright galaxies, we ignore the \akari\ sources within 0.5 arcmin to the edge of those galaxies, which is selected from HyperLEDA database \citep{HyperLEDA} with diameter over 0.5 arcmin in the major axis and total I band magnitude brighter than 15.
The size of the nearby bright galaxy in the direction to a given \akari\ source is estimated assuming an elliptical morphology. 
Finally 72950 out of 918056 \akari\ 90\,$\mu$m sources are selected following the above conditions. 

\subsection{Cross-match of \akari\ sources with SDSS and \wise\ catalogs}
\label{subsec:cross_match}

The 72950 selected \akari\ 90\,$\mu$m sources are then cross-matched with the Sloan Digital Sky Survey (SDSS) and Wide-field Infrared Survey Explorer (\wise) catalogs to identify their optical and MIR counterparts. 
\wise\ possesses deep detection depth ($\sim1$ mJy in w3 band) and high spatial resolution ($\sim6\arcsec$ in w1-w3 bands), and is useful to narrow down the positional uncertainty of the \akari\ FIR sources. 
%We firstly select the candidates of optical and MIR counterparts of the \akari\ sources from the Sloan Digital Sky Survey (SDSS) and Wide-field Infrared Survey Explorer (\wise) catalogs. 

SDSS Data Release 15 (DR15, \citealp{SDSSDR15}) is the third data release of SDSS-IV, which contains the re-calibrated SDSS imaging catalogs, using the hyper-calibration to PanSTARRS-1 implemented by \citet{Finkbeiner2016}. 
%the most current reprocessed imaging and spectra from the SDSS legacy survey and the extended Baryon Oscillation Spectroscopic Survey (eBOSS). The catalog
The limiting magnitudes (95\% completeness for point sources) are 22.0, 22.2, 22.2, 21.3, and 20.5 in \textit{u}, \textit{g}, \textit{r}, \textit{i}, and \textit{z} bands, respectively.
We retrieved model magnitude (called modelMag) in the five bands from SDSS DR15 SkyServer\footnote{http://skyserver.sdss.org/dr15}. 
The model magnitude is measured consistently through the same model profile in all bands, 
which is a good estimate of the total flux for extended sources, and provides the estimate of unbiased colors of galaxies. 
The sources are selected with S\,/\,N\,$>5$ in \textit{i} band (i.e., modelMagerr\_i\,$<$\,0.22). 
In order to avoid objects with unreliable photometry, 
we focus on the objects with the photometric calibration status in \textit{i} band and matches the clean photometry flags\footnote{https://www.sdss.org/dr15/tutorials/flags/}. 
%We also require that the SDSS objects are at least 5 arcmin far from the sky regions with Galactic extinction $E(B-V) > 0.10$, and finally 245182336 objects are selected. 
The SDSS model magnitude was converted to flux following \citet{SDSSEDR}. 
The Galactic extinction in the five bands is corrected with the dust map of \citet{Schlafly2011} and the extinction law of \citet{Fitzpatrick1999}. 
%The Galactic extinction in SDSS pipeline is computed with the dust map of \citet{Schlegel1998}. However, several recent studies reported that the map of \citet{Schlegel1998} over-estimates $E(B-V)$ by about 14\% \citep{Schlafly2011,Yuan2013}. 
%Therefore we corrected the photometry of the selected objects for the Galactic extinction using 86\% of the extinction correction from the original values. 

\wise\ performed an all-sky survey at 3.4 $\mu$m (w1), 4.6 $\mu$m (w2), 12 $\mu$m (w3), and 22 $\mu$m (w4) with angular resolutions of 6.1, 6.4, 6.5, and 12.0 arcsecond, respectively \citep{Wright2010}.
We queried the NASA/IPAC Infrared Science Archive (IRSA) for w1-w4 profile-fit magnitude from the AllWISE Source Catalog 
\citep{Cutri2014}. 
%using a searching radius of 4 arcsecond for the selected SDSS sources. Duplicate matches were allowed in the query. 
In order to obtain reliable entries in the AllWISE catalog, the selection is limited to sources with S\,/\,N\,$\ge5$ in the w3 band. In addition, we only select sources which are with depth-of-coverage at least 5 (w3m\,$\ge5$) and are not flagged as spurious detections of image artifacts in any band (cc\_flags = `0000'). 
%11603456 \wise\ objects are selected following the above conditions and at least 5 arcmin away from the sky regions with Galactic extinction $E(B-V) > 0.10$. 
The \wise\ Vega magnitude was converted to flux densities with the flux corrections and color corrections reported in \citet{Wright2010}. 
We corrected for the Galactic extinction in the w1 and w2 bands using the dust map of \citet{Schlafly2011} and the extinction law of \citet{Indebetouw2005}. %modified \citet{Schlegel1998} dust map \citep{Schlafly2011,Yuan2013} and the ``Cardelli, Clayton \& Mathis'' (CCM) extinction law \citep{Cardelli1989}. 

\begin{figure}
    \begin{center}
    \includegraphics[trim=0 30pt 0 0, width=\columnwidth]{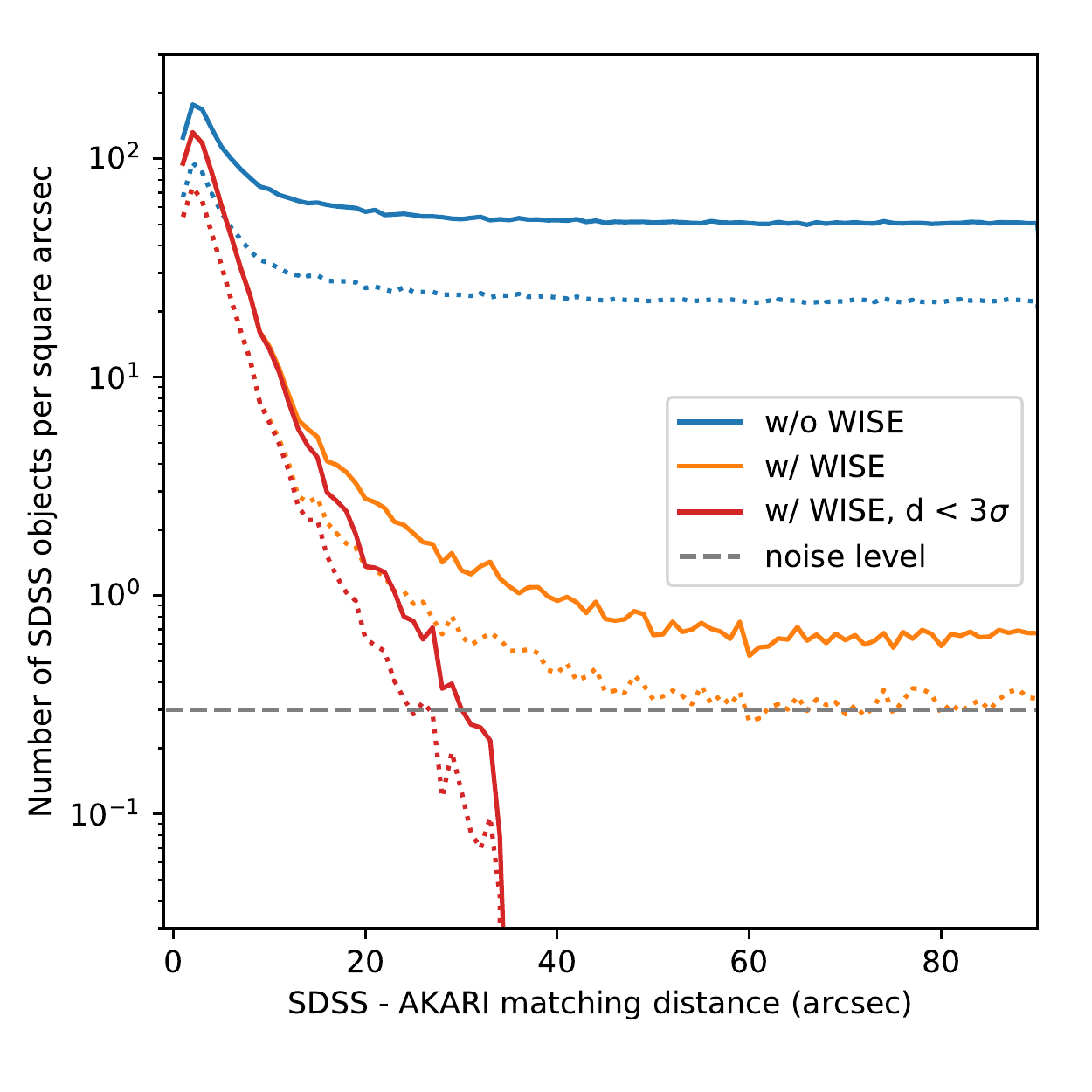}
    \end{center}
    \caption{
    Number of SDSS counterparts per square arcsecond of the entire \akari\ sample as a function of matching radius (blue). 
    The associations which only contain the SDSS objects associated with \wise\ w3 band detections are shown in orange. 
    The SDSS-\wise\ associations within a searching radius of $3\sigma_{\rm pos}$ are shown in red. 
    The solid and dotted curves denote the objects in the entire searching sky area and objects in a rectangular SDSS survey region (RA\,$=120\arcdeg$--$250\arcdeg$, Dec\,$=0\arcdeg$--$60\arcdeg$), respectively. 
    The gray dashed line denotes the the contamination level for the cross-matching, i.e., $n_{\rm noise}*N_{\rm A}$, where $N_{\rm A}=11539$ is the number of \akari\ sources which have at least one SDSS counterparts associated with \wise\ detections in the rectangle region, and $n_{\rm noise}$ is the number of random SDSS-\akari\ associations per square arcsecond per \akari\ sources. 
    }
    \label{fig:match_density}
\end{figure}

Figure \ref{fig:match_density} shows the distribution of matching distance between \akari\ sources and their possible optical counterparts. Duplicate match, i.e., several SDSS objects associated with one \akari\ object, are included in Figure \ref{fig:match_density} (blue curve). In this work we only selected the SDSS objects associated with \wise\ w3 band detections (orange curve in Figure \ref{fig:match_density}) as the candidates of true SDSS-\akari\ associations. 
With a higher resolution ($\sim6\arcsec$ in w1-w3 bands) than \akari\ ($\sim40\arcsec$ in Wide-S band), the \wise\ detections can effectively reduce the contaminations from IR-faint SDSS sources. 
The \wise\ catalog is deep enough ($\sim1$ mJy in w3 band) to detect FIR-dominated starburst galaxies detected in the \akari\ Wide-S band.

\begin{figure}
    \begin{center}
    \includegraphics[trim=0 30pt 0 0, width=\columnwidth]{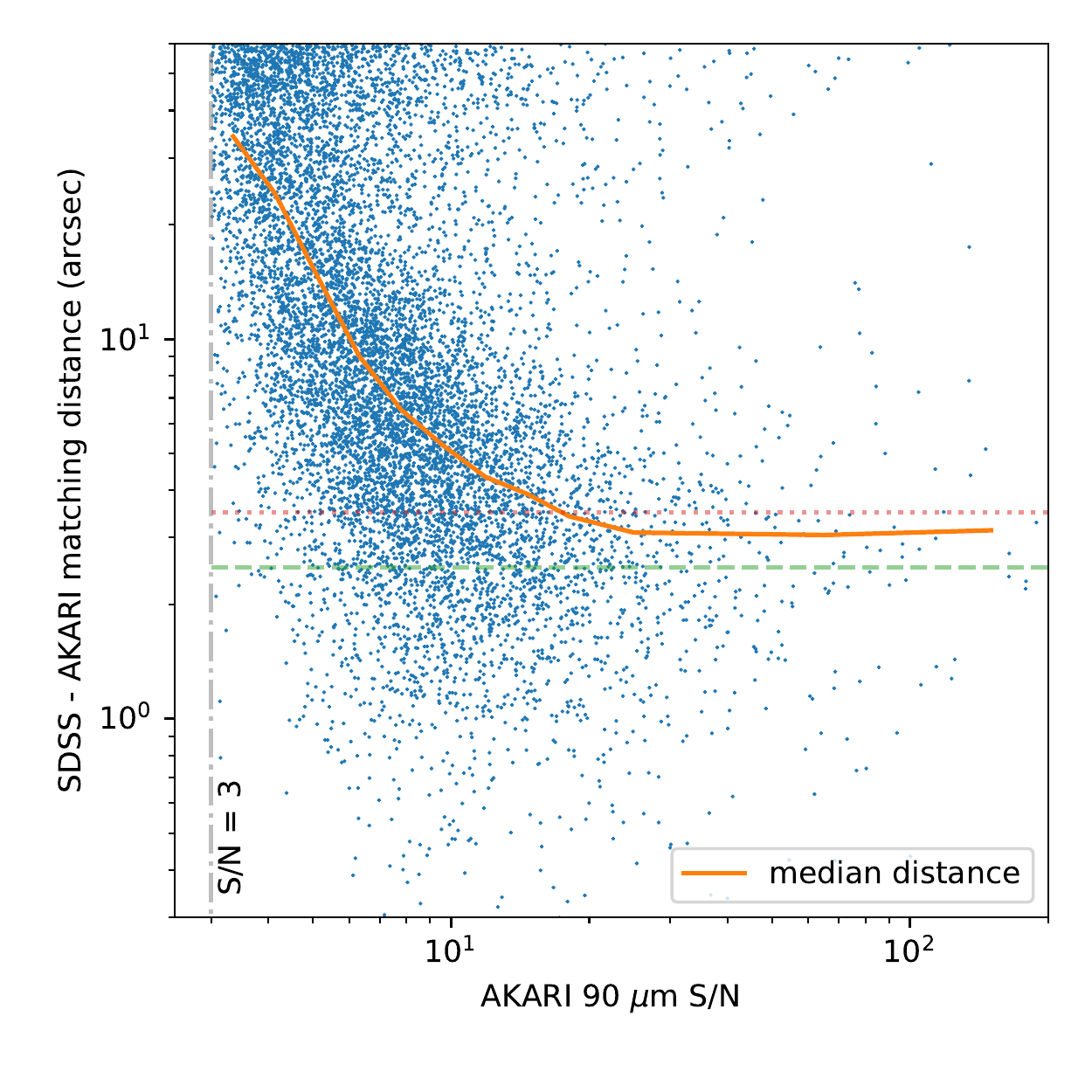}
    \end{center}
    \caption{
    Relationship between \akari\ 90\,$\mu$m S\,/\,N and SDSS-\akari\ matching distance. 
    Only 9544 unique matches (every \akari\ sources has only one SDSS counterpart associated with \wise\ detection) within $60\arcsec$ are shown in the plot (blue dots). 
    The median distance per logarithmic spaced bin are shown in orange solid line, which is used to estimate the positional uncertainty. 
    The horizontal lines show the \akari\ positional error levels in major ($3.5\arcsec$, red dotted) and minor ($2.5\arcsec$, green dashed) reported by Yamamura et al. (2016), respectively. 
    }
    \label{fig:match_poserr}
\end{figure}

The positional uncertainties of \akari\ observations are required to determine a searching radius for the cross-matching. 
Yamamura et al. (2016) cross-matched the FISBSCv2 catalog with 2MASS Redshift Survey Catalog (2MRS, \citealp{2MRS}) within $30\arcsec$ and found that 9128 \akari\ sources have 2MRS counterparts. The positional errors ($3.5\arcsec$ in major and $2.5\arcsec$ in minor axis) recorded in \akari\ catalog is the standard deviation of the positional differences of the 9128 \akari-2MRS associations, which could be a good proxy of the positional uncertainty of bright sources. However, ULIRGs usually appear at flux range close to the detection limit, the positional uncertainties of these faint sources could be larger. 
Figure \ref{fig:match_poserr} shows the distribution of the SDSS-\akari\ matching distance as a function of \akari\ 90\,$\mu$m S\,/\,N with a searching radius of $60\arcsec$. 
Only 9544 unique matches, i.e., only one SDSS objects found within $60\arcsec$ for each \akari\ source, are plotted in Figure \ref{fig:match_poserr}. 
The distribution indicates the positional uncertainty depends on the 90\,$\mu$m S\,/\,N. 
Since the coordinate differences of the matching in RA and Dec ($d_{\rm RA}$ and $d_{\rm Dec}$) follow a Gaussian distribution with mean of zero and variance of $\sigma^2$, the average of the matching distance with the definition, $[d^2_{\rm RA} + d^2_{\rm Dec}]^{1/2}$, in each S\,/\,N bin can be good estimates of $\sigma$.
%which means that we could estimate the positional uncertainty for each \akari\ source with its 90\,$\mu$m S\,/\,N. 
We derive the median distance in each logarithmically spaced S\,/\,N bin (orange curve in Figure \ref{fig:match_poserr}) and the positional error ($\sigma_{\rm pos}$) for each \akari\ source is estimated with linear interpolation, and then employ $3\sigma_{\rm pos}$ as the searching radius for SDSS-\akari\ associations. 
14680 SDSS objects are selected for 12310 \akari\ sources within $3\sigma_{\rm pos}$ radius.

In order to discuss the completeness of the cross-matching for \akari\ catalog, we focus on a rectangular sky region (RA\,$=120\arcdeg$--$250\arcdeg$, Dec\,$=0\arcdeg$--$60\arcdeg$) which is fully covered by SDSS survey, in which 5641 \akari\ sources show SDSS-\wise\ counterparts within $3\sigma_{\rm pos}$ radius. 
Considering the total number of \akari\ sources in the focused region, the $3\sigma_{\rm pos}$ selection covers $5641/11539\simeq49\%$ of the sources in the entire \akari\ catalog. 
The matching ratios are $3922/5477\simeq72\%$ and $1719/6062\simeq28\%$ for \akari\ main and supplemental catalogs, respectively. 
For the remaining 1558 (4345) \akari\ sources without SDSS-\wise\ counterparts in the main (supplemental) catalog, 
205 (296) \akari\ sources do not show any SDSS counterparts within $3\sigma_{\rm pos}$ matching radius; 
158 (342) \akari\ sources do not show any \wise\ counterparts within $3\sigma_{\rm pos}$ matching radius; 
513 (315) and 444 (196) \akari\ sources are removed due to bad image quality in the SDSS \textit{i} band and \wise\ w3 band, respectively; 
35 (96) and 388 (3602) \akari\ sources are removed due to low S\,/\,N of SDSS-\wise\ counterparts in \textit{i} band and w3 band, respectively. 
If we ignore the failed cases due to bad image quality in the \textit{i} band or w3 band, the completeness is 84\% and 31\% for FISBSCv2 main and supplemental catalogs, respectively.
If we employ $5\sigma_{\rm pos}$ selection radius and again ignore the failed cases due to bad image quality in the \textit{i} band or w3 band, 
the completeness of supplemental catalog increases to 39\% but it keeps the same for main catalog. 
The origins of the un-matched \akari\ sources could be the cirrus dust emission heated by remote stars in the Milky Way, or the artificial objects in FISBSCv2 catalog, e.g., side-lobe false sources. 

The reliability of the cross-matching is evaluated in two ways, i.e., the contamination level from matching by chance, and the likelihood of true associations in matching with multiple objects. 
We again focus on the rectangular sky region (RA\,$=120\arcdeg$--$250\arcdeg$, Dec\,$=0\arcdeg$--$60\arcdeg$) to determine the contamination level. 
The contamination level of the cross-matching for the entire \akari\ sample in the focused RA and Dec ranges can be described as $n_{\rm noise}*N_{\rm A}$, where $N_{\rm A}=11539$ is the total number of \akari\ sources in the rectangular region. 
$n_{\rm noise}$ is the number density of random SDSS-\akari\ associations per square arcsecond per \akari\ sources, which is estimated to be $0.3/N_{\rm A}\simeq2\times10^{-5}\, \mathrm{arcsec}^{-2}$ from Figure \ref{fig:match_density}. 
The expectation of random associations for an \akari\ source within a given searching radius ($3\sigma_{\rm pos}$) can be assumed as $9 \pi \sigma_{\rm pos}^2 n_{\rm noise}$. 
If $n_{\rm match}$ SDSS objects with \wise\ detections are found within $3\sigma_{\rm pos}$, for each of the SDSS objects with \wise\ detections the probability of a stochastic match is $9 \pi \sigma_{\rm pos}^2 n_{\rm noise} / n_{\rm match}$, which is used to evaluate the contamination level of a given association. 
97\% of the cross-matched SDSS counterparts within $3\sigma_{\rm pos}$ shows contamination level smaller than 3\%.% (Figure \ref{fig:contamination_level}). 

For multiple SDSS objects associated with a given \akari\ source, the likelihood ratio of each SDSS object can be defined as (e.g., \cite{Hwang2007}): 
\begin{equation}
    LR_{\rm SA} = \frac{\exp{(-\frac{r_{\rm SA}^2}{2\sigma_{\rm SA}^2})}}{N_{i<i_0}},
    \label{equ:LR_SA}
\end{equation}
where $r_{\rm SA}$ is the distance for each SDSS-\akari\ association; 
$\sigma_{\rm SA}^2=\sigma_{\rm SDSS}^2 + \sigma_{AKARI}^2$, where $\sigma_{\rm SDSS}$ and $\sigma_{AKARI}$ are the positional errors of SDSS and \akari\ sources; 
$N_{i<i_0}$ denotes the number of galaxies with \textit{i} band flux brighter than this object with \textit{i} band magnitude $i_0$. 
The probability of a SDSS candidate to be a real counterpart can be defined as $p_{\rm SA}=LR_{\rm SA}/\sum_{i=1}^{i=N}{ LR_{\rm SA, i} }$, where $N$ denotes the number of SDSS candidates associated with a given \akari\ source. 
With \wise\ photometry as an intermediate step to connect SDSS and \akari\ objects, we can also calculate the likelihood ratios for the SDSS-\wise\ matches and \wise-\akari\ matches using similar definitions as:
\begin{equation}
    LR_{\rm SW} = \frac{\exp{(-\frac{r_{\rm SW}^2}{2\sigma_{\rm SW}^2})}}{N_{i<i_0}}, \ \ \ \ 
    LR_{\rm WA} = \frac{\exp{(-\frac{r_{\rm WA}^2}{2\sigma_{\rm WA}^2})}}{N_{\rm w3<w3_0}}, 
    \label{equ:LR_SW_WA}
\end{equation}
where $r_{\rm SW}$ and $r_{\rm WA}$ are the matching distances for each SDSS-\wise\ and \wise-\akari\ association; 
$\sigma_{\rm SW}^2=\sigma_{\rm SDSS}^2 + \sigma_{WISE}^2$ and $\sigma_{\rm WA}^2=\sigma_{WISE}^2+\sigma_{AKARI}^2$ , where $\sigma_{WISE}$ is the positional error of \wise\ photometry; 
$N_{\rm w3<w3_0}$ denotes the number of galaxies with w3 band flux brighter than this object. 
$N_{i<i_0}$ and $N_{\rm w3<w3_0}$ can be estimated from the cumulative functions of magnitude distributions in \textit{i} and w3 bands for the entire SDSS and \wise\ samples. % selected in this work (Figure \ref{fig:SDSS_WISE_dist}). 
The probabilities of SDSS-\wise\ matches and \wise-\akari\ matches can be estimated by normalizing the $LR_{\rm SW}$ and $LR_{\rm WA}$, respectively, i.e., $p_{\rm SW}=LR_{\rm SW}/\sum_{i=1}^{i=N_{\rm SW}}{ LR_{\rm SW, i} }$, where $N_{\rm SW}$ is the number of SDSS candidates associated with a given \wise\ source within $3\arcsec$; and $p_{\rm WA}=LR_{\rm WA}/\sum_{i=1}^{i=N_{\rm WA}}{ LR_{\rm WA, i} }$, where $N_{\rm WA}$ is the number of \wise\ candidates associated with a given \akari\ source within $3\sigma_{\rm pos}$. 
%Note that if several SDSS objects are associated with the same \wise\ counterparts, e.g., nearby interacting galaxies, they share the same weight of \wise\ flux and \wise-\akari\ distance in the calculation of the likelihood ratio. 
%Because of $\sigma_{\rm SDSS} << \sigma_{WISE}$ and $\sigma_{WISE} << \sigma_{AKARI}$, we assume that $\sigma_{\rm SW} \simeq \sigma_{WISE}$ and $\sigma_{\rm WA} \simeq \sigma_{AKARI}$ in the calculation. 
In order to take advantage of the better positional accuracy of \wise\ ($<0.5\arcsec$) to reduce the number of SDSS candidates and enlarge the cross-matched catalog, we employ the selection threshold of $p_{\rm SW}>0.9$ and $p_{\rm WA}>0.9$, instead of $p_{\rm SA}>0.9$. 
%For a given \akari\ source, the proper \wise\ counterpart is selected with $p_{\rm WA}>0.9$, and then the proper SDSS counterpart is selected with $p_{\rm SW}>0.9$. 
%The better positional accuracy of \wise\ ($<0.5\arcsec$) is helpful to reduce the number of SDSS candidates in the crowded region. 
The selection provides a reliable catalog with 11245 SDSS-\wise-\akari\ associations out of 
12310 \akari\ sources with 14680 SDSS counterparts. %, which are shown as green dots in Figure \ref{fig:match_LR}. 
%The distribution of likelihood and contamination level of the entire cross-matched sample within $3\sigma_{\rm pos}$ are shown in (Figure \ref{fig:SDSS_WISE_dist}). Since the contamination level is relatively small, we only set the selection threshold with likelihood, i.e., $p \ge 90\%$, for the duplicate associations. 
%The selection provides a reliable cross-matched catalog with 15362 \akari\ sources. %Figure \ref{fig:match_LR2} 

In addition to the likelihood method with \wise\ detections, the sample is also cross-matched with Faint Images of the Radio Sky at Twenty Centimeters (FIRST) survey catalog\footnote{http://sundog.stsci.edu/first/catalogs/readme\_14dec17.html} of radio sources in 1.4 GHz. 
%using the results from Kimball \& Ivezic (2014). 
759427 FIRST sources are selected with P(S)\,$< 0.1$, where P(S) indicates the probability that the source is spurious, e.g., a sidelobe of a nearby bright source. %The distribution of converted AB magnitude is shown in Figure \ref{fig:SDSS_WISE_dist}. 
Similar to Equation \ref{equ:LR_SW_WA}, we also calculate the likelihood ratio and probabilities for SDSS-FIRST matches ($p_{\rm SF}$) and FIRST-\akari\ matches ($p_{\rm FA}$), respectively. 
Among the entire SDSS-\akari\ associations, 4557 galaxies are identified with FIRST detections, i.e., $p_{\rm SF}>0.9$ and $p_{\rm FA}>0.9$. 
355 out of the 4557 galaxies are ruled out in the above SDSS-\wise-\akari\ identifications with a low likelihood, i.e., $p_{\rm SW}<0.9$ or $p_{\rm WA}<0.9$. 
We add 341 out of the 355 new SDSS-\akari\ associations to the matched catalog, except for 14 conflicting cases, which means that for a given \akari\ source the cross-matching with FIRST observation indicates a different primary SDSS counterpart from the one selected with \wise\ identification 
(i.e., in the conflicting case we choose \wise\ identified sources instead of FIRST identified sources). 
Finally we obtain a cross-matched catalog with 11586 galaxies. 

\begin{figure}
    \begin{center}
    \includegraphics[trim=0 30pt 0 0, width=\columnwidth]{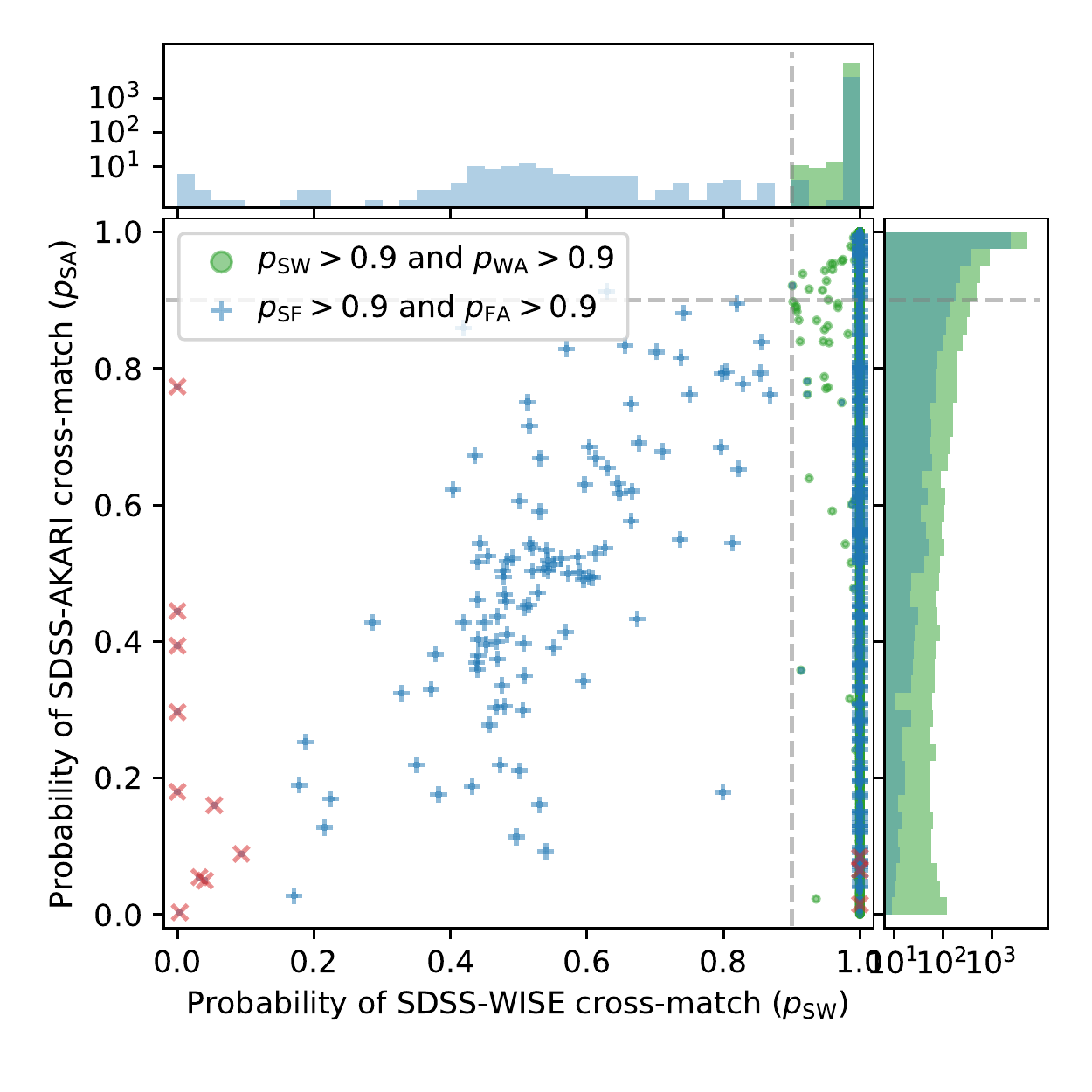}
    \end{center}
    \caption{
    Distribution of probabilities of SDSS-\wise\ matches ($p_{\rm SW}$) and SDSS-\akari\ matches ($p_{\rm SA}$). 
    The green dots and histogram denote the associations selected with $p_{\rm SW}>0.9$ and $p_{\rm WA}>0.9$, where $p_{\rm WA}$ is the probabilities of \wise-\akari\ matches. 
    The blue crosses and histogram show the FIRST identified objects with $p_{\rm SF}>0.9$ and $p_{\rm FA}>0.9$. 
    %The new associations with FIRST detections (not overlapped with green dots) are emphasized with blue circles. 
    The red `$\mathrm{\times}$' markers denote the SDSS objects for which the FIRST detections show different primary sources for \akari\ sources from SDSS-\wise-\akari\ associations. We adopt the \wise\ identified SDSS-\akari\ matches in the conflicting cases, see text for details. 
    }
    \label{fig:match_LR2}
\end{figure}

Among the 11586 sources, 5247 objects are spectroscopically identified in SDSS legacy and BOSS surveys. 
The SDSS spectroscopic objects are selected with median S\,/\,N (snMedian) $>2$ and plate quality of `good' or `marginal', as well as with zWarning\,$< 2^7$ (not no-data for the fiber, bad targeting nor unplugged fiber) and zWarning\,$\ne 1$ (to avoid too little wavelength coverage). 
The SDSS photometric redshift is available for 6119 out of the 6339 sources without spectroscopic observation, which is estimated using the kd-tree nearest neighbor fit method\footnote{https://www.sdss.org/dr12/algorithms/photo-z/}. The redshift information is not available for the remaining 220 sources, which are removed in the later analyses. 

In order to obtain photometric information covering wavelength range as wide as possible, 
we also cross-matched the above catalog with Two Micron All Sky Survey Point Source Catalog (2MASS PSC, \citealp{Skrutskie2006}), 
Infrared Astronomical Satellite Faint Source Catalog (\iras\ FSC, \citealp{Moshir1990}), 
\herschel/PACS Point Source Catalog (\herschel/PACS PSC\footnote{https://doi.org/10.5270/esa-rw7rbo7}), 
and \herschel/SPIRE Point Source Catalog (\herschel/SPIRE PSC\footnote{https://doi.org/10.5270/esa-6gfkpzh}). 
The cross-matching conditions and results are summarized in Table 1. 

\begin{table}[ht]
\centering
\caption{Cross-matching with 2MASS/\iras/\herschel\ Catalogs \label{table: match_number}}
\begin{tabular}{ccc}
\hline
\hline
Catalog & Matching radius & Matched number \\
\hline
2MASS PSC & -$^{1}$ & 9410 (Ks band) \\
\iras\ FSC & $3\sigma_{\iras}$$^{2}$ & 5812 (60\,$\mu$m) \\
\herschel/PACS PSC & $6\arcsec$ (160\,$\mu$m)$^{3}$ & 219 (160\,$\mu$m) \\
\herschel/SPIRE PSC  & $9\arcsec$ (250\,$\mu$m)$^{4}$ & 1096 (250\,$\mu$m) \\
\hline
\multicolumn{3}{l}{$^{1}$\footnotesize{ We adopt the results from SDSS archive.}} \\
\multicolumn{3}{l}{$^{2}$\footnotesize{ $\sigma_{\iras}$ is the length of semi major axis of \iras\ positional elliptical.}} \\
\multicolumn{3}{l}{$^{3,4}$\footnotesize{ The radius is taken as the half of the beam size.}} \\
\end{tabular}
\end{table}

\subsection{Selection of ULIRGs with 2-band estimated $L_{\rm IR}$}
\label{subsec:2_band}

\begin{figure}
    \begin{center}
    \includegraphics[trim=0 20pt 0 20pt, width=\columnwidth]{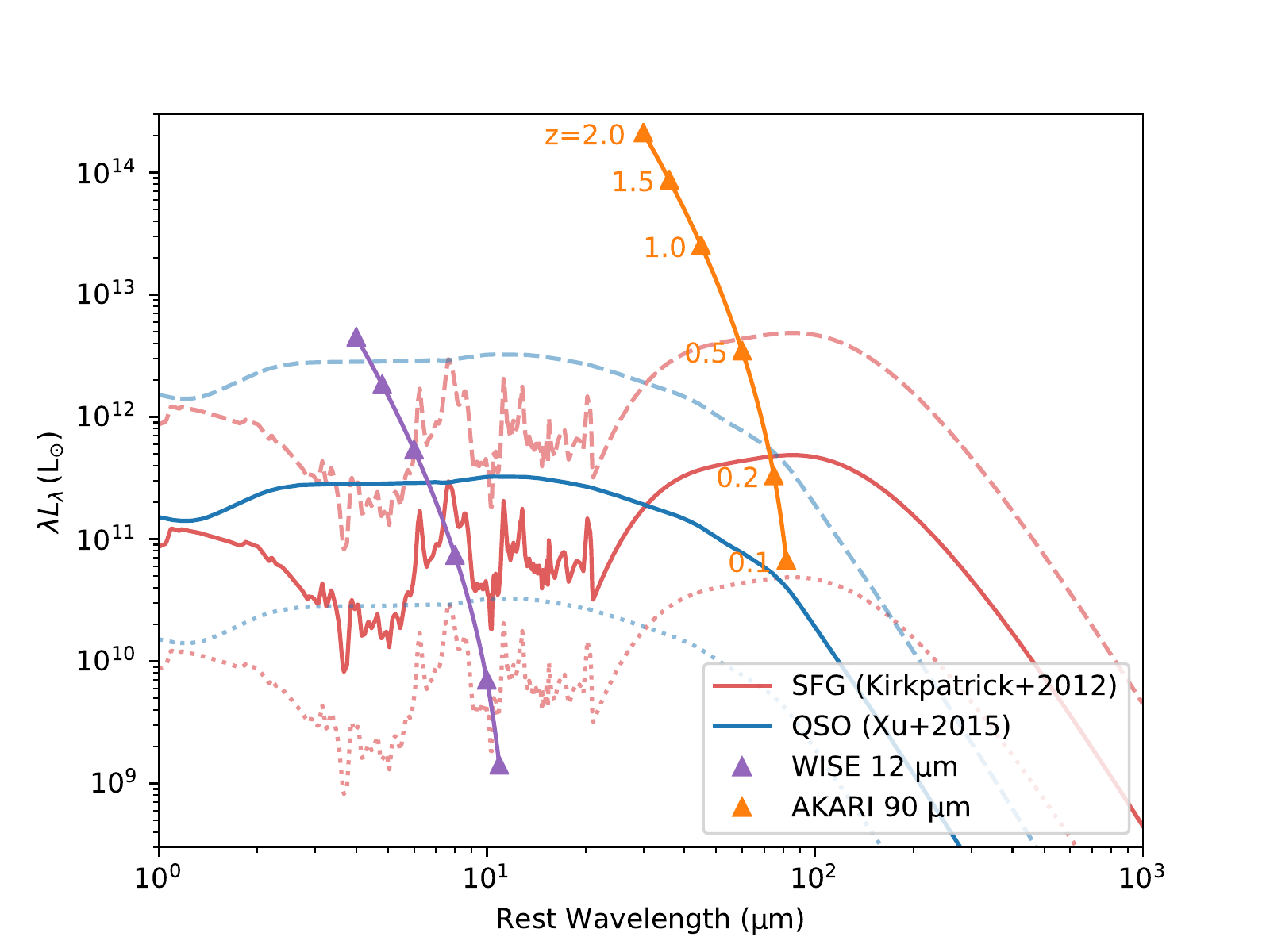}
    \end{center}
    \caption{SED templates of quasar (blue, \citealp{Xu2015}) and starbursts (red, \citealp{Kirkpatrick2012}) used in the 2-band IR luminosity estimation. 
    The thick curves denote the SED templates normalized to integrated luminosity at 1--1000 $\mu$m of ULIRG level ($10^{12}$\,\lsun), while the dotted and dashed curved denote the templates normalized to LIRG ($10^{11}$\,\lsun) and HyLIRG level ($10^{13}$\,\lsun), respectively, 
    The violet and orange triangles show the detection limit of \wise\ w3 band (12\,$\mu$m) and \akari\ Wide-S band (90\,$\mu$m) at redshift from 0.1 to 2.0.
    }
    \label{fig:SED2_example}
\end{figure}

Here we summarize a crude method to select ULIRGs from the cross-matched catalog. 
The total IR luminosity of a galaxy can be crudely estimated with the observed flux densities on the assumption that the galaxy follows an empirically derived spectral energy distribution (SED). In order to estimate the contribution of AGN to the total IR luminosity, we adopt two SEDs, one for starburst component and the other for AGN torus component. The starburst template is taken from the composite SED of 39 star-forming galaxies at $z\sim1$, which was created by 
\citet{Kirkpatrick2012} by stacking \textit{Spitzer} MIR spectroscopy and \herschel\ FIR photometry. The AGN torus template is taken from the typical quasar SED built by 
\citet{Elvis1984}, which was modified by 
\citet{Xu2015} by removing the IR contribution of star formation from the original SED. Both of the two templates are shown in Figure \ref{fig:SED2_example}. We use the fluxes in w3 (12\,$\mu$m) and Wide-S (90\,$\mu$m) bands to evaluate the 
relative contributions of the two templates of the total SED. 
%intensity of both of the two templates by convolving the templates with the transmission curve in each band. 
The IR luminosity of AGN ($L_{\rm AGN}$) and star-formation ($L_{\rm SF}$) can be estimated by solving the following equation: 
\begin{equation}
\begin{split}
    & L_{\rm AGN} \frac{\int S_{\rm AGN}(z) T_{\rm i}\,\mathrm{d}\lambda}{\int T_{\rm i} c/\lambda^2\,\mathrm{d}\lambda} + L_{\rm SF} \frac{\int S_{\rm SF}(z) T_{\rm i}\,\mathrm{d}\lambda}{\int T_{\rm i} c/\lambda^2\,\mathrm{d}\lambda} = 4\pi D_{\rm L}^2(z)F_{\rm i},\\
    \label{equ:SED_2_band_1}
\end{split}
\end{equation}
where $i = 12,\, 90\,\mu\mathrm{m}$ indicates w3 and Wide-S bands; $S_{\rm AGN}(z)$ and $S_{\rm SF}(z)$ means the templates which are redshifted to the observed frame and normalized to unit integrated luminosity at 1--1000 $\mu$m; $T_{\rm i}$ and $F_{\rm i}$ are the transmission curves and fluxes in each band; c is speed of light and $D_{\rm L}(z)$ is the luminosity distance, respectively. 
The spectroscopic redshift is employed for the objects with SDSS spectral observations. 
%For the objects with only imaging data, 
We adopt the SDSS photometric redshifts for the objects with only imaging data.

\begin{figure}
    \begin{center}
    \includegraphics[trim=0 30pt 0 0, width=\columnwidth]{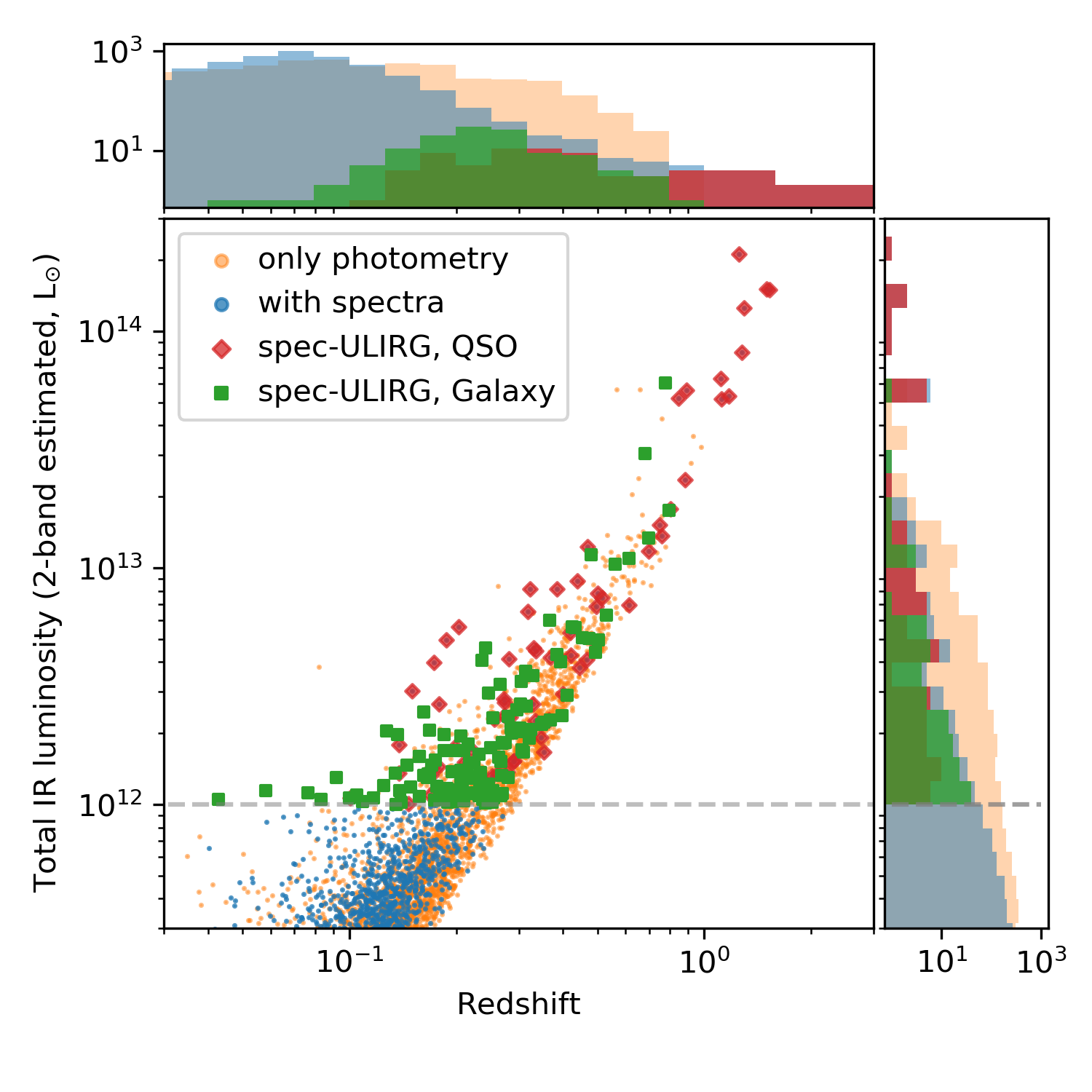}
    \end{center}
    \caption{
    Distribution of the redshift (\textit{x}-axis) and the total IR luminosity (\textit{y}-axis) estimated using the 2-band estimation method. 
    The objects with SDSS spectroscopy are shown with blue dots, and the objects with only photometric identification are shown in orange dots. 
    The ULIRGs identified as quasars by SDSS pipeline are shown with red diamonds, while the ULIRGs identified as galaxies are shown with green squares. 
    The gray dashed line denotes the selection threshold of ULIRGs. Note that the two histograms are plotted in logarithmic grid.}
    \label{fig:SED2_z_LIR}
\end{figure}

The total IR luminosity and AGN contribution to the total IR luminosity can be calculated as:
\begin{equation}
\begin{split}
    & L_{\rm IR} = L_{\rm AGN} + L_{\rm SF},\ \ \ f_{\rm AGN} = L_{\rm AGN} / L_{\rm IR}, 
    \label{equ:SED_2_band_2}
\end{split}
\end{equation}
and the results are shown in Figure \ref{fig:SED2_z_LIR}. % and Figure \ref{fig:SED2_fAGN_LIR}. 
Within the entire cross-matched catalog, 1077 galaxies are identified as ULIRG with the threshold $L_{\rm IR} \ge 10^{12}$\,\lsun. In order to cover the NIR emission of AGN torus, the luminosity is integrated in the wavelength range of 1--1000\,$\mu$m, which is 
extended toward short wavelength range compared to the typically adopted wavelength range of 8-1000\,$\mu$m. 
195 out of the 1077 ULIRGs are spectroscopically identified in SDSS legacy and BOSS surveys with median S\,/\,N\,$>2$. 
%, in which the galaxy SDSS J073735.22+353621.0 is removed with visual check, due to possible contamination by cosmic ray in the spectrum. 
%The remaining 196 ULIRGs show an average snMedian (median S\,/\,N of the spectrum) of 13, in which 8 galaxies show snMedian of 2--3. 
For the 8 galaxies with $2<$\,S\,/\,N\,$<3$, we bin the observed spectra by two adjacent pixels in later analyses. 
All of the observed spectra are corrected for Galactic dust extinction with the Cardelli-Clayton-Mathis (CCM) extinction law \citep{Cardelli1989} and the dust map updated by \citet{Schlafly2011}. 

% \begin{figure}
%     \begin{center}
%     \includegraphics[trim=0 30pt 0 0, width=0.4\columnwidth]{XMC_ULIRG_fAGN_LIR.png}
%     \end{center}
%     %\caption{
%     \red{REMOVE}
%     Distribution of the total IR luminosity (\textit{y}-axis) and AGN contribution to the total IR luminosity (\textit{x}-axis) estimated using the 2-band estimation method. The legends are the same as those in Figure \ref{fig:SED2_z_LIR}. 
%     %}
%     \label{fig:SED2_fAGN_LIR}
% \end{figure}

The majority of the ULIRGs ($\sim90\%$) show $f_{\rm AGN} \lesssim 0$, indicating that their IR emission is dominated by star formation activity. 
A negative $f_{\rm AGN}$ suggests that the starburst template overestimates the MIR luminosity compared to the observed SED
%(Figure \ref{fig:SED2_example_negLAGN})
, which could be due to the higher fraction of polycyclic aromatic hydrocarbon (PAH) component of the template
, or due to the residual AGN contribution in starburst template (with average of about 10\%, \citealp{Kirkpatrick2012}). 
We use the $f_{\rm AGN}$ from 2-band estimation with fixed template for illustrative purposes only. 
The $f_{\rm AGN}$ estimated with a detailed SED analysis will be discussed in Section \ref{subsec:results_AGN}. 

\begin{figure}
    \begin{center}
    \includegraphics[trim=0 30pt 0 0, width=\columnwidth]{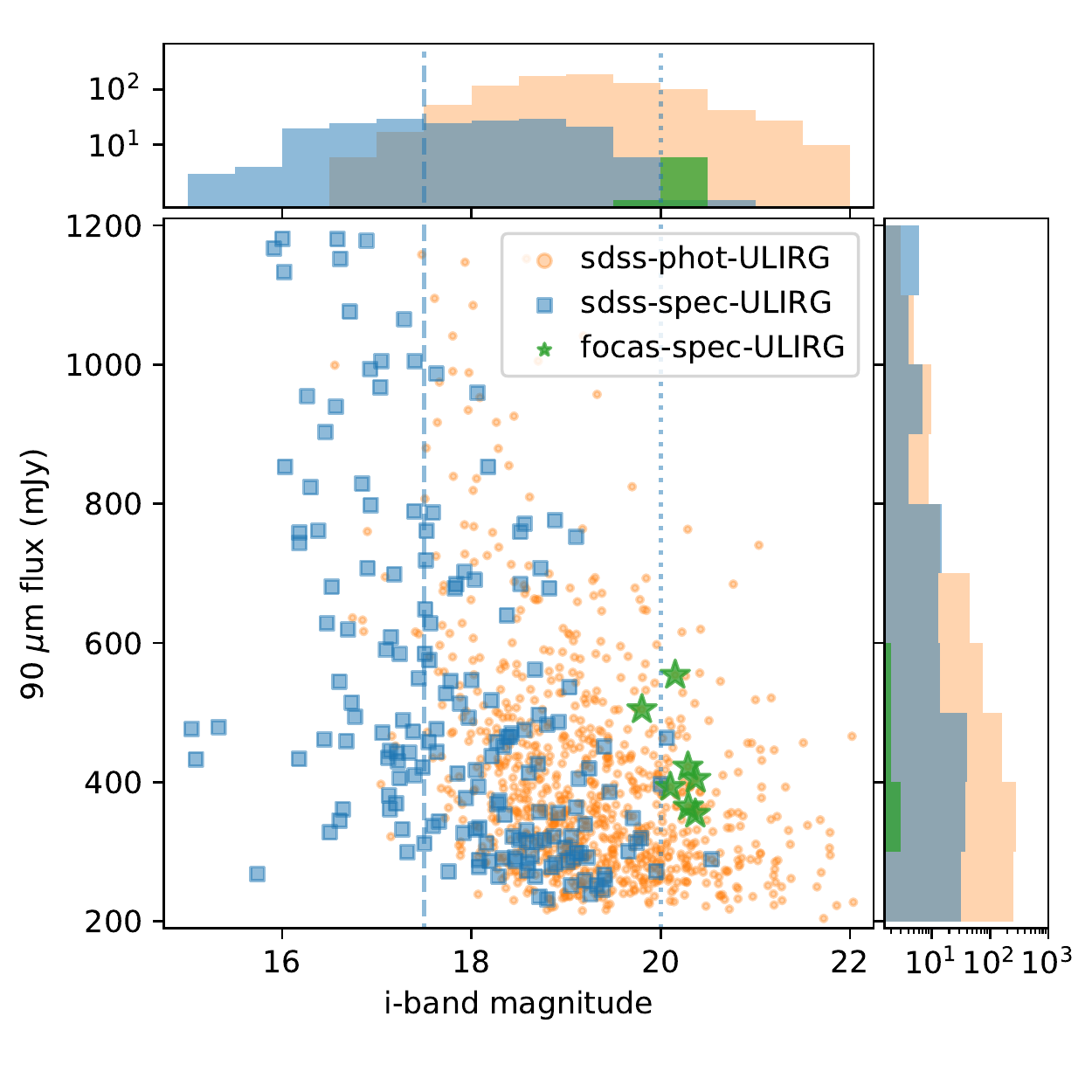}
    \end{center}
    \caption{Distribution of \textit{i} band magnitude and 90\,$\mu$m flux of \akari-selected ULIRGs, where the ULIRGs with SDSS spectroscopy are shown in blue squares, and those with only photometric identification are shown in orange dots. 
    The green stars denote the ULIRGs spectroscopically identified by FOCAS. 
    %The gray histograms denote the overlapped region of blue and orange histograms. 
    Note that the histograms are plotted with numbers in logarithmic grid.}
    \label{fig:ULIRG_flux_dist2}
\end{figure}

We construct a 90\,$\mu$m flux limited ULIRG sample through the above cross-matching and luminosity estimation. 
However, the flux limit of SDSS spectroscopic survey results in incompleteness of the sample of spectroscopically identified ULIRGs. 
The distribution of \textit{i} band magnitude and 90\,$\mu$m flux of the ULIRG sample are shown in Figure \ref{fig:ULIRG_flux_dist2}, in which the blue dashed and dotted lines denote the flux limit of SDSS legacy spectroscopic survey ($i<17.5$) and BOSS galaxy survey ($i<20.0$). The completeness of the optically bright subsample ($i<17.5$) is 78\% (83/106), while that of the optically-faint sample ($17.5<i<20.0$) is 12\% (112/971). 

In order to understand the properties of extremely optically-faint ULIRGs at intermediate redshifts ($0.5<z<1$), we are conducting an optical follow-up program for \akari-selected ULIRGs with $20.0<i<21.0$ using FOCAS on the Subaru telescope. Seven objects are observed in a service program in S17A (S17A0216S, PI:Masayuki Akiyama). The details of the data reduction are described in \citet{Chen2019}. 
Finally we use all of the 202 spectroscopically identified ULIRGs (195 from SDSS and BOSS, 7 from the FOCAS observation) as the final sample for later analyses. %, for which the completeness of about 19\% (202/1077). 

%%%%%%%%%%%%%%%%%%%%%%%%%%%%%%%%%%%%%%%%%%%%%%%%%%%%%%%%%%%%%%%%%%%%%%%%%%%%%%%%%%%%%%%%%%%%%%%%%%%%%%%%%%%%%%%%%%%%%%%%%%%%%
%%%%%%%%%%%%%%%%%%%%%%%%%%%%%%%%%%%%%%%%%%%%%%%%%%%%%%%%%%%%%%%%%%%%%%%%%%%%%%%%%%%%%%%%%%%%%%%%%%%%%%%%%%%%%%%%%%%%%%%%%%%%%
%%%%%%%%%%%%%%%%%%%%%%%%%%%%%%%%%%%%%%%%%%%%%%%%%%%%%%%%%%%%%%%%%%%%%%%%%%%%%%%%%%%%%%%%%%%%%%%%%%%%%%%%%%%%%%%%%%%%%%%%%%%%%
%%%%%%%%%%%%%%%%%%%%%%%%%%%%%%%%%%%%%%%%%%%%%%%%%%%%%%%%%%%%%%%%%%%%%%%%%%%%%%%%%%%%%%%%%%%%%%%%%%%%%%%%%%%%%%%%%%%%%%%%%%%%%

\section{A self-consistent spectrum-SED decomposition Method}
\label{sec:chap2_method}

In order to determine the properties of the outflowing gas, the stellar population, the star formation rate, as well as the AGN luminosity, 
we develop a method to connect the optical spectral fitting and SED decomposition to obtain a self-consistent result. 
For simplicity, we briefly introduce the spectral fitting method in Section \ref{subsec:spectral_fitting}, and then introduce the SED decomposition as well as the spectrum-SED connection in Section \ref{subsec:SED_decomp}. A more detailed description of the method can be found in Appendix \ref{sec:Appendix_specfit} and \ref{sec:Appendix_sedfit}.  

\subsection{Optical spectral fitting}
\label{subsec:spectral_fitting}

The optical spectral fitting code is modified from Quasar Spectral Fitting package (\texttt{QSFit}, v2.0.0, 
\citealp{Calderone2017}\footnote{http://qsfit.inaf.it}). 
\texttt{QSFit} is an IDL program based on MPFIT \citep{Markwardt2009}, which uses the Levenberg-Marquardt technique to solve the least-squares problem. 
The original code of \texttt{QSFit} is specialized to the analysis of spectra of quasar-dominated systems, e.g., power-law continuum with broad Balmer lines. 
We modified the codes to support the fitting with host galaxy component with multiple stellar populations as well as the automatic separation between quasar spectra and host galaxy spectra under a given threshold. 
The fitting procedure is also optimized for the detection of outflowing gas emission lines. 

The model used to fit the observed spectrum is a collection of several components, which can be classified into three categories: (1) AGN continua, e.g., power-law continuum; (2) host galaxy continua, e.g., stellar continuum; (3) emission lines, e.g, \ha\ and \oiii. The AGN continua consist of a power-law continuum representing the radiation originating from the accretion disc, a Balmer continuum with blending high order Balmer emission lines (H11-H50), and a blending iron emission line pseudo-continuum. 
In order to model the star formation history (SFH) of a galaxy in the spectral fitting, we introduce two stellar populations: one is an underlying main stellar population (MS) and the other represents the ongoing starburst population (SB). 
The stellar library of \citet{Bruzual2003} with a \citet{Salpeter1955} initial mass function (IMF) and a solar metallicity is adopted in this work. 
The SFH of MS and SB components is described with star formation rate (SFR) as a function of time: 
$\mathrm{SFR}_{\rm MS}(t)=\mathrm{SFR}_{\rm MS}(t_0)\exp{(-(t-t_0)/\tau_{\rm SF})}$ and $\mathrm{SFR}_{\rm SB}(t)=\mathrm{constant}$, respectively. 
In the fitting procedure we accounts for most of the emission lines which could be covered by the observed spectrum from \hbox{Ly$\alpha$} to \hbox{[S\sc iii] 9531\AA}. The majority of the emission lines are described with a combination of
a narrow profile to account for the emission from \hii\ region and\,/\,or AGN narrow line region (NLR), and a broad profile to represents the outflowing gas and\,/\,or the emission line from AGN broad line region (BLR). 
In order to calculate uncertainties of the best-fit parameters, we use Monte Carlo resampling method. 
30 mock spectra are generated for each observed spectrum and similar fitting processes are performed for the mock spectra. 
Hence the uncertainty of a given parameter can be estimated from the distribution of the best-fit values of the mock spectra. 
The details of the fitting procedure are explained in Appendix \ref{sec:Appendix_specfit_process}. 

\begin{figure}
    \begin{center}
    \includegraphics[trim= 0 10pt 0 0, width=\columnwidth]{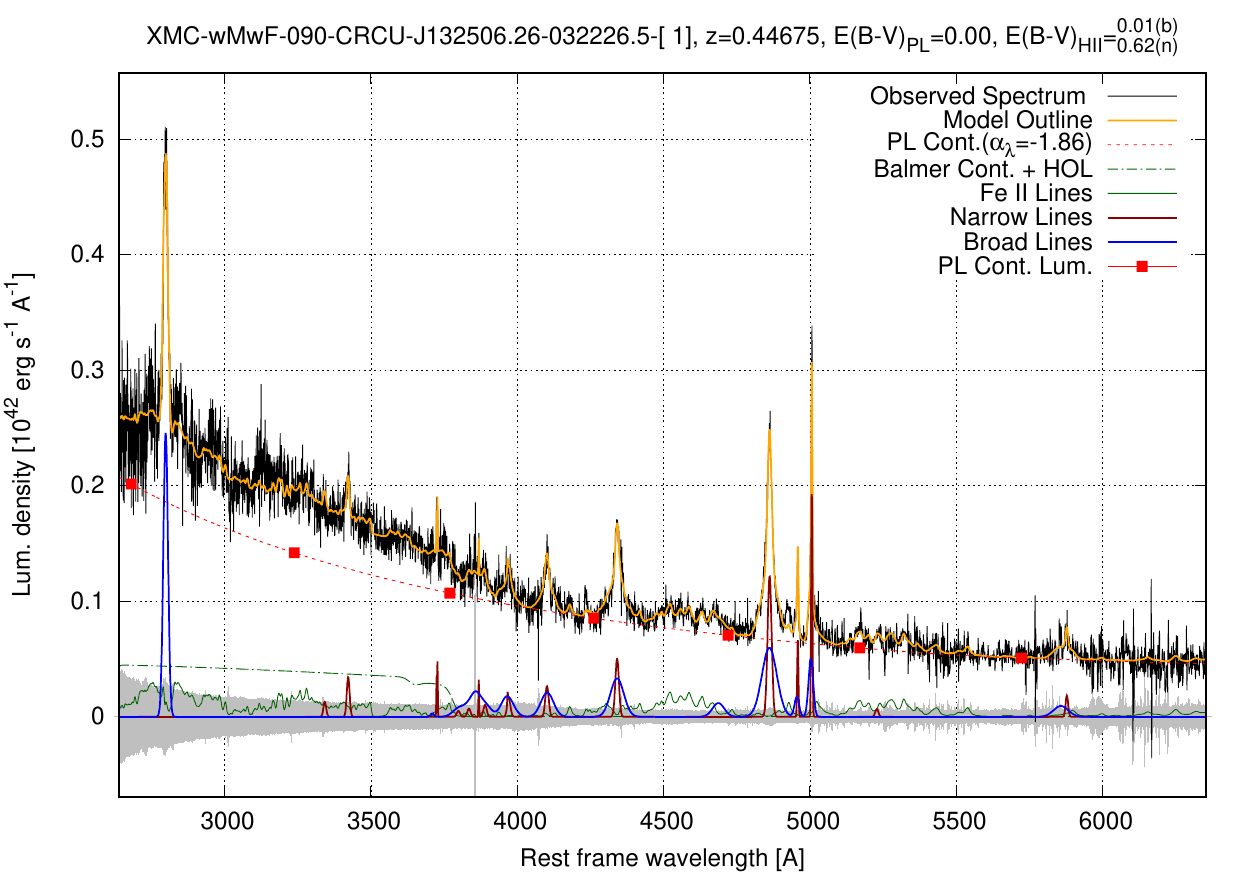}
    \includegraphics[trim= 0 10pt 0 0, width=\columnwidth]{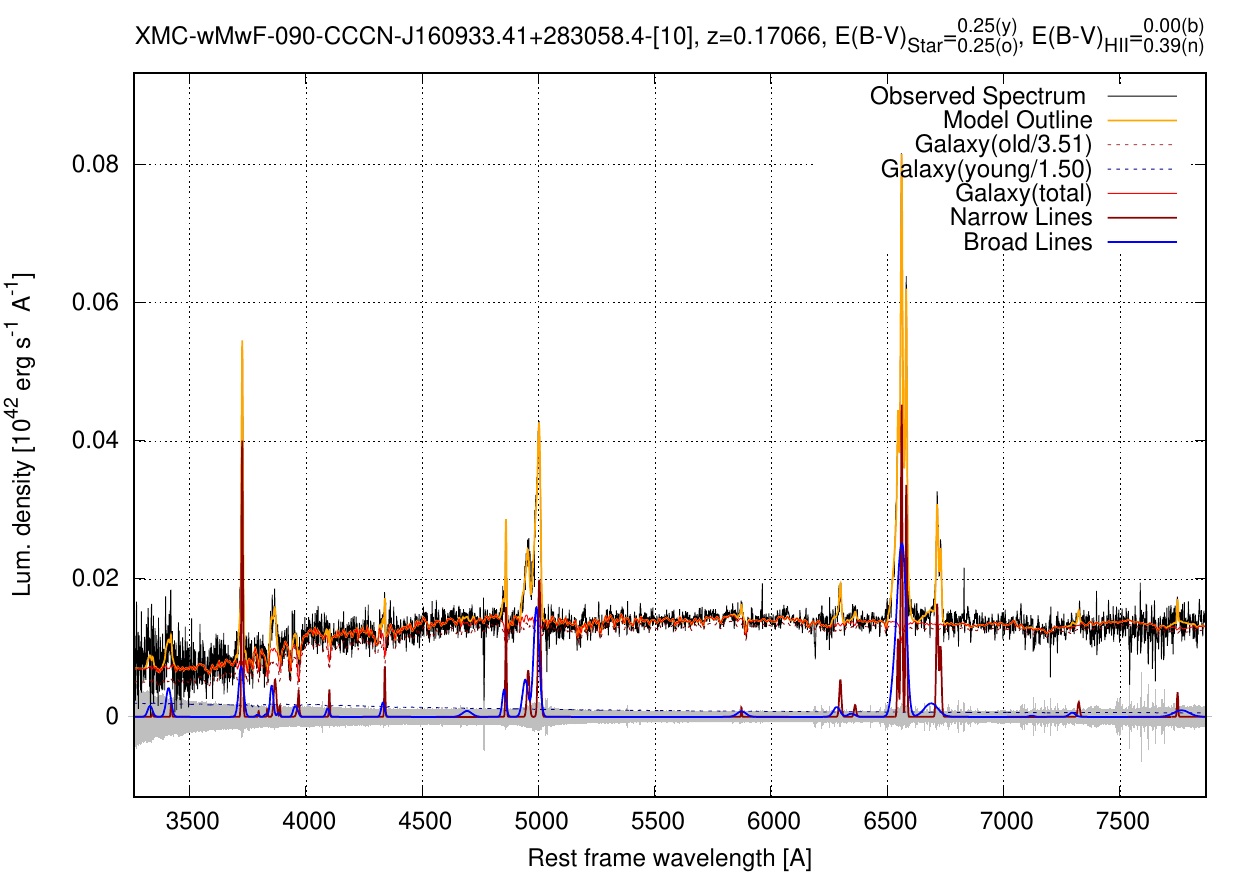}
    \end{center}
    \caption{
    Example of a quasar dominated optical spectrum (J132506.26-032226.5, upper) and a host galaxy dominated optical spectrum (J160933.41+283058.4, lower). 
    The observed spectrum is shown in black and the measurement error is shown in gray at $y=0$. 
    The power-law continuum is shown in red dotted line. 
    The green dash-dotted and solid curve denote the Balmer continuum and iron pseudo-continuum, respectively. 
    The red and blue dotted curves denote the stellar continua of MS and TSB components, respectively. 
    The red and blue solid curves show the narrow and broad components of emission lines, respectively. 
    }
    \label{fig:Spec_QSO}
\end{figure}

We identify the optical spectra as quasar or host galaxy (hereafter HG) dominated for each ULIRG in the sample. 
The quasar dominated spectra are separated from the HG dominated ones using two indicators of AGN BLR feature. 
The first BLR indicator is the sum of equivalent width (EW) of observed \feii\ and Balmer (pseudo-) continua, i.e., $\mathrm{EW}_{\rm FeII+BAC}=\mathrm{EW}_{\rm FeII}+\mathrm{EW}_{\rm BAC}$, to the underlying AGN power-law and stellar continua. 
The second BLR indicator is the 80\% width of the line profile for permitted lines, i.e., $w_{80,\rm permitted}=w_{80, \rm H\beta},\, z\le 0.8$ and $w_{80,\rm permitted}=w_{80, \rm MgII},\, z > 0.8$, where the choice of \hb\ and \mgii\ depends on the wavelength coverage. 
A spectrum is identified as quasar dominated (e.g., Figure \ref{fig:Spec_QSO}, upper) if it shows significant Balmer and iron (pseudo-) continua, i.e., $\mathrm{EW}_{\rm FeII+BAC}\ge 100$\,\AA\ (e.g., \citealp{Boroson1992}), or shows significantly broad permitted lines, i.e., $w_{80,\rm permitted}\ge 3000$\,\kms. 
Since the identification is based on BLR features, we also check the fitting quality with and without power-law continuum if the spectrum only show weak BLR features, i.e., $\mathrm{EW}_{\rm FeII+BAC}< 100$\,\AA\ and $w_{80,\rm permitted} < 3000$\,\kms.  
If the $\chi^2$ without power-law continuum exceeds at least three times of $\chi^2$ with power-law continuum, i.e., the spectral fitting fails without power-law continuum, we classify the spectrum belongs to a `weak BLR quasar'
\footnote{A possible explanation is that the extinction of the BLR is much heavier than the accretion disc. It can be also due to the beaming of the nuclear component, e.g., BL Lac. }
, otherwise the spectrum is host galaxy dominated (e.g., Figure \ref{fig:Spec_QSO}, lower). 

% \begin{figure}
%     \begin{center}
%     \includegraphics[trim= 0 10pt 0 0, width=\columnwidth]{QSFit_HG_J1609.pdf}
%     \end{center}
%     \caption{
%     Example of host galaxy dominated optical spectrum for a ULIRG, J160933.41+283058.4. %J090307.84+021152.1. %082445.64+153944.1. 
%     The red and blue dotted curves denote the best-fit continuum of MS and TSB, respectively. 
%     Other legends are the same as those in Figure \ref{fig:Spec_QSO}. }
%     \label{fig:Spec_HG}
% \end{figure}

\begin{figure*}
    \begin{center}
    \includegraphics[trim= 0 30pt 0 0, width=0.8\textwidth]{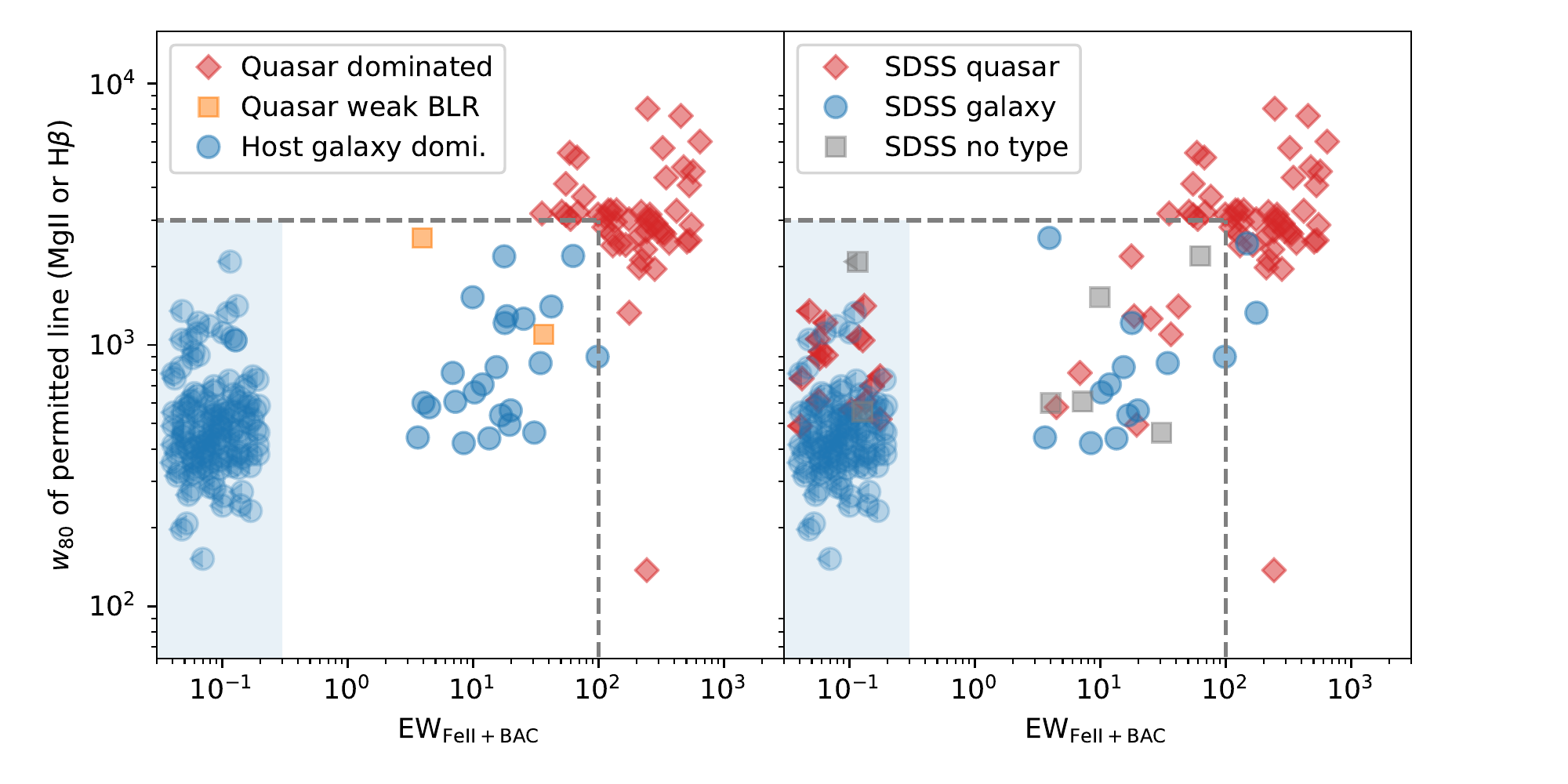}
    \end{center}
    \caption{Spectral identification for \akari-selected ULIRGs. 
    The left panel shows the checking results according to the features of AGN BLR, 
    in which the \textit{x}-axis denotes the sum of equivalent width (EW) of observed \feii\ and Balmer (pseudo-) continua, i.e., $\mathrm{EW}_{\rm FeII+BAC}=\mathrm{EW}_{\rm FeII}+\mathrm{EW}_{\rm BAC}$, to the underlying AGN power-law and stellar continua; 
    the \textit{y}-axis denotes the 80\% width of permitted lines, i.e., $w_{80,\rm permitted}=w_{80, \rm H\beta},\, z\le 0.8$ and $w_{80,\rm permitted}=w_{80, \rm MgII},\, z > 0.8$, where the choice of \hb\ and \mgii\ depends on the wavelength coverage. 
    A spectrum is identified as quasar dominated (red diamonds) if it shows $\mathrm{EW}_{\rm FeII+BAC}\ge 100$\,\AA\ or $w_{80,\rm permitted}\ge 3000$\,\kms. Orange squares denote two spectra with weak BLR features, however a power-law continuum is required to obtain a good spectral fitting. The pure galaxies are shown with blue circles. See text for details. 
    Note that for galaxies in the blue hatched region no \feii\ nor Balmer (pseudo-) continua are required in the fitting, i.e., $\mathrm{EW}_{\rm FeII+BAC}=0$. 
    Random shift in \textit{x}-axis is added for such galaxies for the purpose of illustration. 
    The right panel shows the identifications from SDSS archive with the same axes as those in the left panel. 
    }
    \label{fig:Spec_type}
\end{figure*}

Finally 53 out of the 202 spectroscopically observed ULIRGs are identified as quasar dominated, in which 2 ULIRGs are identified as `weak BLR quasar', and the rest 149 ULIRGs are identified as HG dominated. The identification results and the comparison with SDSS archived spectral classification are shown in Figure \ref{fig:Spec_type}. 
23 ULIRGs which are classified as quasars by SDSS pipeline are re-identified as HG dominated by this fitting process. The different SDSS classification for 17 ULIRGs could be due to the mis-identification of the broad \oiii\ emission line with FWHM over 1000 \kms. 
Since the broad permitted lines and iron pseudo-continuum usually blend with the surrounding forbidden lines, e.g., \oiiiblong, making it difficult to determine the properties of outflow; and the presence of power-law continuum also affects the decomposition of stellar components, in the later analyses with an SED decomposition we only focus on the 149 ULIRGs with HG dominated optical spectra.

%%%%%%%%%%%%%%%%%%%%%%%%%%%%%%%%%%%%%%%%%%%%%%%%%%%%%%%%%%%%%%%%%%%%%%%%%%%%%%%%%%%%%%%%%%%%%%%%%%%%%%%%%%%%%%%%%%%%%%%%%%%%%
%%%%%%%%%%%%%%%%%%%%%%%%%%%%%%%%%%%%%%%%%%%%%%%%%%%%%%%%%%%%%%%%%%%%%%%%%%%%%%%%%%%%%%%%%%%%%%%%%%%%%%%%%%%%%%%%%%%%%%%%%%%%%

\subsection{Multi-band SED decomposition with optical spectral fitting results}
\label{subsec:SED_decomp}

Several fitting procedures have been developed in order to model the multi-band SEDs of galaxies, e.g., \texttt{MAGPHYS} \citep{daCunha2008} and \texttt{CIGALE} \citep{Noll2009, Boquien2019}. These codes usually invoke a huge (e.g., $\sim10^5$) library of stellar SED to fit the UV-optical radiation of stars and to obtain the physical information such as stellar mass and star formation rate. In order to take advantage of the results from the spectral fitting discussed above, e.g., stellar continuum decomposition and extinction, in this work we develop a new SED fitting code to obtain a self-consistent result between spectral and SED fitting analyses.  

A widely adopted strategy of multi-band SED decomposition is so-called `energy balance' which requires the conservation between the attenuated primary radiation from stars and\,/\,or AGN, and the re-emitted emission by the dust surrounding the primary emitters. 
We firstly consider the dust absorption and re-emission in host galaxies. The real conditions and properties of dust in the galaxies can be very complicated (e.g., \citealp{Galliano2008, Galliano2018}). 
In the fitting procedure we employ a two-components model following the assumption in the optical spectral fitting, i.e., an old main stellar component (MS) with exponential SFH, and an on-going starburst component (SB) with constant SFR.  
The optically observed SB components represents the young stars for which the natal clouds have been destroyed and migrate to the diffuse interstellar-medium (ISM) dust. For simplicity, hereafter we name the young stellar population embedded in optically-thin diffuse ISM as `transparent starburst' (TSB) component, which corresponds to the young stellar population in the optical spectra, and the young stars which are almost fully absorbed by optically-thick natal clouds in optical band as `attenuated starburst' (ASB). 
We assume the ASB component has the same stellar population as the TSB component, which can be estimated from optical spectral fitting. 
The TSB component is assumed to be embedded in diffuse ISM dust with the same extinction for MS component. 
Extinction of $A_{\rm V}=100$ is assumed for ASB component for all the galaxies, which corresponds to a typical dense collapsing cloud with spacial scale of 0.2 pc, $n_{\rm H2}=2\times10^5$\ccm, and V-band opacity of $3\times10^3$\,m$^2$/Kg \citep{Shu1987}. 

Here we employ the dust emission SED using 
The Heterogeneous Evolution Model for Interstellar Solids (THEMIS\footnote{https://www.ias.u-psud.fr/themis/THEMIS\_model.html}) dust model \citep{Jones2013,Kohler2014} updated by \citet{Nersesian2019}. 
The diffuse ISM dust heated by MS and TSB component is considered to be exposed in ambient starlight with a constant intensity, while the dust surrounding the young stars (ASB) is exposed in a power-law distributed starlight intensity, i.e., $\mathrm{d}M/\mathrm{d}U \propto U^{-\alpha}$ 
(e.g., \citealp{Dale2002, Dale2014, Draine2007}). 

In addition to the dust surrounding stars, the AGN can also contribute to the IR SED of the galaxy by the thermal emission from a thick layer of dust surrounding an accretion disc. The UV-optical radiation from accretion disk is absorbed by the torus and then re-emitted as torus thermal radiation. 
In this work, we employ the SKIRTor\footnote{https://sites.google.com/site/skirtorus/} torus model developed by 
\citet{Stalevski2012} and updated in \citet{Stalevski2016} using 3D Monte Carlo radiative transfer code SKIRT 
\citep{Baes2003, Baes2011}. The SKIRTor model consist of two-phase medium, i.e., a large number of high-density clumps embedded in a smooth dusty component of low density grains. 
As explained in Section \ref{subsec:spectral_fitting}, only torus emission models of type-2 AGN (
for which the optical spectrum is dominated by galaxy component) are used in the procedeeding analyses.  
%In the typical SED fitting codes, e.g., \texttt{CIGALE}, the luminosity of AGN is usually estimated by integrating the best-fit torus template assuming an isotropic radiation. However, both of the observational and simulation results show that the torus is an asymmetric structure (e.g., \citealp{Stalevski2017}). The SED of torus seen in face-on direction (type-1 case) has a bluer IR color than the SED seen in edge-on direction, because the dust in the inner region, which is heated to temperature up to 1000-1500\,K (dominating radiation around 3\,$\mu$m), can be directly observed in the line of sight of the face-on direction. 
The observed flux of the torus emission ($F_{\rm torus}$) can be estimated from the IR SED decomposition. 
Then the intrinsic luminosity of the AGN primary source ($L_{\rm AGN}$), i.e., the accretion disc, can be estimated from $F_{\rm torus}$
within the framework of SKIRTor torus model 
using the energy conservation equation (see Equation \ref{equ:EB_AGN} in Appendix \ref{sec:Appendix_sedfit_connection}). 
The most important parameter in the conversion from $F_{\rm torus}$ to $L_{\rm AGN}$ 
is the half opening angle $\Theta$ (or the covering factor of the torus). 
Since the torus is usually considered to be optically thick, half opening angle $\Theta$ determines the fraction of the primary radiation of AGN, which is absorbed and re-emitted by dusty torus. 
Recently, the study on X-ray selected AGN reported that the typical covering factor is about 0.6 \citep{Stalevski2016, Mateos2017, Ichikawa2019}, which corresponds to $\Theta\sim30\arcdeg$. 
%The relationship between $L_{\rm AGN, pri}$ and $L_{\rm torus}^{\rm iso}$ with $\Theta=30\arcdeg$ is shown in Figure \ref{fig:AGN_Torus_OA} (blue and cyan dots). In addition, we also plot the results with a smaller $\Theta=10\arcdeg$, which are shown in green and yellow-green dots in Figure \ref{fig:AGN_Torus_OA}. The comparison suggests that the energy conversion ratio from accretion disk to torus increases by nearly 10 times from $\Theta=10\arcdeg$ to $\Theta=30\arcdeg$. 
Since $\Theta$ highly affects the estimation of $L_{\rm AGN}$ and it is hard to be determined only with limited number of IR photometric data points, it is a proper choice to fix it to the typical value from the literature. 
The fitting with the fixed $\Theta=30\arcdeg$ results in an average 
absorption ratio of AGN primary radiation, i.e., $L_{\rm torus} / L_{\rm AGN}$, of 0.22. 

%With $\Theta=30\arcdeg$, we can derive the average ratio $4\pi D_{\rm L}^2 F_{\rm torus}/L_{\rm AGN, pri}\sim0.15$, which is consistent with the empirical relationship between the MIR and bolometric luminosities of AGN (e.g., \citealp{Ichikawa2014}); 
%and the ratio for the real anisotropic torus within SKIRTor model, $L_{\rm torus}^{\rm aniso}/L_{\rm AGN, pri}\sim0.21$. 
%With $\Theta=30\arcdeg$, we can derive the average ratio $4\pi D_{\rm L}^2 F_{\rm torus}/L_{\rm AGN, pri}\sim0.15\pm0.07$, which is consistent with the empirical relationship between the MIR and bolometric luminosities of AGN (e.g., \citealp{Ichikawa2014}); the ratio for the real anisotropic torus within SKIRTor model is $L_{\rm torus}^{\rm aniso}/L_{\rm AGN, pri}=0.21\pm0.01$. 
%Due to the degeneracy between inclination and torus half-opening angel ($\Theta$), and the limited number of data points, we fix $\Theta=30\arcdeg$ in according to the average dust covering factor of X-ray selected AGNs (e.g., \citealp{Ichikawa2019}). 

We consider the energy conservation in each emitter-absorber system including all of the primary radiation components, i.e., MS, TSB, ASB, and AGN (accretion disk), as well as the dust re-emitted components. i.e., ISM dust heated by MS and TSB, BC heated by ASB, and torus heated by AGN (See Equation \ref{equ:EB_HG}, \ref{equ:EB_AGN} in Appendix \ref{sec:Appendix_sedfit_connection} for details). 
Note that in the integration of absorbed luminosity by dust, we set a lower limit of wavelength of 912\AA, with an underlying assumption that most of the photons with energy higher than 13.6 eV are absorbed by the photoionized Hydrogen and Helium gas in \hii\ region, rather than attenuated by the dust \citep{Draine2007}. 

In order to connect the spectral fitting results to SED fitting process, we adopt the best-fit stellar continua and dust extinction as the input of SED fitting to constrain the stellar population. 
Two additional parameters, i.e., $C_{\rm ape, MS}$ and $C_{\rm ape, TSB}$, are employed to correct for the aperture-loss of optical spectroscopy for extended MS and TSB components, respectively. 
The best-fit emission line profiles from spectral fitting are also used to represent their contribution to the broad-band photometry. 
The average of $C_{\rm ape, MS}$ and $C_{\rm ape, TSB}$ is used to correct for the aperture-loss of narrow lines from \hii\ region and AGN NLR, and broad lines from outflowing gas. 
No aperture correction is considered for AGN continuum components and BLR emission lines. 

\begin{figure}
    \begin{center}
    \includegraphics[width=\columnwidth]{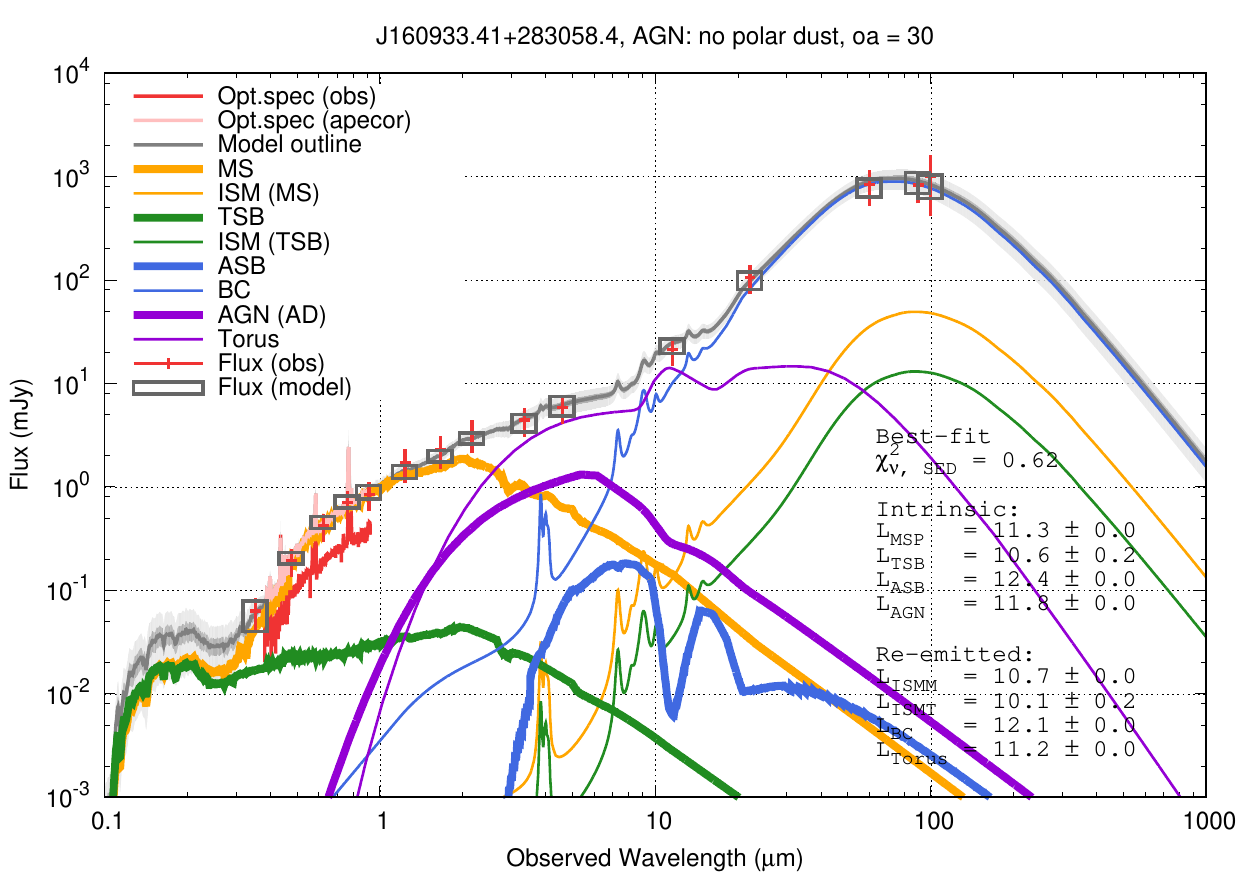}
    \end{center}
    \caption{
    Best-fit SED decomposition for a ULIRG, J160933.41+283058.4. %J090307.84+021152.1. %082445.64+153944.1. 
    The red and pink curve denote the observed (Figure \ref{fig:Spec_QSO}, lower) and aperture corrected optical spectrum, respectively. 
    The observed fluxes in each band are shown in red crosses with the error bar showing the $3\sigma$ measurement errors, 
    while the best-fit flux values marked using gray open squares. 
    The outline of the best-fit SED is shown in gray curve, while the SEDs of extinct primary radiation (i.e., MS, TSB, ASB, and AGN (accretion disc) and the dust re-emitted emission (i.e., ISM, BC, and, Torus) are shown in thick and thin curves, respectively, 
    The same color is used for the primary source and the related dust component. 
    }
    \label{fig:SSFIT_example}
\end{figure}

Following the fitting strategy of \texttt{CIGALE}, we add an additional error (10\% of the flux) to each photometric point to take into account uncertainties in the models in the minimization of $\chi^2$. 
Figure \ref{fig:SSFIT_example} shows one example of the best-fit SED model. 
In order to calculate uncertainties of the best-fit parameters, we use Monte Carlo resampling method. 
For each ULIRG, 100 mock photometric observations are generated by adding the random noise to the observed fluxes in each band. 
The random noise is Gaussian distributed with the amplitude from the measurement error in each band. 
The same fitting processes are employed for the mock photometric data and the scatters of the best-fit values are taken as the uncertainties of the parameters. 

%%%%%%%%%%%%%%%%%%%%%%%%%%%%%%%%%%%%%%%%%%%%%%%%%%%%%%%%%%%%%%%%%%%%%%%%%%%%%%%%%%%%%%%%%%%%%%%%%%%%%%%%%%%%%%%%%%%%%%%%%%%%%
%%%%%%%%%%%%%%%%%%%%%%%%%%%%%%%%%%%%%%%%%%%%%%%%%%%%%%%%%%%%%%%%%%%%%%%%%%%%%%%%%%%%%%%%%%%%%%%%%%%%%%%%%%%%%%%%%%%%%%%%%%%%%
%%%%%%%%%%%%%%%%%%%%%%%%%%%%%%%%%%%%%%%%%%%%%%%%%%%%%%%%%%%%%%%%%%%%%%%%%%%%%%%%%%%%%%%%%%%%%%%%%%%%%%%%%%%%%%%%%%%%%%%%%%%%%
%%%%%%%%%%%%%%%%%%%%%%%%%%%%%%%%%%%%%%%%%%%%%%%%%%%%%%%%%%%%%%%%%%%%%%%%%%%%%%%%%%%%%%%%%%%%%%%%%%%%%%%%%%%%%%%%%%%%%%%%%%%%%
%%%%%%%%%%%%%%%%%%%%%%%%%%%%%%%%%%%%%%%%%%%%%%%%%%%%%%%%%%%%%%%%%%%%%%%%%%%%%%%%%%%%%%%%%%%%%%%%%%%%%%%%%%%%%%%%%%%%%%%%%%%%%
%%%%%%%%%%%%%%%%%%%%%%%%%%%%%%%%%%%%%%%%%%%%%%%%%%%%%%%%%%%%%%%%%%%%%%%%%%%%%%%%%%%%%%%%%%%%%%%%%%%%%%%%%%%%%%%%%%%%%%%%%%%%%

\section{Properties of \akari-selected ULIRGs from spectrum-SED decomposition}
\label{sec:chap2_results}

%\subsection{Total IR luminosity and bolometric luminosity}
%\label{subsec:results_LIR_LBol}

The total IR (1--1000 $\mu$m) luminosity, i.e., $L_{1\textup{--}1000}$, can be estimated with the best-fit SED models. 
125 out of the 149 galaxies with galaxy dominated optical spectra (Section \ref{subsec:spectral_fitting}) 
show $L_{1\textup{--}1000}\ge 10^{12}$\,\lsun, 
indicating that the 2-band ULIRG selection method presented in Section \ref{subsec:2_band} is effective to select ULIRGs
\footnote{If we use the luminosity integrated in 8--1000 $\mu$m range, i.e., $L_{8\textup{--}1000}$, the number of selected ULIRGs is 111.}. 
The remaining 24 galaxies are luminous-LIRGs with a median integrated luminosity of $L_{1\textup{--}1000}= 10^{11.9\pm0.1}$\,\lsun. 
We keep all of the 149 galaxies in later discussions. 

The bolometric luminosity of the entire galaxy can be defined as: 
\begin{equation}
\begin{split}
    L_{\rm Bol, Gal} &= L_{\rm star} + L_{\rm AGN}, \\
    & = C_{\rm ape, MS}L_{\rm MS}^{\rm\,sp} + C_{\rm ape, TSB}L_{\rm TSB}^{\rm\,sp} + L_{\rm ASB} + L_{\rm AGN}, 
    \label{equ:Lum_Bol}
\end{split}
\end{equation}
where $L_{\rm star}$ is the total stellar bolometric luminosity; 
$C_{\rm ape, MS}L_{\rm MS}^{\rm\,sp}$ and $C_{\rm ape, TSB}L_{\rm TSB}^{\rm\,sp}$ are the aperture corrected luminosity of stars in MS and TSB components; 
$L_{\rm ASB}$ shows the radiation of ASB, i.e., the starbursts hidden in optical band, estimated from the FIR dust emission after reducing the contribution of MS and TSB components; 
$L_{\rm AGN}$ denotes the bolometric luminosity of AGN, which is estimated from the best-fit torus SED 
(see Appendix \ref{sec:Appendix_sedfit_connection} for details). 
$L_{\rm Bol, Gal}$ describes the sum of the primary (intrinsic) UV-optical-NIR radiation from stars in host galaxy and the accretion disk of AGN. 
The average $L_{\rm Bol, Gal}$ of the 149 selected galaxies is $10^{12.6\pm0.3}$\,\lsun. 
%, with the ratio $L_{1\textup{--}1000}/L_{\rm Bol, Gal}$ from 13\% for AGN-dominated cases to 50\% for SF-dominated cases (red and blue dotted lines in Figure \ref{fig:LTIR_LBol}, respectively). 
%The ratio $L_{1\textup{--}1000}/L_{\rm Bol, Gal}$ reflect the fraction of the radiation absorbed and re-emitted by dust in the entire galaxy. 
We report the esitmation of AGN luminosities in Section \ref{subsec:results_AGN} 
and the properties of the host galaxies, e.g., stellar mass and SFR, in Section \ref{subsec:results_HG}. 
The outflow detection is reported in Section \ref{subsec:results_outflow}. 

% \begin{figure}
%     \begin{center}
%     \includegraphics[trim=0 30pt 0 0, width=\columnwidth]{LTIR_LBol_hist2.pdf}
%     \end{center}
%     %\caption{
%     \red{MOVE to text}
%     Distributions of the total IR luminosity integrated in 1--1000 $\mu$m range ($L_{1\textup{--}1000}$, \textit{x}-axis), and the bolometric luminosity of the entire galaxy ($L_{\rm Bol, Gal}$, \textit{y}-axis) for the 149 \akari-selected IR bright galaxies. 
%     The blue shadow region denotes the requirement of luminosity for ULIRGs, which covers 125 galaxies in the sample. 
%     $L_{\rm Bol, Gal}$ shows the sum of the primary (intrinsic) UV-optical-NIR radiation from stars in host galaxy and the accretion disk of AGN (Equation \ref{equ:Lum_Bol}).
%     The blue circles and orange stars denote the galaxies spectroscopically identified by SDSS and FOCAS, respectively, while the blue histograms reflect the sums of SDSS and FOCAS identified objects. 
%     The markers circled in red shows 7 AGN-dominated ULIRGs ($f_{\rm AGN} > 0.75$, $f_{\rm AGN}=L_{\rm AGN, pri}/L_{\rm Bol, Gal}$). 
%     The red and blue dotted lines denote the AGN-dominated ($L_{1\textup{--}1000}/L_{\rm Bol, Gal}=13\%$) and SF-dominated ($L_{1\textup{--}1000}/L_{\rm Bol, Gal}=50\%$) cases, respectively. 
%     %}
%     \label{fig:LTIR_LBol}
% \end{figure}

\subsection{AGN activities in the \akari-selected ULIRGs}
\label{subsec:results_AGN}

\begin{figure}
    \begin{center}
    \includegraphics[trim=0 20pt -10pt 0, width=\columnwidth]{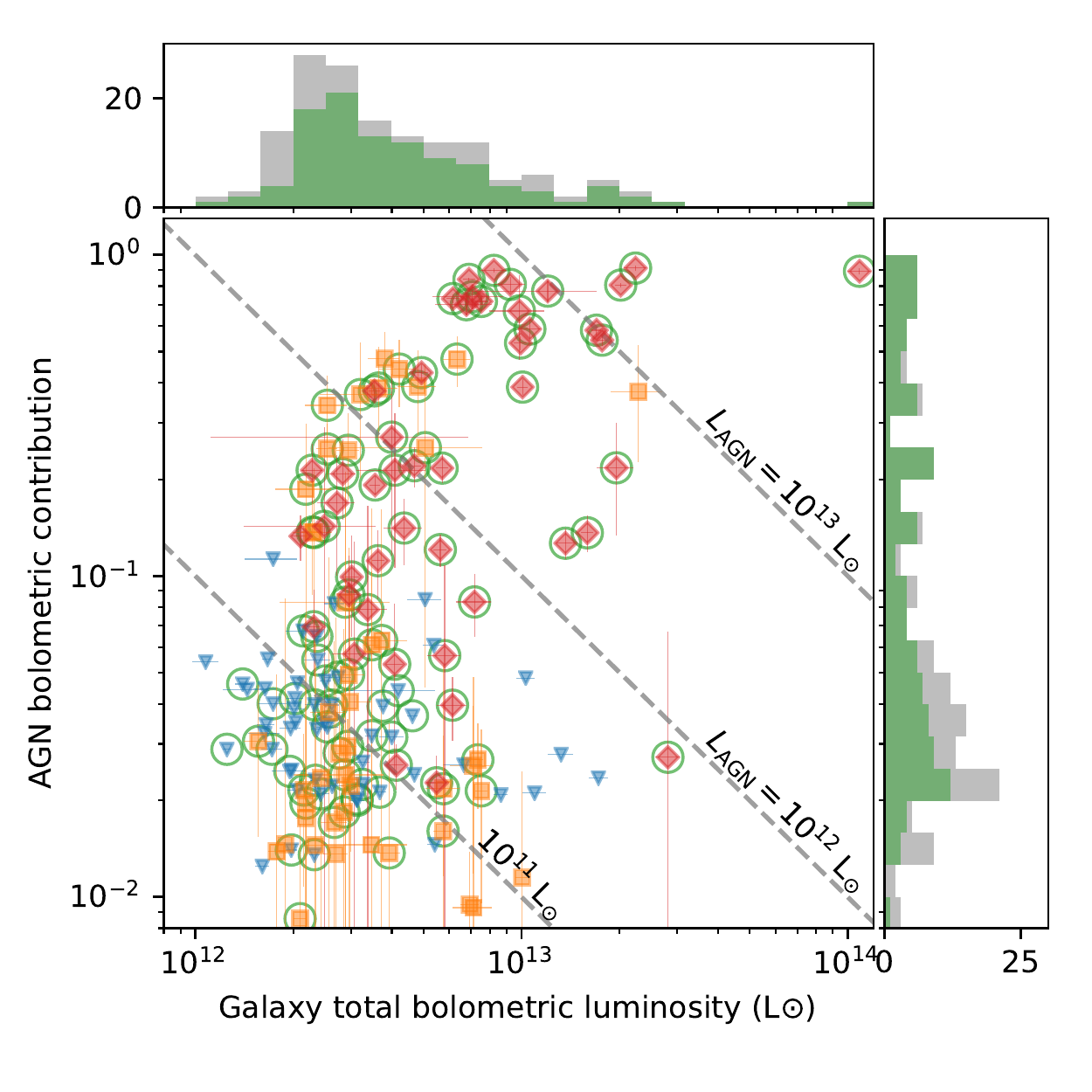}
    \end{center}
    \caption{
    AGN bolometric contribution ($f_{\rm AGN}=L_{\rm AGN}/L_{\rm Bol, Gal}$, \textit{y}-axis) as a function of the total bolometric luminosities of the galaxy ($L_{\rm Bol, Gal}$, \textit{x}-axis) of the \akari-selected ULIRGs. 
    The red diamonds denote the galaxies (47/149) with the AGN component detection ratio $\beta_{\rm AGN}>3$ in the w1--w3 bands in the SED fitting. 
    The orange squares denote the galaxies (46/149) with $1<\beta_{\rm AGN}<3$ in the w1--w3 bands. 
    %, for which the estimated $f_{\rm AGN}$ is considered as the upper limit. 
    The blue triangles show the galaxies (56/149) with $\beta_{\rm AGN}<1$, for which we estimate the upper limit of $f_{\rm AGN}$ by scaling the mean AGN SED from the 47 galaxies with $\beta_{\rm AGN}>3$ (see Section \ref{subsec:results_AGN} for detailed discussion). 
    The gray histograms show the distributions for all of the 149 galaxies. 
    The green circled markers and the green histograms denote that they are selected using the 90\%-confidence criterion of \citet{Assef2018} method.
    The gray dashed lines denote the positions of AGN bolometric luminosities of $10^{11}$, $10^{12}$, and $10^{13}$\,\lsun, respectively.  
    }
    \label{fig:FracAGN_LBol}
\end{figure}

With the bolometric luminosity of AGN primary emission ($L_{\rm AGN}$) and the total bolometric luminosity of the galaxy ($L_{\rm Bol, Gal}$), we can define the AGN bolometric energy contribution as $f_{\rm AGN}=L_{\rm AGN}/L_{\rm Bol, Gal}$. The distributions of $f_{\rm AGN}$ and $L_{\rm Bol, Gal}$ are shown in Figure \ref{fig:FracAGN_LBol}. 
In order to check the reliability of the detection of AGN in the galaxies, 
we evaluate the detection ratio of the MIR excess at rest 3--10\,$\mu$m (w1--w2 bands for galaxies at $z\sim0$, and w2--w3 bands for galaxies at $z\sim1$), which is the most significant feature of AGN in the IR SED fitting. 
The detection ratio in one band, e.g., w1, can be calculated as $\beta_{\rm AGN,w1}=F_{\rm AGN, w1}/\sigma_{\rm w1}$, 
where $F_{\rm AGN, w1}$ is the flux contributed from the AGN component, 
and $\sigma_{\rm w1}=\sigma_{\rm obs,w1}+10\% F_{\rm obs,w1}$, where $F_{\rm obs,w1}$ and $\sigma_{\rm obs,w1}$ are the observed flux and measurement error in the w1 band. The 10\% of the observed flux is included to take into account the uncertainty in the fitting model (see Section \ref{subsec:SED_decomp}). 

We consider the AGN component is significantly detected if the galaxy shows $\beta_{\rm AGN, MIR}>3$ in any one band among w1, w2, and w3 bands. 
47 galaxies are selected with the threshold (red diamonds in Figure \ref{fig:FracAGN_LBol}), which show a median bolometric contribution of 20\% with S\,/\,N ($f_{\rm AGN}/\sigma_{\rm fAGN}$) of 7. 
46 galaxies in the remaining sample show $\beta_{\rm AGN, MIR}>1$ in one band among w1--w3, 
for which the AGN is less significantly detected with a median $f_{\rm AGN}$ of 3\% and S\,/\,N of 2 
(orange squares in Figure \ref{fig:FracAGN_LBol}). 
% \red{
% Hereafter we consider the estimated $f_{\rm AGN}$ and $L_{\rm AGN}$ of the 46 galaxies as their upper bounds. 
% }
The rest 56 galaxies show $\beta_{\rm AGN, MIR}<1$ in all the bands among w1--w3, 
which means the AGN component is not detected in those galaxies (blue triangles in Figure \ref{fig:FracAGN_LBol}). 
In order to estimate the upper limits of $f_{\rm AGN}$ and $L_{\rm AGN}$ for the 56 galaxies, 
we firstly estimate the averaged AGN SED from the best fit results of the 47 galaxies with $\beta_{\rm AGN, MIR}>3$, which are logarithmically averaged after normalization to unit AGN bolometric luminosity. 
The averaged AGN SED is then scaled until it exceeds 1$\sigma$ uncertainty as shown above in any bands among w1--w3 for a galaxy, 
and the $f_{\rm AGN}$ and $L_{\rm AGN}$ corresponding to the scaled AGN SED is employed as the upper limits for the galaxy.  

We compare our AGN detection method with 
the AGN selection criteria with 90\% reliability of \citet[hereafter A18]{Assef2018}, which also traces the AGN MIR excess but using the w1-w2 color. 
For the ULIRGs with significant AGN MIR excess, i.e., $\beta_{\rm AGN, MIR}>3$, the two methods match well and 46 out of the 47 ULIRGs are selected with the A18 criteria. 
The matching ratios of A18 selection are 67\% (31/46) and 48\% (27/56) for the galaxies with $1<\beta_{\rm AGN, MIR}<3$ and $\beta_{\rm AGN, MIR}<1$, respectively, 
for which the blue w1-w2 color can be mainly contributed by the radiation of stellar direct component 
or the bright PAH 3.3 $\mu$m emission line in the SED decomposition here 
\citep[e.g.,][]{Mateos2012, Ichikawa2017a}.
% for which the blue w1-w2 color can be mainly contributed by the radiation of dust heated by intense star formation activity, e.g., the predominant photodissociation region (PDR) component or the bright PAH 3.3$\mu$m emission line. 
% 104 of the 149 galaxies match the requirements of the A18 method.
% Among the 104 galaxies, 86 ULIRGs show $f_{\rm AGN}\ge0.001$ in this method, indicating that the AGN detection in our method is basically consistent with the results from the A18 method. 
% %, and the matching rate 86/104\,$\sim$\,83\% is roughly consistent with the confidence level (reliability) of A18 method, i.e., 90\%. 
% The remaining 18 galaxies are identified as no AGN in this method. For the 18 galaxies, the blue w1-w2 color is mainly contributed by the radiation of dust heated by intense star-formation activity, e.g., the predominant photodissociation region (PDR) component or the bright PAH 3.3$\mu$m emission line. 

\begin{figure}
    \begin{center}
    \includegraphics[trim=0 20pt -10pt 0, width=\columnwidth]{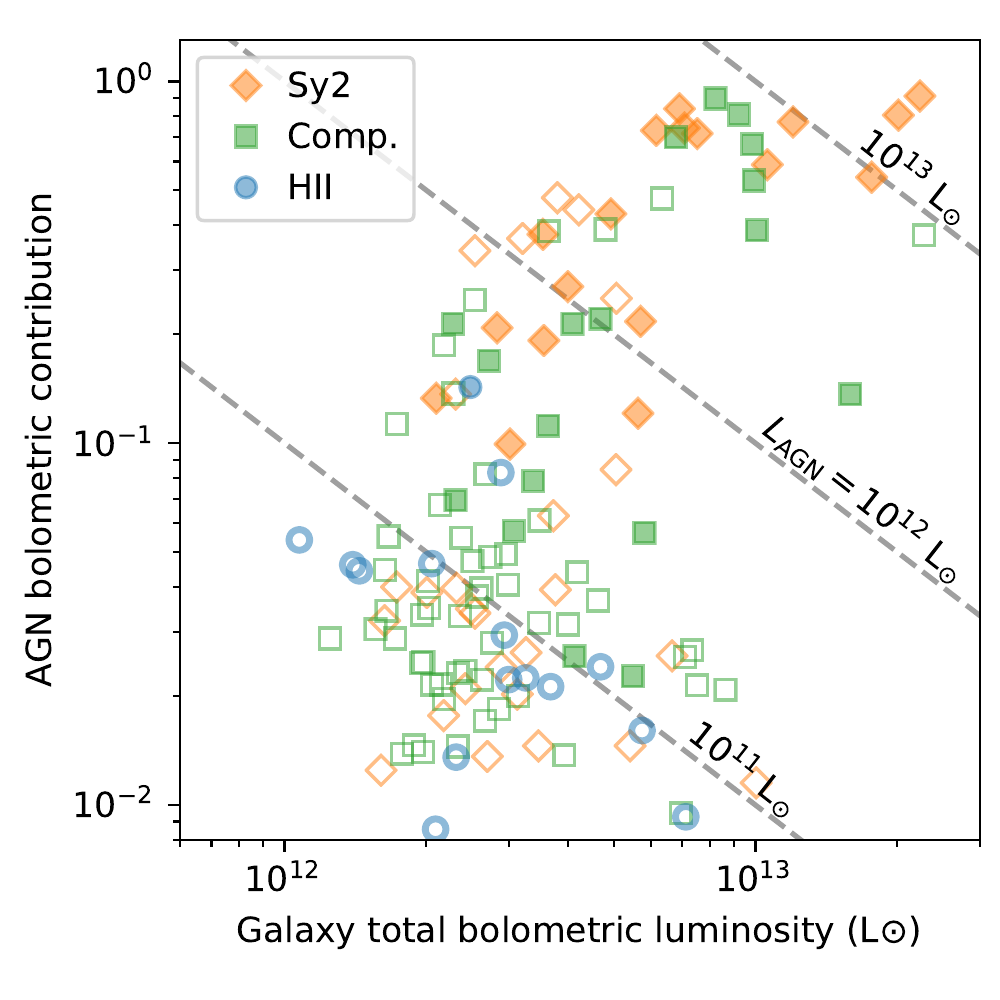}
    \end{center}
    \caption{
    AGN bolometric contribution ($f_{\rm AGN}$) as a function of the total bolometric luminosities of the galaxy ($L_{\rm Bol, Gal}$) 
    of 131 \akari-selected ULIRGs with BPT diagram classifications. 
    The Seyfert 2 (orange diamonds), composite (green squares) and star-forming (\hii, blue circles) galaxies are separated 
    using the flux ratios of the narrow components of emission lines by \citet{Kewley2001} and \citet{Kauffmann2003} curves. 
    Filled markers show the galaxies with significantly detected AGNs in the MIR bands ($\beta_{\rm AGN, MIR}>3$ in Figure \ref{fig:FracAGN_LBol}). 
    Open markers denote the galaxies with less significantly detected AGNs ($1<\beta_{\rm AGN, MIR}<3$) 
    and the galaxies with only upper limits of AGN luminosities ($\beta_{\rm AGN, MIR}<1$). 
    The gray dashed lines denote the positions of AGN bolometric luminosities of $10^{11}$, $10^{12}$, and $10^{13}$\,\lsun, respectively.  
    }
    \label{fig:FracAGN_LBol_BPT}
\end{figure}

Another widely adopted method to identify AGNs (Seyfert 2 galaxies) is the classification with the Baldwin-Phillips-Terlevich (BPT) diagram 
\citep{Baldwin1981}. 131 of the 149 galaxies have the spectra which cover the wavelength range of \oiii, \hb, \ha, and \nii\ lines. We require that for each line the fraction of bad pixels (e.g., contaminated by bright night sky lines) should be smaller than 25\% from the 10th to 90th percentiles of the narrow line profile. 
The classification is performed using the narrow lines, and the numbers of Seyfert 2, composite, and star-forming (\hii) galaxies are 44, 72, and 15, respectively (Figure \ref{fig:FracAGN_LBol_BPT}). 
%The stellar bolometric luminosity ($L_{\rm star}$, the sum of intrinsic radiation of old and young stars) versus AGN bolometric luminosity ($L_{\rm AGN}$) with the BPT identifications are shown in Figure \ref{fig:BPT_LumAGN_LumSF}. 
The mean $f_{\rm AGN}$ of \hii, composite, and Seyfert 2 galaxies are 4\%, 13\%, and 26\%, respectively, suggesting the trend that as AGN becomes luminous, it begins to dominate the ionization of the gas. 
The majority of the \hii\ ULIRGs (14/15) show $f_{\rm AGN}<10\%$; 
however, the Seyfert 2 galaxies cover a wide range from SF-dominated ($f_{\rm AGN}\sim5\%$) to AGN-dominated ($f_{\rm AGN}\sim50\%$) cases. 
One explanation for the SF-dominated ULIRGs but with Seyfert 2-like ionization could be that 
the \hii\ regions ionized by stellar light is more dusty than the AGN NLR. 
Since the ionization in \hii\ regions is dominated by young O/B type stars, the line emission from the ionized gas could be severely extinct by the dense dust surrounding the young stars (e.g., compact \hii\ region, \citealp{Stephan2018}). 
Therefore although the star formation activity could dominate the bolometric luminosity of the galaxy (e.g., low $f_{\rm AGN}$), it may only show a small contribution to the gas ionization observed in optical band. 

\begin{figure}
    \begin{center}
    \includegraphics[trim=0 20pt -10pt 0, width=\columnwidth]{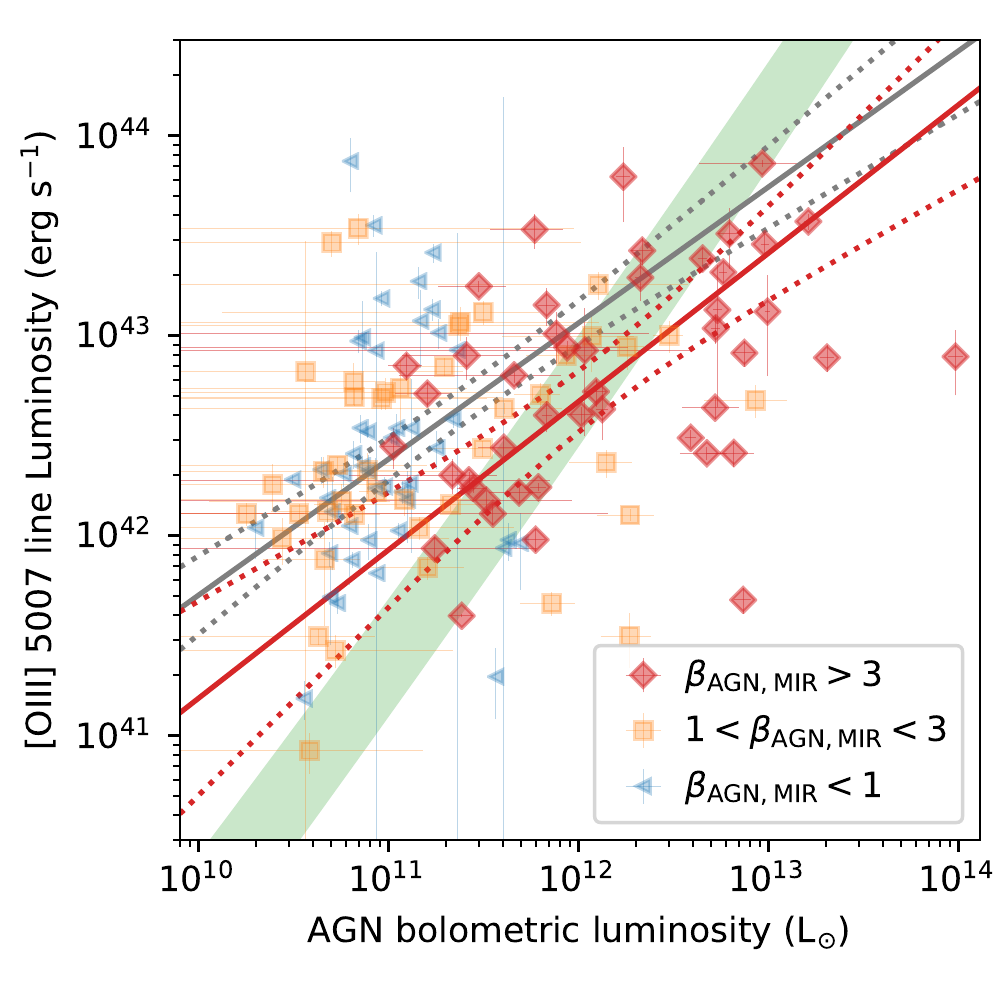}
    \end{center}
    \caption{
    Corrected \oiii\ luminosity vs. AGN bolometric luminosity for 136 galaxies with \oiii\ line detection. 
    The red, orange, and blue markers correspond to the detection ratios of AGN MIR excess (see Section \ref{subsec:results_AGN} and Figure \ref{fig:FracAGN_LBol}). 
    Note that for the galaxies with no AGN detection ($\beta_{\rm AGN, MIR}<1$), 
    upper limit of $L_{\rm AGN}$ is used for plotting. 
    The gray solid line is the ordinary least-squares bisector fitting function for the entire sample 
    and red solid line for the sources with significantly detected AGNs (red diamonds). 
    The \citet{Schmitt1985} binning method is used to account for the upper limit of $L_{\rm AGN}$ for the ULIRGs with no AGN detection 
    in the fitting for the entire sample. 
    The dotted lines denote the 95\% confidence bands of the fitting lines, respectively. 
    The green hatched regions denote the empirical $L_{\rm [OIII]}$-$L_{\rm AGN}$ functions for Type-2 AGNs using the $L_{\rm [OIII]}$-$L_{\rm 2-10\,keV}$ functions of \citet{Berney2015} and \citet{Ueda2015}, and the X-ray bolometric corrector of \citet{Marconi2004}. 
    %The orange dots and blue squares denote the observed $L_{\rm [OIII]}^{\rm obs}$ and corrected $L_{\rm [OIII]}^{\rm c(e+a)}$, respectively. 
    %Both extinction and aperture-loss are corrected for blue dots. 
    %The dots covered by gray shadow mean galaxies without AGN detection in the SED decomposition, and these galaxies are not used in the linear fitting (the \textit{x}-axis is shifted for illustration). 
    %gray dashed line shows the empirical relationship, i.e., $L_{\rm AGN}\sim600L_{\rm [OIII]}$, from \citet{Lamastra2009} and \citet{Kauffmann2009}. 
    %gray dots show the galaxies for which the \nii\ and \ha\ lines are out of the spectra coverage. 
    } 
    \label{fig:LumOIII_LumAGN}
    %\ref{ADD results from only narrow lines}
\end{figure}

Finally we show the relationship between $L_{\rm AGN}$ and the luminosity of \oiiiblong\ emission line. With a relatively high ionization potential (IP, 35.12 eV), the intensity of \oiii\ line is usually considered to reflect the AGN activity 
\citep[e.g.,][]{Kauffmann2003}. 
%We consider both of the directly observed luminosity from optical spectral fitting ($L_{\rm [OIII]}^{\rm obs}$), and the luminosity after correction for extinction and aperture-loss ($L_{\rm [OIII]}^{\rm c(e+a)}$). 
The dust extinction is estimated with Balmer decrement, or extinction of stellar continuum when the Balmer lines are too weak (see Table \ref{tab:Ebv_tying} in Appdendix \ref{sec:Appendix_specfit}). 
We use the average of $C_{\rm ape, MS}$ and $C_{\rm ape, TSB}$ to correct for the aperture-loss of \oiii\ line. 
Note that if the \oiii\ line originates from compact NLR (e.g, $<\sim 0.1$\,kpc), 
there is no need to correct \oiii\ line for aperture-loss and thus, 
the corrected $L_{\rm [OIII]}$ could be considered as some upper limit in such case. 
%Therefore the value of $L_{\rm [OIII]}^{\rm c(e+a)}$ could be considered as some upper limit, and $L_{\rm [OIII]}^{\rm obs}$ as the lower limit. 
The results for 136 galaxies with 
high quality detections of \oiii\ line 
(measurement S\,/\,N over 5 
%$L_{\rm [OIII]}^{\rm obs}/\sigma_{\rm [OIII]}^{\rm obs}>5$ 
and the fraction of bad pixels smaller than 25\% from the 10th to 90th percentiles of the entire \oiii\ line profile) 
are shown in Figure \ref{fig:LumOIII_LumAGN}. 
%for $L_{\rm [OIII]}^{\rm obs}$ and $L_{\rm [OIII]}^{\rm c(e+a)}$ respectively. 
%32 galaxies without AGN detection in SED decomposition are excluded in the linear fitting. 
%In order to avoid the effect of extreme values on the fitting result, we also use a 95\% criterion selection for the rest 104 galaxies, i.e., the galaxies with top 2.5\% and low 2.5\% values of either $L_{\rm AGN}$ or $L_{\rm [OIII]}$ are rejected, and finally 93 galaxies remains for the linear fit. 

In order to determine the underlying functional relation between $L_{\rm [OIII]}$ and $L_{\rm AGN}$ which are considered as two independent variables, 
we employ an ordinary least-squares bisector fit after normalizing with the value ranges in both of $x$- and $y$-axes in logarithmic grid \citep{Isobe1990}. 
The Schmitt binning method \citep{Schmitt1985} in \texttt{ASURV} software\footnote{A Python package in http://python-asurv.sourceforge.net} \citep{Feigelson1985,Isobe1986,Lavalley1992} 
is used to account for the estimated upper limit of $L_{\rm AGN}$ for the ULIRGs with barely AGN detection (i.e., $\beta_{\rm AGN, MIR}<1$). 

The Spearman correlation coefficient ($r_{\rm s}$) and the $p$-value of the hypothesis testing are applied to describe the correlation between $L_{\rm [OIII]}$ and $L_{\rm AGN}$. 
%, which is defined as 
%\begin{equation}
%\begin{split}
%    \rho_{\mathrm{pcc}X,Y} = \frac{\mathrm{E}[(X-\mu_{X})(Y-\mu_{Y})]}{\sigma_{X}\sigma_{Y}} ,  
%    \label{equ:PCC}
%\end{split}
%\end{equation}
%where $X=L_{\rm AGN}$ and $Y=L_{\rm [OIII]}$; $\mu$ and $\sigma$ are the mean and standard deviation, respectively; E denotes expectation. 
%$\rho_{\rm pcc}=\rm1$ for ideal linear relationship; $\rho_{\rm pcc}=0$ means no linear relationship. 
%For the observed $L_{\rm [OIII]}^{\rm obs}$, we obtain 
%$\log L_{\rm [OIII]}^{\rm obs}=0.24(\pm0.07)\,\log L_{\rm AGN} + 38.5(\pm0.83)$ and $\rho_{\rm pcc}=0.33$, 
%where $L_{\rm [OIII]}^{\rm obs}$ in the unit of erg\,s$^{-1}$ and $L_{\rm AGN}$ is in the unit of \lsun. 
%For the corrected $L_{\rm [OIII]}^{\rm c(e+a)}$, 
We obtain\footnote{Note that the Schmitt binning method requires an arbitrary choice of the number of binning. 
The shown results in Equation \ref{equ:LOIII_LAGN} are derived using 10 bins in both of $x$- and $y$-axes. 
We also employ the regression methon with binning size from 8 to 20, and the derived fitting slope slightly changes from 0.64 to 0.70. 
We adopt the regression results with binning size of 10 throughout the paper. }
\begin{equation}
\label{equ:LOIII_LAGN}
\log L_{\rm [OIII]}=0.68\,(\pm0.06)\,\log L_{\rm AGN} + 34.90\,(\pm0.63), 
\end{equation}
with $r_{\rm s}=0.41$ and $p$-value of $7\times10^{-7}$. 
If only the 46 ULIRGs with significantly detected AGN ($\beta_{\rm AGN, MIR}>3$) is considered, 
the regression function is\footnote{We employ the original bisector linear regression instead of the Schmitt binning method for the ULIRGs with significantly detected AGNs.} 
\begin{equation}
\log L_{\rm [OIII]}=0.74\,(\pm0.07)\,\log L_{\rm AGN} + 33.78\,(\pm0.89), 
\end{equation}
with $r_{\rm s}=0.49$ and $p$-value of $6\times10^{-4}$. 
The coefficients indicate a correlation between estimated $L_{\rm [OIII]}$ and $L_{\rm AGN}$, suggesting the consistency of the determination of AGN's activity from the \oiii\ emission line and the IR SED decomposition. 

The slope of the fitting function does not change between the entire sample and the sources only with significantly detected AGN. 
We also compare the derived $L_{\rm [OIII]}$-$L_{\rm AGN}$ regression functions 
to the empirical correlations for Type-2 AGNs in the literature \citep{Berney2015, Marconi2004, Ueda2015}. 
The ULIRG sample shows a shallower $L_{\rm [OIII]}$-$L_{\rm AGN}$ slope than that of typical Type-2 AGNs 
(e.g., 0.7 vs. 1.1, see Figure \ref{fig:LumOIII_LumAGN}), 
which could be due to the higher contamination of \oiii\ emission from star-forming regions in the ULIRGs especially with low AGN bolometric luminosity.
%If we consider the $L_{\rm AGN}/L_{\rm [OIII]}$ as the bolometric correction for $L_{\rm [OIII]}$, then for a AGN with $L_{\rm AGN}=10^{12}$\lsun, the estimated $L_{\rm AGN}/L_{\rm [OIII]}^{\rm c(e+a)}$ is 900, which is close to the \oiii\ bolometric correction in the literature, e.g., 600–-800, for AGNs with $L_{\rm AGN}\sim10^{12}$\lsun\ \citep{Kauffmann2009, Lamastra2009}. 

\subsection{Stellar mass and star formation rate}
\label{subsec:results_HG}

The stellar decomposition in the optical spectral fitting results in the best-fit stellar population templates, which provides the mass-to-luminosity ratio of stars, and the luminosity of MS and TSB components, thus the stellar masses $M_{\rm MS}^{\rm\,sp}$ and $M_{\rm TSB}^{\rm\,sp}$ can be estimated. However, $M_{\rm MS}^{\rm\,sp}$ and $M_{\rm TSB}^{\rm\,sp}$ only denote the masses of components within the $3\arcsec$ fibers of SDSS spectroscopy observations ($2\arcsec$ for BOSS spectroscopy), not the masses of the whole galaxies, which should be corrected to the SDSS photometric fluxes of the entire galaxies. The aperture correction factors $C_{\rm ape, MS}$ and $C_{\rm ape, TSB}$ can be estimated in the SED fitting process, and the best-fit results show averages of $C_{\rm ape, MS}=1.9\pm0.6$ and $C_{\rm ape, TSB}=2.0\pm1.8$ for the entire sample. The mass of stars in the ASB component can be estimated from the reproduced intrinsic bolometric luminosity ($L_{\rm ASB}$) with the assumption that the ASB has the same stellar population (and the same mass-to-luminosity ratio) as TSB components, i.e., $M_{\rm ASB}=L_{\rm ASB}(M_{\rm TSB}^{\rm\,sp}/L_{\rm TSB}^{\rm\,sp})$. Therefore the total stellar mass of the galaxy can be described as:
\begin{equation}
    M_{\rm star, tot} = C_{\rm ape, MS}\,M_{\rm MS}^{\rm\,sp} + (C_{\rm ape, TSB} + \frac{L_{\rm ASB}}{L_{\rm TSB}^{\rm\,sp}})\,M_{\rm TSB}^{\rm\,sp}. % + L_{\rm ASB} \frac{M_{\rm TSB}^{\rm\,sp}}{L_{\rm TSB}^{\rm\,sp}}.  
    \label{equ:Mass_stars_tot}
\end{equation}
% REMOVE THIS
% Since the SED decomposition with the uncertainty estimation is performed with the fixed best-fit results of the spectral fitting, in order to show the scatters of best-fit spectral model in the final results, the uncertainty of $M_{\rm star, tot}$ is calculated using uncertainty transfer equation: 
% \begin{equation}
% \begin{split}
%     \left(\mathrm{d}M_{\rm star, tot}\right)^2 =& \left(\mathrm{d}M_{\rm star, tot}^{\rm SED}\right)^2 
%     + C_{\rm ape, MS} \left(\mathrm{d}M_{\rm MS}^{\rm\,sp}\right)^2 \\
%     &+ C_{\rm ape, TSB} \left(\mathrm{d}M_{\rm TSB}^{\rm\,sp}\right)^2 
%     + L_{\rm ASB} \left(\mathrm{d}\frac{M_{\rm TSB}^{\rm\,sp}}{L_{\rm TSB}^{\rm\,sp}}\right)^2, 
%     \label{equ:Mass_stars_tot_unc}
% \end{split}
% \end{equation}
% where $\mathrm{d}M_{\rm star, tot}^{\rm SED}$ is the scatter estimated from Monte Carlo simulation in SED fitting with $M_{\rm MS}^{\rm\,sp}$, $M_{\rm TSB}^{\rm\,sp}$ and $L_{\rm TSB}^{\rm\,sp}$ fixed to the best-fit values in spectral fitting; $\mathrm{d}M_{\rm MS}^{\rm\,sp}$, $\mathrm{d}M_{\rm TSB}^{\rm\,sp}$ and $\mathrm{d} (M_{\rm TSB}^{\rm\,sp}/L_{\rm TSB}^{\rm\,sp})$ are the scatters estimated from Monte Carlo simulation in spectral fitting. Note that because the variables in the right side of Equation \ref{equ:Mass_stars_tot_unc} are not absolutely independent, the $\mathrm{d}M_{\rm star, tot}$ calculated from the transfer equation should be considered as some upper bound. 
The total current star formation rate can be similarly described as the sum of the SFRs of the three stellar components:
\begin{equation}
\begin{split}
    \mathrm{SFR}_{\rm star, tot} =\ & \mathrm{SFR}_{\rm MS} + \mathrm{SFR}_{\rm TSB} + \mathrm{SFR}_{\rm ASB} \\ 
    =\ & C_{\rm ape, MS}\,\mathrm{SFR}_{\rm MS}^{\rm\,sp} + (C_{\rm ape, TSB} + \frac{L_{\rm ASB}}{L_{\rm TSB}^{\rm\,sp}})\,\mathrm{SFR}_{\rm TSB}^{\rm\,sp}, % \\
      %& + L_{\rm ASB} \frac{\mathrm{SFR}_{\rm TSB}^{\rm\,sp}}{L_{\rm TSB}^{\rm\,sp}},  
    \label{equ:SFR_stars_tot}
\end{split}
\end{equation}
where $\mathrm{SFR}_{\rm MS}^{\rm\,sp}$ and $\mathrm{SFR}_{\rm TSB}^{\rm\,sp}$ mean the simultaneous star formation rates of MS and TSB components estimated from the optical spectral fitting with assumed exponential SFH. 
%The uncertainty of $\mathrm{SFR}_{\rm star, tot}$ can be calculated using a similar method with Equation \ref{equ:Mass_stars_tot_unc}. 

\begin{figure}
    \begin{center}
    \includegraphics[trim=0 20pt -10pt 0, width=\columnwidth]{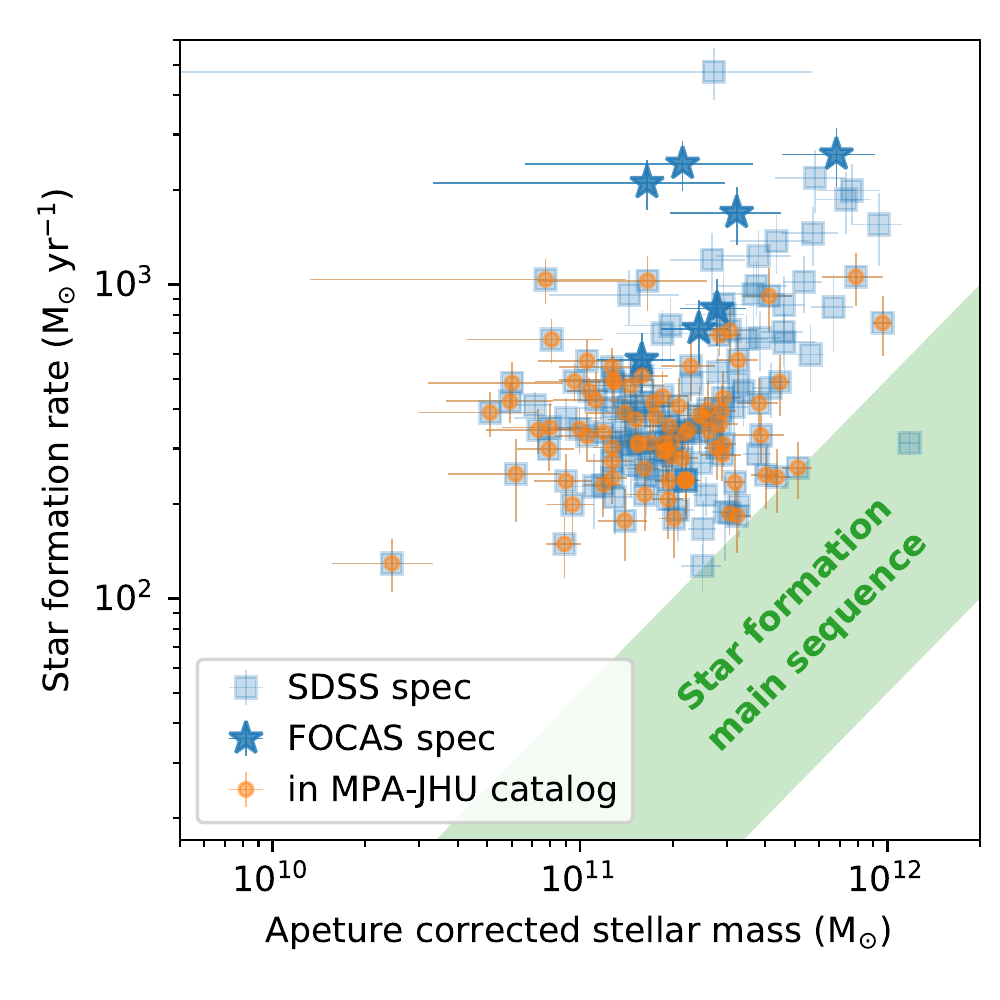}
    \end{center}
    \caption{
    SFR versus total stellar mass ($M_{\rm star}$) for \akari-selected ULIRGs. The SFR is estimated using SFH from spectral fitting results and modified with SED decomposition (Equation \ref{equ:SFR_stars_tot}). The $M_{\rm star}$ denotes the total mass of old and young stellar populations, which is estimated using SFH from spectral fitting results and corrected for aperture-loss with SDSS photometry (Equation \ref{equ:Mass_stars_tot}). 
    The blue squares and stars denote the galaxies spectroscopically identified by SDSS and FOCAS, respectively, while orange dots mark the galaxies which are reported in the sample of MPA-JHU group. 
    The green shadow region denotes the star-forming main-sequence at $z=0$--$0.5$ \citep{Peng2014}.
    }
    \label{fig:SFR_Mstar}
\end{figure}

Figure \ref{fig:SFR_Mstar} shows the SFR-$M_{\rm star}$ diagram for the \akari-selected ULIRGs. 
The results indicating a population of massive galaxies with average $M_{\rm star}=4.1\,(\pm1.3) \times10^{11}$\msun, which locate above the star formation main sequence ($z=0$--$0.5$, \citealp{Peng2014}), 
with an average SFR\,$=390\,(\pm290)$\,\sfrunit. 12 ULIRGs have SFR\,$>$\,1000\,\sfrunit, in which J115458.02+111428.8 ($z=0.798$, $L_{1\textup{--}1000}=1.28\times10^{13}$\lsun) shows the largest SFR close to $5000$\,\sfrunit. 
%The best-fit spectral and SED decomposition results of the object are shown in Figure \ref{fig:example_starburst}. 

70 galaxies in the total sample are covered by the galaxy catalog of MPA-JHU group\footnote{https://www.sdss.org/dr12/spectro/galaxy\_mpajhu/} (shown in orange dots in Figure \ref{fig:SFR_Mstar}). 
We compare the estimation of $M_{\rm star}$ for the 70 galaxies. 
The $M_{\rm star}$ in MPA-JHU catalog was estimated using SDSS photometry after correcting for the contribution of emission lines using the optical spectra. 
Our results show consistency with the MPA-JHU with an average factor of about 1.7.
% We compare the estimation of $M_{\rm star}$ and SFR for the 70 galaxies, and the results of the comparison are shown in Figure \ref{fig:Mstar_check} and \ref{fig:SFR_check}, respectively. For both $M_{\rm star}$ and SFR reported by MPA-JHU group, we choose the median values of the probability distribution function (PDF) as the estimates, and use the half width from the 16th to the 84th percentile of the PDF as the uncertainty. The $M_{\rm star}$ in MPA-JHU catalog was estimated using SDSS photometry after correcting for the contribution of emission lines using the optical spectra. We compare both of the results estimated based on fiber magnitudes (MPA-JHU) / fiber spectra (this work), and the results using model magnitudes (MPA-JHU) / spectra corrected for aperture-loss (this work). Both of the two comparison show consistent with an average factor of about $1.8\sim1.9$. 
The discrepancy could be due to the different IMF used in the analysis. Compared to the IMF of \citet{Kroupa2001} used by MPA-JHU group, the IMF of \citet{Salpeter1955} employed in this work contains larger number of low mass stars and could yield out larger mass-to-luminosity ratio (e.g., \citealp{Cappellari2012}), which leads to a higher estimate of stellar mass.  
\citet{Akiyama2008} found that for a typical SED with rest U-V color of 1.0, the ratio of estimated stellar masses using IMF and \citet{Salpeter1955} and \citet{Kroupa2001} is 1.8, which is fully consistent with the observed offset in this work.

However, the SFR of the 70 galaxies estimated in this work is 10--100 times higher than the results reported by MPA-JHU, 
which is estimated using optical emission lines luminosity \citep{Brinchmann2004} and the optical galaxy photometry \citep{Salim2007}. 
The large difference is due to the heavy dust obscuration of star formation regions in ULIRGs and thus, 
the SFR only based on optical detection can be significantly underestimated. 
% However, the comparison for SFR shows quite different results. We also compare both of the fibre SFR and total SFR (orange and blue dots in Figure \ref{fig:SFR_check}, respectively). The fibre SFR in MPA-JHU catalog is estimated using emission lines luminosity 
% \citep{Brinchmann2004} and the total SFR is estimated using the galaxy photometry following 
% \citet{Salim2007}. 
% The SFR reported in this work is measured from the assumed SFH form (fiber SFR) and corrected to match the FIR flux (total SFR). 
% Both of the fiber SFR and total SFR in our results are with higher values compared to the results in MPA-JHU catalog, i.e., 20:4 for fiber SFR, and 300:10 for total SFR (all in unit of \sfrunit). 
% The difference between the estimates of SFR in the two works indicates that for dust obscured galaxies like ULIRGs, the SFR estimated from pure optical method could be severely underestimated. 
In order to check the reliability of the SFR from the SED modeling, 
which considers fully obscured starburst to explain the observed FIR flux, 
we also compare with the results directly estimated from the integrated IR luminosity $L_{8\textup{--}1000}$, which is estimated from the best-fit SED model, and the empirical relationship determined by \citet{Kennicutt1998}: 
\begin{equation}
\begin{split}
    \mathrm{SFR}\, (\mathrm{M_{\rm \odot}\, yr^{-1}})=4.5\times10^{-44} L_{8\textup{--}1000}\, (\mathrm{\rm erg\,s^{-1}}), 
    \label{equ:SFR_stars_8_1000}
\end{split}
\end{equation}
which is obtained with an assumption of 
%was yields out for 
continuous starbursts of age 10--100 Myr with the IMF of \citet{Salpeter1955}. 
%The results and comparison are shown in Figure \ref{fig:SFR_check_2}. 
The SFR in this work is consistent with that from the \citet{Kennicutt1998} relation with a factor of 1.8. 
%show a nearly constant ratio (with the linear slope close to 1.0) of about 1.8. 
The higher estimated values from the SED fitting could be due to the fact that we ignore the higher energy photons ($>13.6$\,eV) in the dust extinction, which requires more young stars (higher SFR) to explain the FIR dust emission in the energy conservation equations (Equation \ref{equ:EB_HG} in Appendix \ref{sec:Appendix_sedfit_connection}). 

%%%%%%%%%%%%%%%%%%%%%%%%%%%%%%%%%%%%%%%%%%%%%%%%%%

\subsection{Outflow velocity of ionized gas}
\label{subsec:results_outflow}

\begin{figure}
    \begin{center}
    \includegraphics[width=\columnwidth]{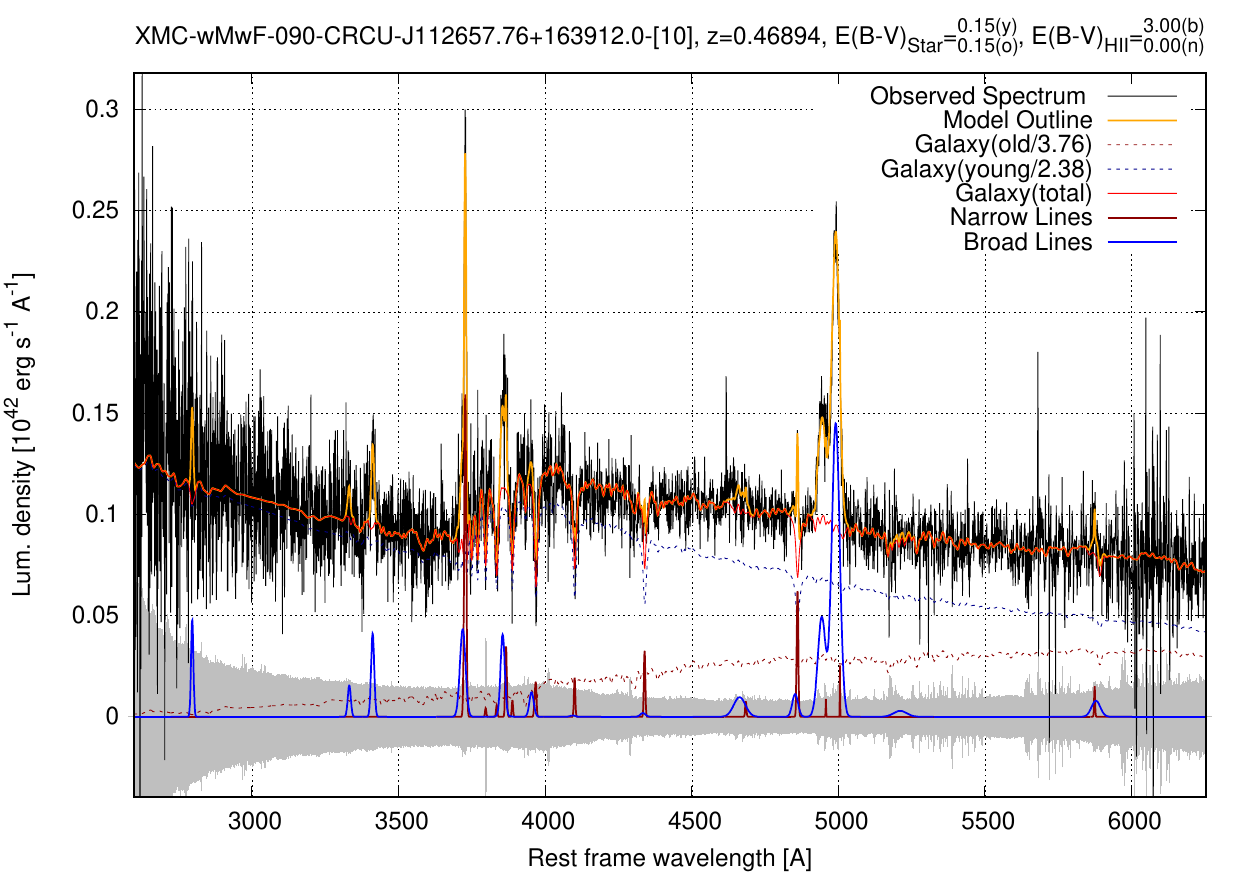}
    \includegraphics[width=\columnwidth]{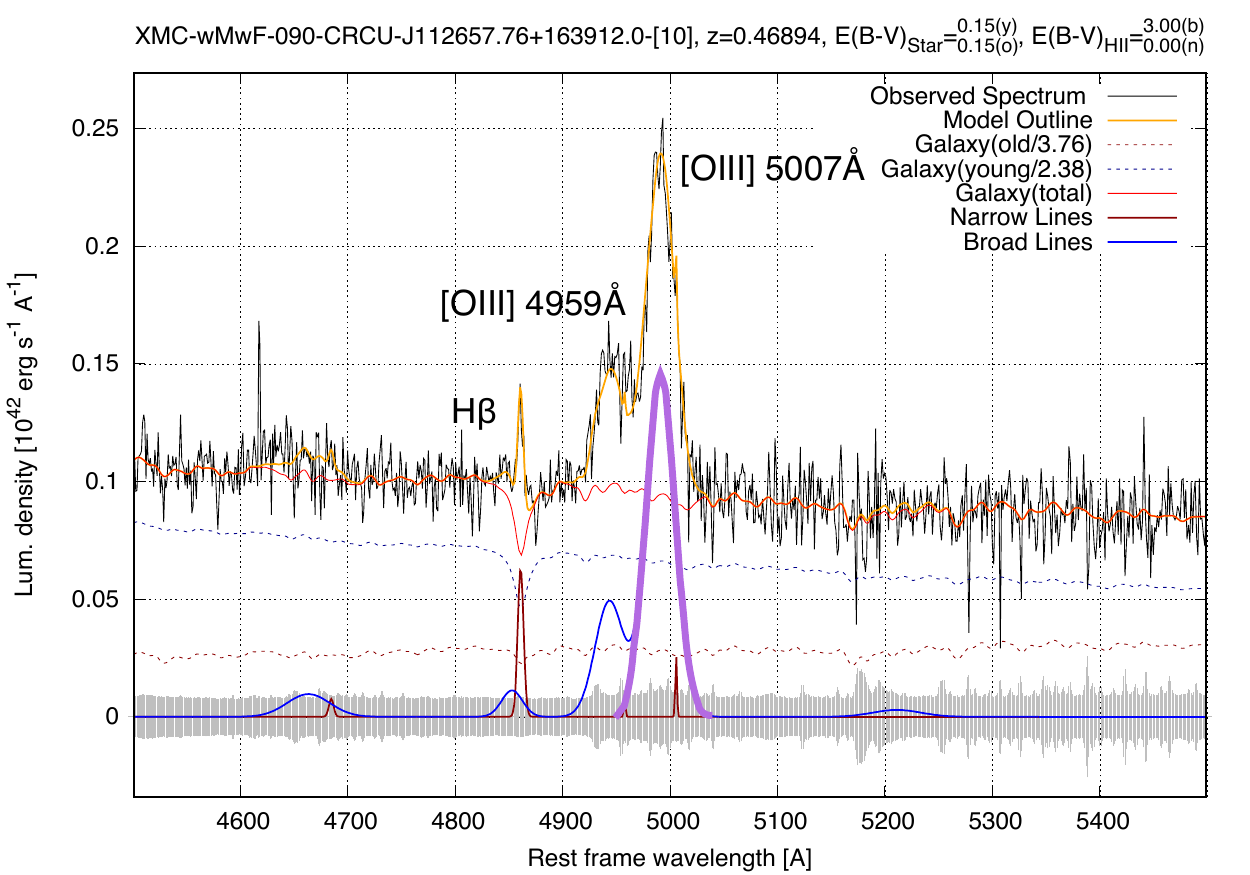}
    \end{center}
    \caption{
    Optical spectral fitting (top) and zoom-in around \oiiiblong\ line (bottom) for J112657.76+163912.0, a ULIRG with outflow velocity of $v_{\rm out}=2061$\,\kms. 
    The broad component of its \oiiiblong\ line is shown in violet thick curve. 
    Other legends are the same as those in Figure \ref{fig:Spec_QSO}. }
    \label{fig:Spec_OIII}
\end{figure}

\begin{figure}
    \begin{center}
    \includegraphics[trim=0 20pt -10pt 0, width=\columnwidth]{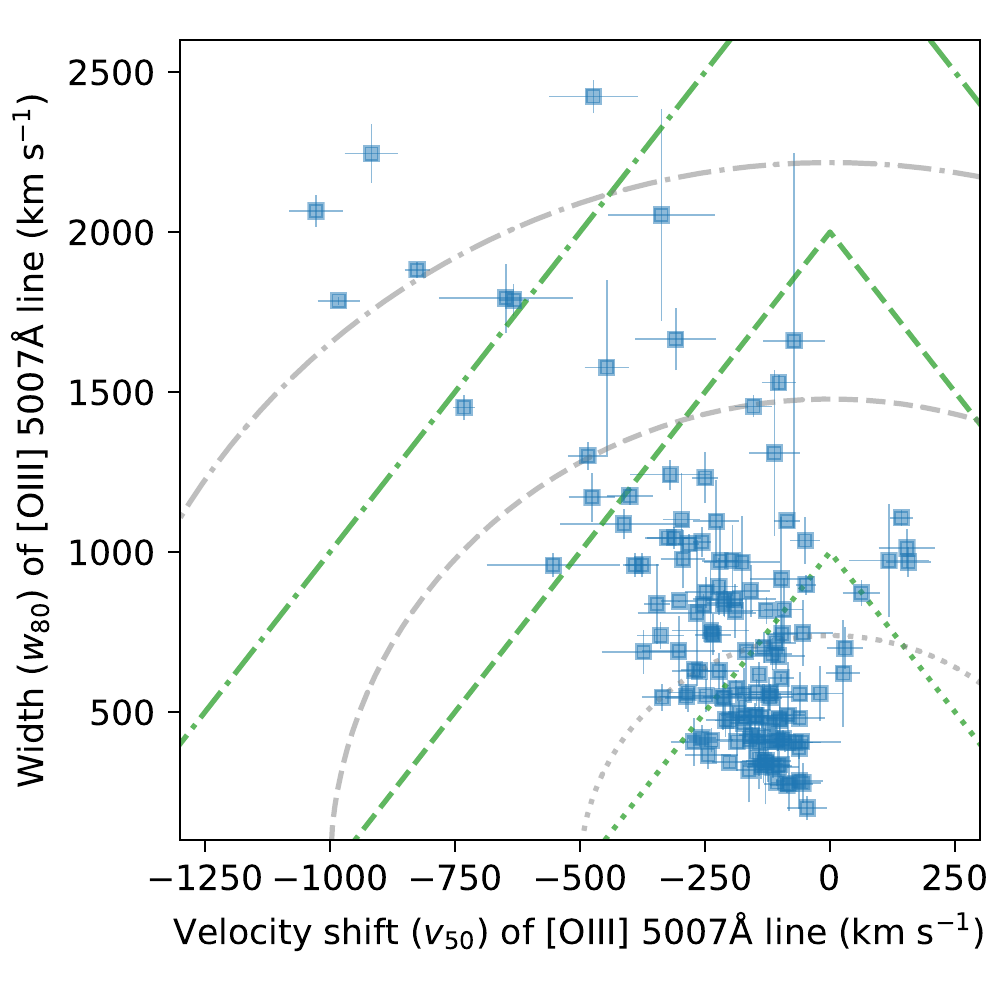}
    \end{center}
    \caption{
    Velocity shifts ($v_{50}$) and widths ($w_{80}$) of 136 \akari-selected ULIRGs with \oiiiblong\ detection. 
    %Blue squares denotes the 136 ULIRGs with S\,/\,N\,$>5$ for observed \oiii\ flux, while orange dots show 10 ULIRGs with S\,/\,N\,$<5$. 
    The green lines denote the outflow velocity with the definition (this work), $v_{\rm out} = |v_{50}|+w_{80}/2$, of 500 (dotted), 1000 (dashed), and 1500 (dash-dotted) \kms, respectively. 
    The gray lines denote the outflow velocity with kinetic-energy definition \citep[e.g.,][]{Harrison2014}, $v_{\rm out}=[v_{50}^2 + 3(w_{80}/2.56)^2]^{1/2}$, of 500 (dotted), 1000 (dashed), and 1500 (dash-dotted) \kms, respectively. 
    }
    \label{fig:Velocity_shift_disp}
\end{figure}

The \oiiiblong\ emission line kinematics is widely used as an indicator of the outflowing ionized gas (an example shown in Figure \ref{fig:Spec_OIII}). 
In order to describe the multi-component profiles, following \citet{Zakamska2014} and \citet{Perna2015}, 
we employ the velocity defined from the normalized cumulative distribution: 
\begin{equation}\label{equ:w80_chap2}
    F=\int_{-\infty}^{v_{\rm F}}f(v')\;\mathrm{d}v' \,/\, \int_{-\infty}^{\infty}f(v')\;\mathrm{d}v',
\end{equation}
where $f(v)$ is the flux per velocity unit from the best fit spectrum, $F$ equals to the fraction of flux with velocity $v\le v_{\rm F}$. 
$|v_{50}|$ is the 50\% flux velocity of the total \oiii\ line profile (narrow + broad components), which denotes the median velocity shift.
The width comprising 80 percent of the total flux, i.e., $w_{80}\equiv v_{90}-v_{10}$, is adopted to denote the width of the entire line profile, which is equivalent to 2.563 times of the standard deviation for a single Gaussian profile. 
Figure \ref{fig:Velocity_shift_disp} shows the measured velocity shifts ($v_{50}$) and widths ($w_{80}$) of the \oiii\ lines for 136 ULIRGs with 
%$L_{\rm [OIII]}^{\rm obs}/\sigma_{\rm [OIII]}^{\rm obs}>5$ ($\sigma_{\rm [OIII]}^{\rm obs}$ is the scatter of estimated $L_{\rm [OIII]}^{\rm obs}$) 
S\,/\,N\,$>5$ for observed \oiii\ flux. 
The measured $w_{80}$ is correlated with $v_{50}$, indicating that the outflowing gas with a faster velocity shift tends to be more turbulent. 

128 out of the 136 ULIRGs show \oiii\ emission line blueshifted in relative to the stellar continuum, i.e., $v_{50}<0$, indicating that the ionized gas moves towards the observer. 
Due to the obscuration of the dusty stellar disk to the emission in the side far from the observer, we consider the observed \oiii\ emission line is dominated by the outflowing gas in the side near to the observer. 
The rest 8 ULIRGs show redshifted \oiii\ line ($v_{50}>0$) with 3 ULIRGs show $v_{50}>100$\,\kms, which could be due to predominant inflowing gas, or a large inclination of the outflow direction compared to the perpendicular direction of the stellar disk.  

Estimating the bulk outflow velocity from the observed line velocities is not straightforward, because the conversion depends on the geometry of the outflowing gas and the spatial distribution of velocity \citep{Liu2013, Fiore2017}. \citet{Veilleux2005} defined the maximum velocity of the outflow as $v_{\rm out} \simeq |v_{\rm shift}|+\mathrm{FWHM}/2$ (see also \citealp{Rupke2005, Westmoquette2012, Arribas2014}); while other researchers adopted a partly different definition with a higher weight of line width, i.e., $v_{\rm out} \simeq |v_{\rm shift}|+\mathrm{FWHM}$ \citep{Rupke2013, Fiore2017}. We employ the first definition assuming $v_{\rm shift}=v_{50}$ but replace FWHM with $w_{80}$ following \citet{ManzanoKing2019}. 
%The results of the estimated outflow velocity ($v_{\rm out} = |v_{50}|+w_{80}/2$) are shown in Figure \ref{fig:Velocity_kine_vout}. 
We also compare the outflow velocity with the values from another definition $v_{\rm out}=[v_{50}^2 + 3(w_{80}/2.56)^2]^{1/2}$, which is usually used in the estimation of the kinetic energy \citep{Harrison2014}, and find that the two definitions yield out consistent results with average ratio of $1.0\pm0.1$. %The comparison between the outflow velocities from the two definitions are also shown in Figure \ref{fig:Velocity_kine_vout}. 

\begin{figure}
    \begin{center}
    \includegraphics[trim=0 20pt -10pt 0, width=\columnwidth]{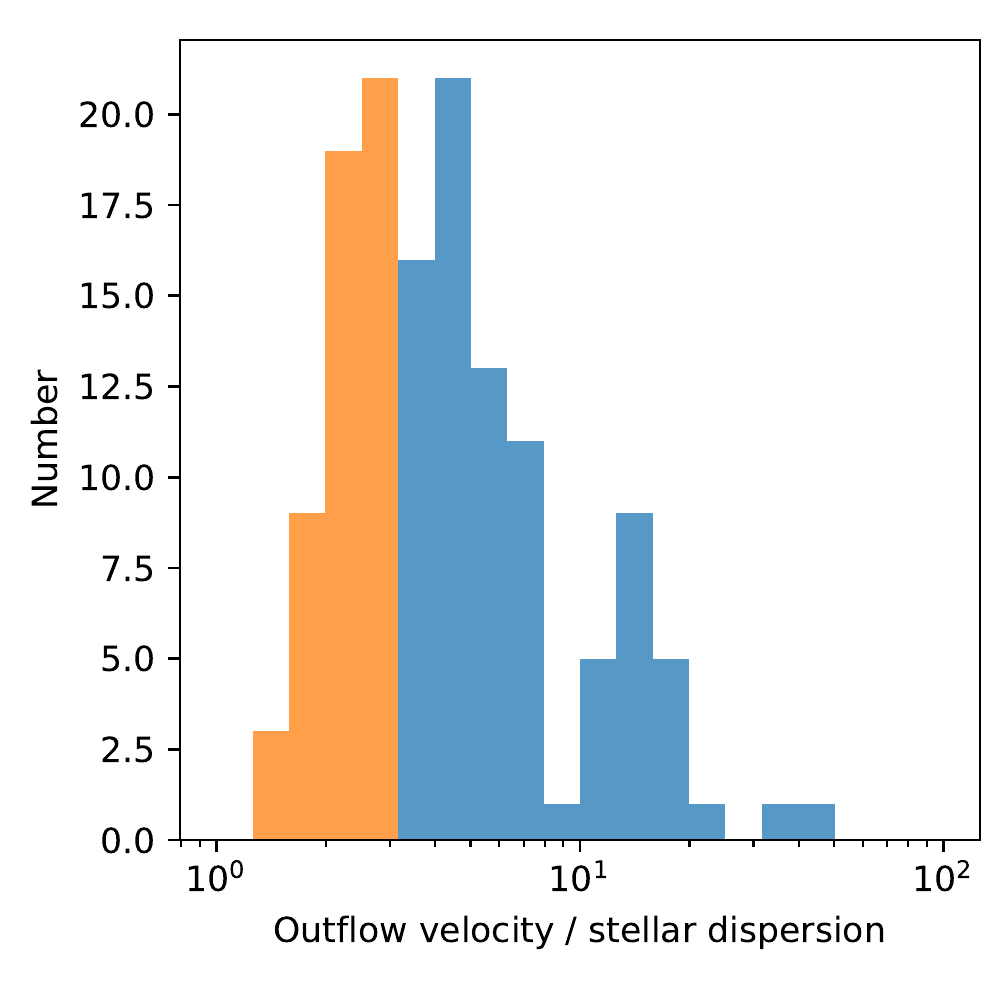}
    \end{center}
    \caption{
    Histogram of the ratio $v_{\rm out}/\sigma_{\rm star}$, where $v_{\rm out}$ is the outflow velocity and $\sigma_{\rm rot}$ is the velocity dispersion of stellar continuum. 
    Blue and orange histograms denote galaxies with $v_{\rm out}/\sigma_{\rm star}>3$ and $v_{\rm out}/\sigma_{\rm star}<3$, respectively. 
    }
    \label{fig:Velocity_roat_hist}
\end{figure}

Finally we compare the estimated outflow velocity ($v_{\rm out} = |v_{50}|+w_{80}/2$) with the rotation velocity ($\sigma_{\rm star}$) of the stellar disk. $\sigma_{\rm star}$ is estimated as the dispersion (FWHM\,$/2.35$) in the absorption line features of the stellar continuum (see Appendix \ref{sec:Appendix_specfit} for details). 
Among the 136 ULIRGs with \oiii\ detection, 87 ULIRGs show $v_{\rm out}/\sigma_{\rm rot}>3$ (Figure \ref{fig:Velocity_roat_hist}), 
suggesting that the outflow velocity exceeds the rotation velocity of the stellar disk in most of the ULIRGs. 

% \begin{figure}
%     \begin{center}
%     \includegraphics[trim=0 20pt -10pt 0, width=\columnwidth]{Velocity_ratio_vout.pdf}
%     \end{center}
%     \caption{
%     \red{CHANGE to Histogram}
%     Outflow velocity (\textit{x}-axis, $v_{\rm out} = |v_{50}|+w_{80}/2$) vs. $v_{\rm out}/\sigma_{\rm star}$ ratio (\textit{y}-axis), where $\sigma_{\rm rot}=\mathrm{FWHM_{cont}}/2.35$ is the velocity dispersion of stellar continuum. 
%     The green dashed and dotted lines denote the limit of the line width of stellar absorption lines in the spectral fitting, i.e., $100<\mathrm{FWHM_{cont}}<500$\,\kms. 
%     Other legends are the same as those used in Figure \ref{fig:Velocity_shift_disp}. 
%     }
%     \label{fig:Velocity_roat_vout}
% \end{figure}

\section{Discussions}
\label{sec:chap2_discussions}

\subsection{Relationship between outflow velocity and AGN activity}
\label{subsec:discuss_vout_LAGN}

\begin{figure*}
    \begin{center}
    \includegraphics[trim=0 20pt -10pt 0, width=2\columnwidth]{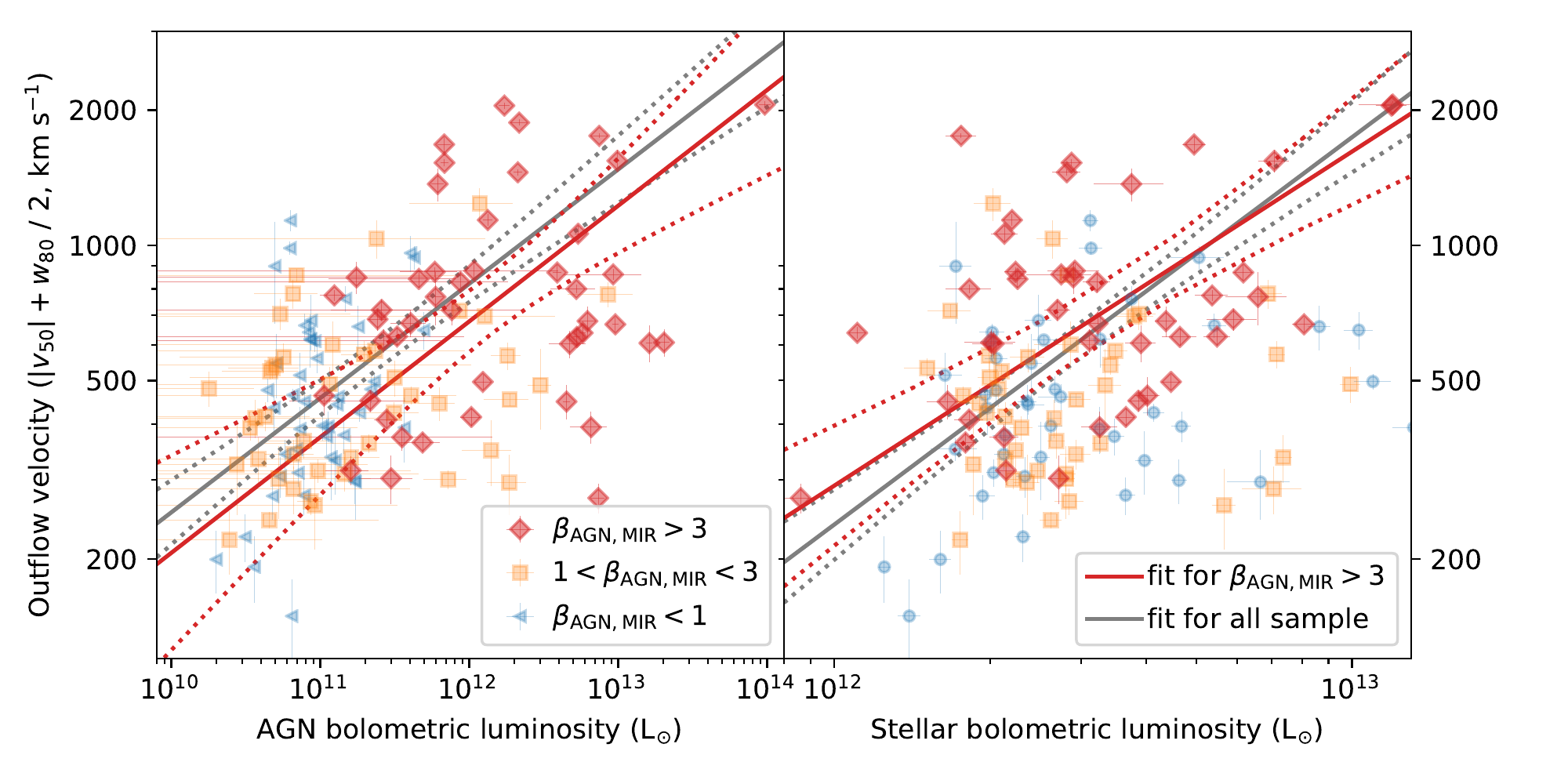}
    \end{center}
    \caption{
    \oiii\ outflow velocity ($v_{\rm out}$) vs. AGN bolometric luminosity ($L_{\rm AGN}$, left) 
    and star formation bolometric luminosity ($L_{\rm SF}$, right)
    for 136 galaxies with \oiii\ line detection. 
    The red, orange, and blue markers correspond to the detection ratios of AGN MIR excess (see Section \ref{subsec:results_AGN} and Figure \ref{fig:FracAGN_LBol}). 
    Note that for the galaxies with no AGN detection ($\beta_{\rm AGN, MIR}<1$), the shown $L_{\rm AGN}$ is the estimated upper limit. 
    The gray solid line is the ordinary least-squares bisector fitting function for the entire sample 
    and red solid line for the sources with significantly detected AGNs (red diamonds). 
    The dotted lines denote the 95\% confidence bands of the fitting lines, respectively. 
    }
    \label{fig:vout_Lum2}
\end{figure*}

\begin{figure}
    \begin{center}
    \includegraphics[trim=0 20pt -10pt 0, width=\columnwidth]{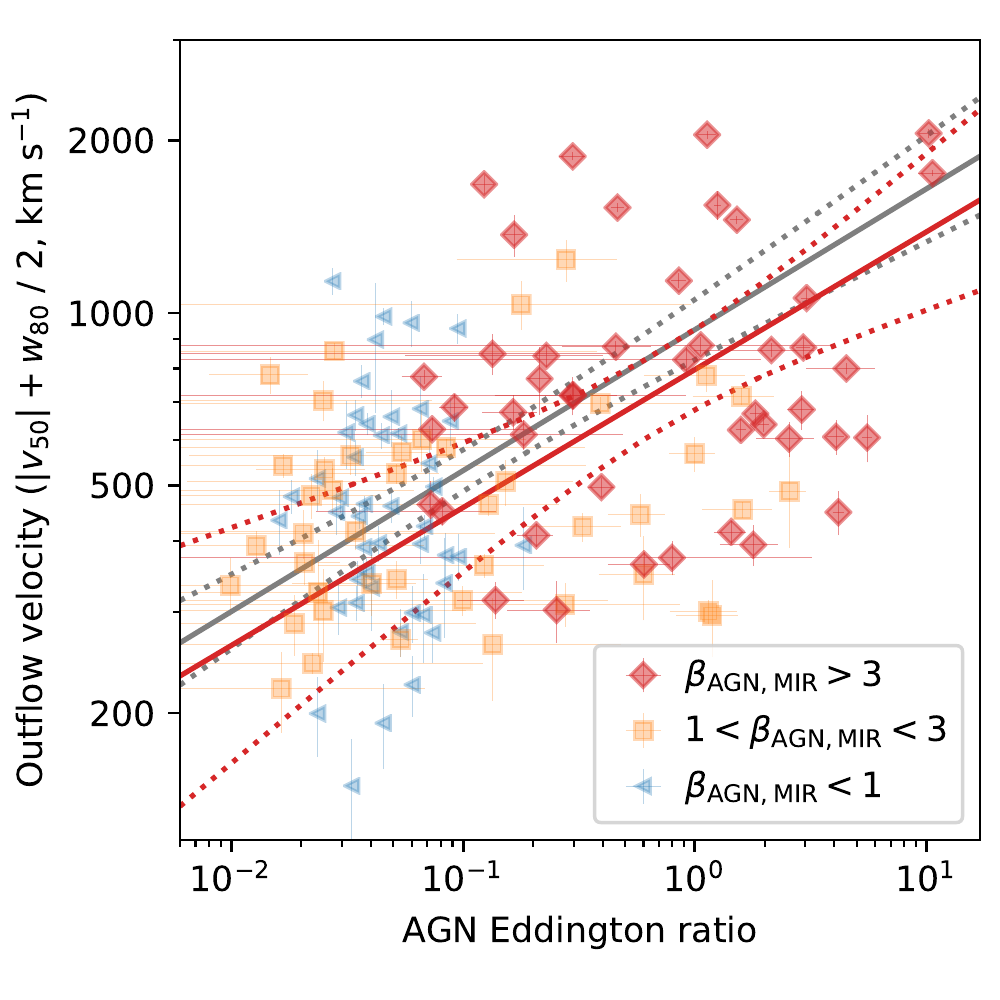}
    \end{center}
    \caption{
    \oiii\ outflow velocity ($v_{\rm out}$) vs. AGN Eddington ratio ($\lambda_{\rm Edd}$).
    %The linear fit function is $\log v_{\rm out}= 0.09(\pm0.02)\,\log \lambda_{\rm Edd} + 2.80(\pm0.03)$ with $\rho_{\rm pcc}=0.35$. 
    The legends are the same as in Figure \ref{fig:vout_Lum2}. 
    }
    \label{fig:vout_EddAGN}
\end{figure}

The estimated outflow velocity ($v_{\rm out}$) and AGN bolometric luminosity ($L_{\rm AGN}$) of 136 ULIRGs with \oiii\ detection are shown in Figure \ref{fig:vout_Lum2} (left panel). 
The mean $v_{\rm out}$ of 54 ULIRGs with $L_{\rm AGN}>3\times10^{11}$\,\lsun\ is $820\pm460$\,\kms, which is larger than the mean $v_{\rm out}$, $480\pm200$\,\kms, of the remaining 82 ULIRGs with $L_{\rm AGN}<3\times10^{11}$\,\lsun, 
indicating a positive correlation between outflow velocity and AGN luminosity. 
%Similar to the test between $L_{\rm [OIII]}$ and $L_{\rm AGN}$ (Section \ref{subsec:results_AGN}), 
We perform an ordinary least-squares bisector fit between $v_{\rm out}$ and $L_{\rm AGN}$ 
after normalizing with the value ranges in both of $x$- and $y$-axes in logarithmic grid 
for the 136 galaxies.
The \citet{Schmitt1985} binning method is used to account for the estimated upper limit of $L_{\rm AGN}$ for the ULIRGs with barely AGN detection (i.e., $\beta_{\rm AGN, MIR}<1$).
%$L_{\rm [OIII]}^{\rm obs}/\sigma_{\rm [OIII]}^{\rm obs}>5$ and $L_{\rm AGN}>10^9$\,\lsun.  
%93 out of the 103 galaxies are selected within the 95\% criterion to avoid the effect of extreme values (Section \ref{subsec:results_AGN}). 
The derived regression function is 
\begin{equation}
    \log v_{\rm out}= 0.25\,(\pm0.02)\,\log L_{\rm AGN} - 0.14\,(\pm0.23), 
\end{equation} 
with $r_{\rm s}=0.46$ and $p$-value of $2\times10^{-8}$. 
The fitting method is also employed for $v_{\rm out}$ and AGN bolometric contribution ($f_{\rm AGN}$), and the regression function is 
\begin{equation}
    \log v_{\rm out}= 0.33\,(\pm0.03)\,\log f_{\rm AGN} + 3.13\,(\pm0.04), 
\end{equation} 
with $r_{\rm s}=0.37$ and $p$-value of $1\times10^{-5}$. 
We also investigate the association between $v_{\rm out}$ and the AGN Eddington ratio ($\lambda_{\rm Edd}$, Figure \ref{fig:vout_EddAGN}). 
In order to estimate the Eddington luminosity, we derive the mass of central black hole ($M_{\rm BH}$) using total stellar mass ($M_{\rm star, tot}$) and the empirical $M_{\rm BH}$-$M_{\star}$ relationship in the local universe \citep{Reines2015}: 
\begin{equation}
    \log_{10}(\frac{M_{\rm BH}}{10^8\ \rm{M_{\odot}} }) = 1.05 \times \log_{10}(\frac{M_{\rm star}}{10^{11}\ \rm{M_{\odot}} })-0.55.
\end{equation} 
Then we obtain 
\begin{equation}
    \log v_{\rm out}= 0.25\,(\pm0.02)\,\log \lambda_{\rm Edd} + 2.90\,(\pm0.04), 
\end{equation} 
with $r_{\rm s}=0.34$ and $p$-value of $5\times10^{-5}$. 

Comparing the coefficients among the relations, i.e., $v_{\rm out}$-$L_{\rm AGN}$, $v_{\rm out}$-$f_{\rm AGN}$, and $v_{\rm out}$-$\lambda_{\rm Edd}$, we find that the outflow velocity is more tightly associated with the absolute power of AGN ($L_{\rm AGN}$) than the relative intensity ($f_{\rm AGN}$ and $\lambda_{\rm Edd}$). 
These results suggest an intense AGN activity (e.g., $L_{\rm AGN}>3\times10^{11}$\,\lsun, $f_{\rm AGN}>0.1$, $\lambda_{\rm Edd}>0.1$) is necessary to blow out a fast outflowing wind (e.g., $v_{\rm out}>1000$\,\kms). 
However, the galaxies with a powerful AGN but showing moderate outflow ($v_{\rm out}\sim500$\,\kms) 
imply that the existence of a luminous AGN could not always invoke a strong outflow.

The bisector fit is also carried out only for the 46 galaxies with 
obvious AGN MIR excess feature ($\beta_{\rm AGN, MIR}>3$), 
and the regression results are shown in red lines in Figure \ref{fig:vout_Lum2} and \ref{fig:vout_EddAGN} for the correlations, respectively.
Compared to the results for the entire sample, the regression functions for the 46 galaxies show weaker correlations (e.g., $r_{\rm s}=0.26$ vs.\,0.46 for $v_{\rm out}$-$L_{\rm AGN}$ coefficients), which is partly due to the lack of faint AGN sources (e.g., $L_{\rm AGN}<10^{11}$\,\lsun) among the 46 galaxies. 

Finally we test the association between $v_{\rm out}$ and stellar bolometric luminosity ($L_{\rm star}$; Figure \ref{fig:vout_Lum2}, right panel). 
The bisector fit function is 
\begin{equation}
    \log v_{\rm out}= 0.86\, (\pm0.07)\,\log L_{\rm SF} - 7.99\, (\pm0.82), 
\end{equation} 
with $r_{\rm s}=0.29$ and $p$-value of $7\times10^{-4}$. 
The coefficient of $v_{\rm out}$-$L_{\rm SF}$ relation, 0.29, is smaller than $r_{\rm s}=0.46$ of $v_{\rm out}$-$L_{\rm AGN}$ relation, 
implying that the outflow velocity is more tightly associated with AGN activity. 
% \cyan{
% The $v_{\rm out}$-$L_{\rm SF}$ regression function for the 46 galaxies with reliably detected AGN ($\beta_{\rm AGN, MIR}>3$, red line in Figure \ref{fig:vout_Lum2} right panel) lies over the fitting line for the entire sample, which indicates again the correlation between outflow velocity and AGN intensity. 
% }

% \begin{figure}
%     \begin{center}
%     \includegraphics[trim=0 20pt -10pt 0, width=\columnwidth]{Velocity_LumSF.pdf}
%     \end{center}
%     \caption{
%     \oiii\ outflow velocity ($v_{\rm out}$) vs. star formation bolometric luminosity ($L_{\rm SF}$). 
%     %For the 93 ULIRGs with AGN detection (blue squares), the linear fit function is $\log v_{\rm out}= 0.09(\pm0.10)\,\log L_{\rm SF} + 1.68(\pm1.27)$ with $\rho_{\rm pcc}=0.09$ (blue solid line). 
%     %For the total 123 ULIRGs with \oiii\ flux S\,/\,N\,$>5$ and 95\% criterion (blue squares and gray triangles), the linear fit function is $\log v_{\rm out}= 0.12(\pm0.08)\,\log L_{\rm SF} + 1.22(\pm1.00)$ with $\rho_{\rm pcc}=0.13$ (green dashed line). 
%     The legends are the same as in Figure \ref{fig:vout_LumAGN}. 
%     }
%     \label{fig:vout_LumSF}
% \end{figure}

%The correlation between outflow velocity and AGN activity suggests AGN’s capability to affect the evolution of the host galaxy. It is consistent with the merger induced evolutionary scenario that powerful outflows are originated as ULIRGs evolve from weak-AGN phase to luminous-AGN phase, which can blow out the gas and dust, terminate the star formation in the galaxy and finally transfer the ULIRG to quiescent elliptical galaxy. 

\subsection{Co-existence of fast outflows and intense starbursts}
\label{subsec:discuss_vout_SFR}

In the evolutionary scenario of massive galaxies, the AGN-driven outflows are required to be powerful enough to expel the gas completely from the host galaxy potentials to effectively terminate star formation in the host galaxies and remove the fuel for the proceeding accretion onto the black holes (e.g., \citealp{DiMatteo2005, Hopkins2008}). 
In order to check whether the fast outflowing gas are capable of escaping from the gravitational potential of the host galaxy and the embedded halo, we estimate the escape velocity in different spatial scales. 

\begin{figure}
    \begin{center}
    \includegraphics[trim=0 20pt -10pt 0, width=\columnwidth]{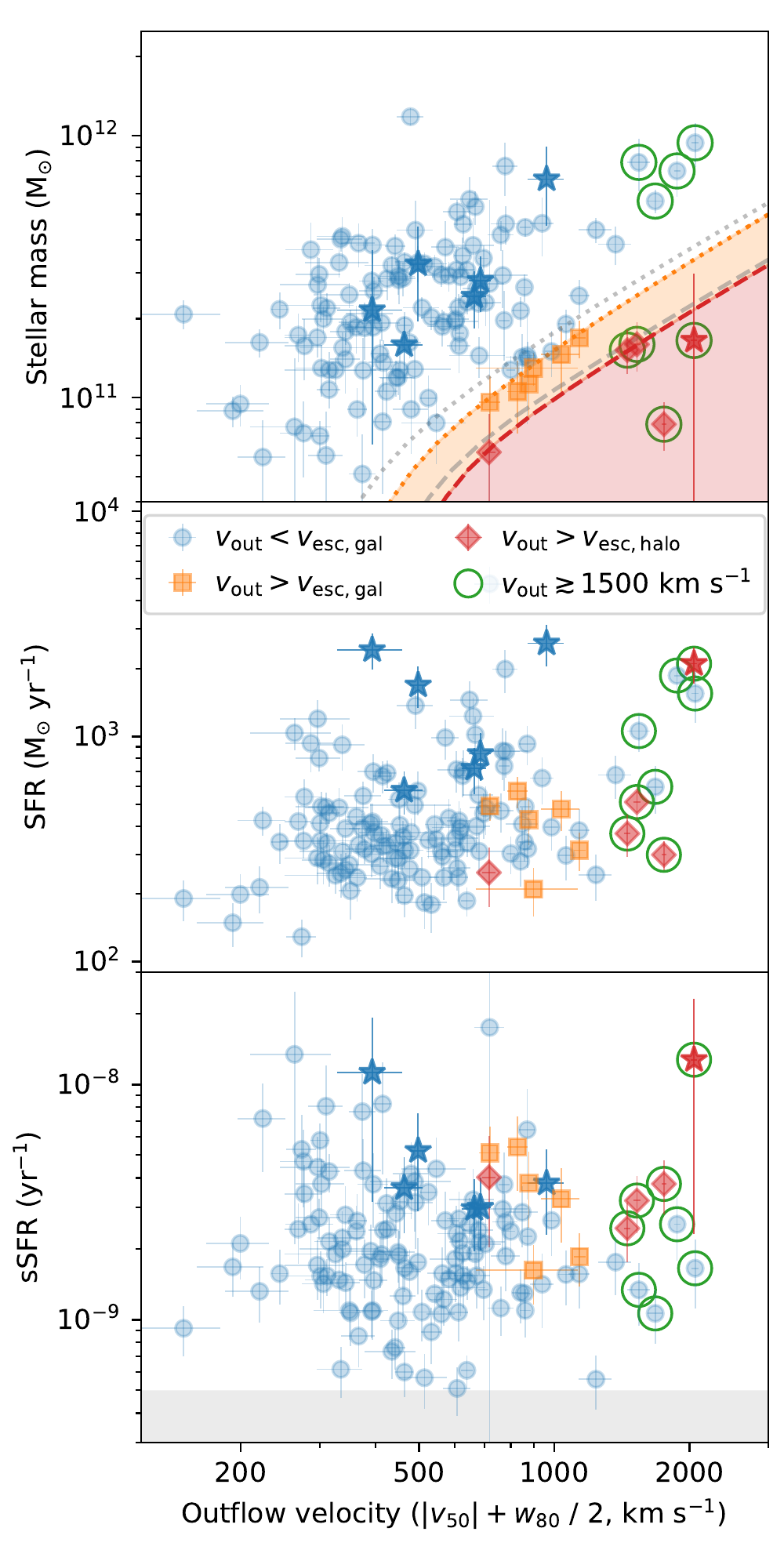}
    \end{center}
    \caption{
    \textbf{Upper:} 
    Stellar mass of the host galaxy vs. outflow velocity ($v_{\rm out}$). 
    The galaxies with $v_{\rm out} > v_{\rm esc, halo}$, $v_{\rm esc, gal} < v_{\rm out} \le v_{\rm esc, halo}$ and $v_{\rm out} \le v_{\rm esc, gal}$ are shown in red diamonds, orange squares, and blue filled circles, respectively. 
    The star markers denote the galaxies spectroscopically identified by FOCAS. 
    The green open circles mark the eight galaxies with extremely fast outflow ($v_{\rm out} \gtrsim 1500$\,\kms). 
    %gray dots denote the galaxies with detection of \oiii\ in low S\,/\,N. 
    The red dashed and orange dotted curves show the escape velocity of the halo and the galaxy, i.e., $v_{\rm esc, halo}$ and $v_{\rm esc, gal}$, respectively, starting from the galaxy center for a galaxy at $z=0.23$, the average redshift of the sample. 
    The gray dashed and dotted curves show the same estimates but starting from the galaxy outskirt region, i.e., approximately 5 kpc from the center. 
    %$0.01\,r_{\rm vir}$, where $r_{\rm vir}$ is the virial radius of the halo ($\sim500$\,kpc for a galaxy with $M_{\rm star}=10^{11}$\,\msun). 
    \textbf{Middle:} 
    SFR vs. $v_{\rm out}$ for \akari-selected ULIRGs, with the same legends as upper panel. 
    \textbf{Bottom:} 
    sSFR vs. $v_{\rm out}$ with the same legends. 
    The gray hatched region denotes the star-forming main sequence ($z=0$--$0.5$, \citealp{Peng2014})
    }
    \label{fig:vout_Mass}
\end{figure}

Firstly we model the gravitational potential profile using the combination of bulge, disk, and dark matter halo components. 
The bulge and disk components are assumed to follow de Vaucouleurs law and exponential law, respectively \citep[e.g.,][]{Sofue2009}.
We adopt the average bulge-to-disk ratio of local ULIRGs, i.e., 1.5 \citep{Veilleux2006}, to estimate the masses of bulge and disk from the total stellar mass (Equation \ref{equ:Mass_stars_tot}). 
The radii of bulge and disk components are estimated using the empirical relationship between radius and mass of local early-type and late-type galaxies, respectively \citep{Shen2003}. 
The halo mass is estimated using stellar mass and the galaxy to halo mass ratio, which is a function of the galaxy stellar mass, from the multi-epoch abundance matching simulation \citep{Moster2013}. 
The gravitational potential at a given distance ($r$) from the center, $\Psi_{\rm halo}(r)$, can be calculated using a spherical 
Navarro-Frenk-White (NFW) potential profile \citep{Navarro1996,Lokas2001}. 
%Then the virial velocity of the halo is estimated following Mo \& White (2002). The circular velocity ($v_{\rm cir}(r)$) at a given distance ($r$) from the center can be calculated using the virial velocity and a spherical NFW potential profile (Lokas \& Mamon 2001). 
Then the total gravitational potential can be calculated as 
$\Psi(r)=\Psi_{\rm bulge}(r)+\Psi_{\rm disk}(r)+\Psi_{\rm halo}(r)$. 
We define the escape velocity as the minimum velocity required for the gas to travel from the galaxy center to a outer radius $r_{\rm out}$, i.e., 
$v_{\rm esc} = \sqrt{2|\Psi(r_{\rm in}=0)-\Psi(r_{\rm out})|}$. 
Then the escape velocity of the halo, i.e., $v_{\rm esc, halo}$, is calculated at $r_{\rm out}=r_{\rm vir}$, where $r_{\rm vir}$ is the virial radius of the halo ($\sim500$\,kpc for a galaxy with $M_{\rm star}=10^{11}$\,\msun; see also Section 5.2.2 of \citealp{Harrison2012}). 
The escape velocity of the galaxy, i.e., $v_{\rm esc, gal}$, is calculated for $r_{\rm out}=0.1\,r_{\rm vir}$. 
Note that both of $v_{\rm esc, halo}$ and $v_{\rm esc, gal}$ could be overestimated since we assume a starting point at the center of the galaxy, 
the proportion of the overestimation is 10\%--15\% if we consider a starting point at the galaxy outskirt region, i.e., $r_{\rm in}=0.01\,r_{\rm vir}$ (5\,kpc with $M_{\rm star}=10^{11}$\,\msun) from the galaxy center (gray curves in Figure \ref{fig:vout_Mass}, upper panel). 
The 136 ULIRGs with \oiii\ detections can be divided into three groups: 
(1) 5 ULIRGs with $v_{\rm out} > v_{\rm esc, halo}$; 
(2) 6 ULIRGs with $v_{\rm esc, gal} < v_{\rm out} \le v_{\rm esc, halo}$; 
(3) 125 ULIRGs with $v_{\rm out} \le v_{\rm esc, gal}$ 
(Figure \ref{fig:vout_Mass}). 

The comparison between the outflow and escape velocity suggests that five ULIRGs possess outflows which are fast enough to escape from the gravitational potential of the host halos, and the other six ULIRGs show the capability to escape from the host galaxies. 
However, those galaxies possess a comparable or even higher SFR and specific star formation rate (sSFR) compared to the ULIRGs with moderate outflows ($v_{\rm out} \le v_{\rm esc, gal}$; see Figure \ref{fig:vout_Mass}, middle and bottom panels). 
The co-existence of the strong outflow and vigorous starburst possibly suggests that the star formation has not yet been suppressed, even though the powerful wind show the capability to blow the gas out of the intergalactic environment ($v_{\rm out} > v_{\rm esc, halo}$). 
It is probable that the wind is clumpy, which could pass through the ISM without a severe influence on the star-forming region. 

A positive AGN feedback, i.e., outflow compresses the ISM and enhances star formation, could be an alternative possibility. 
\citet{Ishibashi2013} discussed the scenario that the AGN feedback triggers star formation and contribute to the build up of the outskirt region as the gas being pushed out by the outflow. 
Their scheme predicts that the triggered SFR (about 100\,\sfrunit) scales with the outflow velocity. 
A direct evidence for star formation within galactic outflow itself in a mergering system, IRAS F23128-5919, has been reported by \citet{Maiolino2017}. The SFR in the outflow, 15--30\,\sfrunit, can contribute to about 25\% of the total SFR in the galaxy. 
A spatially resolved spectroscopical survey to map the outflow and star formation activity 
is required to determine if a much higher SFR (300--3000 \sfrunit) can be enhanced by 
AGN-drived wind in the \akari-selected ULIRGs.

\subsection{No clear evidence of the evolution path from SF-dominated to AGN-dominated ULIRGs}
\label{subsec:discuss_evolution_path}

\begin{figure}[t]
    \begin{center}
    \includegraphics[trim=0 20pt -10pt 0, width=\columnwidth]{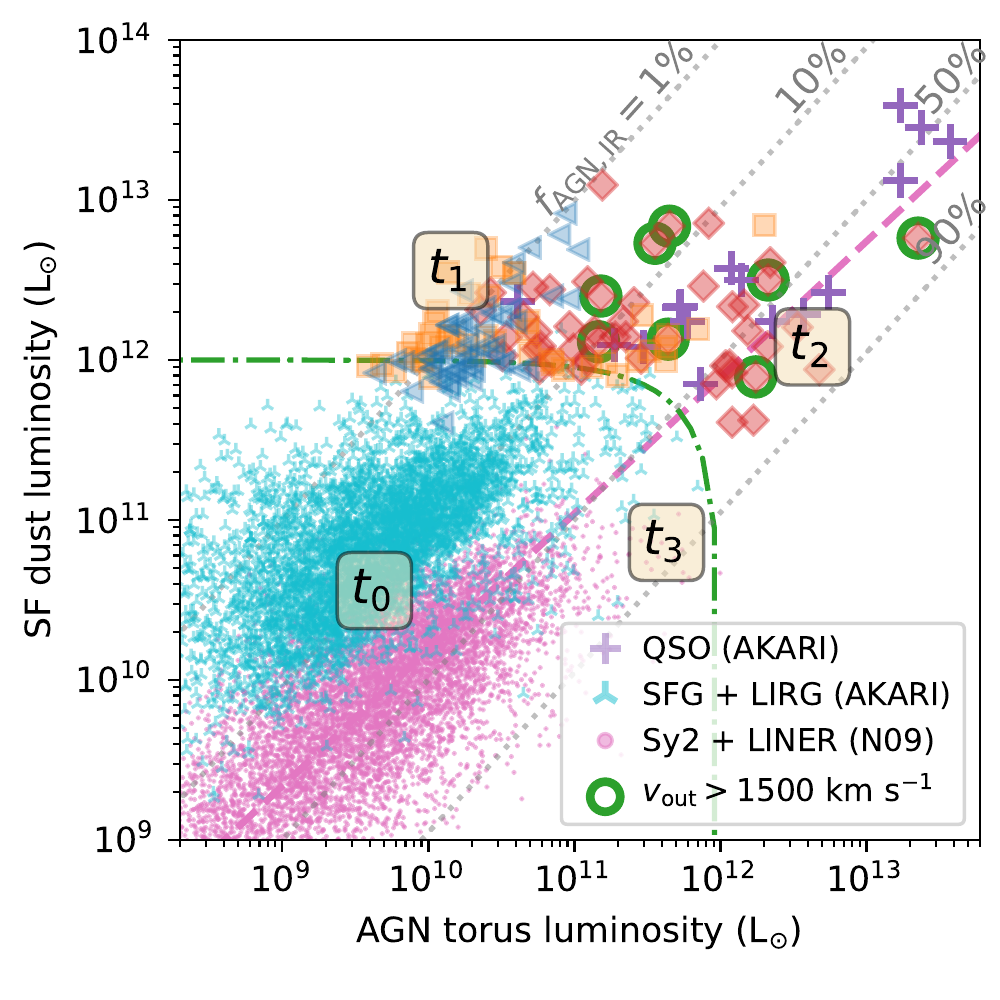}
    \end{center}
    \caption{
    Star formation dust luminosity ($L_{\rm SF, dust}$) vs. AGN torus luminosity ($L_{\rm torus}$) of \akari-selected ULIRGs. 
    The red diamonds, orange squares, and blue filled circles denote to the ULIRGs with AGN MIR excess ratio 
    $\beta_{\rm AGN, MIR}<1$, $1<\beta_{\rm AGN, MIR}<3$, and $\beta_{\rm AGN, MIR}>3$, respectively
    (see Section \ref{subsec:results_AGN} and Figure \ref{fig:FracAGN_LBol}). 
    The green open circles show the eight ULIRGs with $v_{\rm out} \gtrsim 1500$\,\kms. 
    The green dash-dotted curve shows the rough selection threshold for ULIRGs ($L_{\rm IR, total} \simeq L_{\rm SF, dust} + L_{\rm torus} \ge 10^{12}$\,\lsun). 
    The violet corsses show the \akari-selected luminous Type-1 quasars (see Section \ref{subsec:spectral_fitting}). 
    The cyan markers ($\Yup$) denote the normal star-forming galaxies and LIRGs selected from \akari\ sample with $L_{1\textup{--}1000}<10^{12}$\,\lsun. 
    The purple dots with the dashed fitting line show the local Seyfert 2 galaxies sample of \cite{Netzer2009}. 
    See Section \ref{subsec:discuss_evolution_path} for the estimation method of $L_{\rm SF, dust}$ and $L_{\rm torus}$ for the reference quasars and galaxies. 
    The gray dotted lines denote the positions with AGN IR contribution ($f_{\rm AGN, IR}=L_{\rm torus}/(L_{\rm torus}+L_{\rm SF, dust})$) of 1\%, 10\%, 50\%, and 90\%, respectively. 
    }
    \label{fig:evolution_path}
\end{figure}

\begin{figure}[t]
    \begin{center}
    \includegraphics[trim=0 20pt -10pt 0, width=\columnwidth]{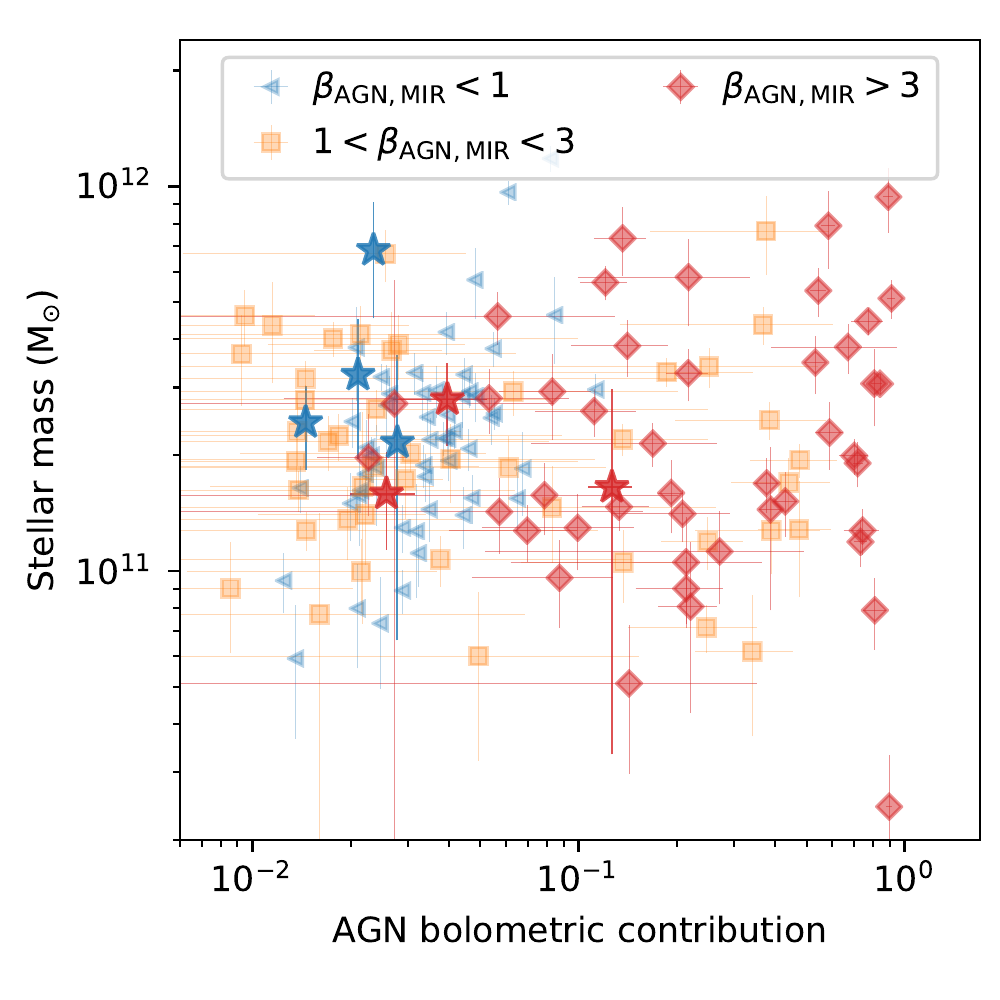}
    \end{center}
    \caption{
    Total stellar mass ($M_{\rm star, tot}$) versus AGN bolometric contribution ($f_{\rm AGN}$) of \akari-selected ULIRGs. 
    The red, orange, and blue markers correspond to the detection ratios of AGN MIR excess (see Section \ref{subsec:results_AGN} and Figure \ref{fig:FracAGN_LBol}). 
    The stars denote the galaxies spectroscopically identified by FOCAS, 
    and the other type markers show the galaxies spectroscopically identified by SDSS. 
    The result suggests that there is no apparent difference in stellar mass between SF-dominated and AGN-dominated ULIRGs, 
    i.e., non-correlation between $M_{\rm star, tot}$ and $f_{\rm AGN}$. 
    %The legends are the same as in Figure \ref{fig:FracAGN_LBol} and \ref{fig:vout_Lum2}. 
    }
    \label{fig:Mstar_FracAGN}
\end{figure}

In the commonly accepted scenario of the evolution of massive galaxies (e.g., \citealp{Sanders1988, Hopkins2008, Netzer2009, Casey2014, Ichikawa2014}), ULIRGs plays an intermediate role in the transition from normal star-forming galaxies to quiescent elliptical galaxies that host luminous quasars, as they evolve from a cold (red MIR-FIR color), SF-dominated phase to a warm (blue MIR-FIR color), AGN-dominated phase. 
%The evolutionary path is usually described with the $L_{\rm SF}$-$L_{\rm AGN}$ diagram as shown in Figure \ref{fig:evolution_path}. 
Figure \ref{fig:evolution_path} shows the star formation dust luminosity 
($L_{\rm SF, dust}=L_{\rm ISM}+L_{\rm BC}$, where $L_{\rm ISM}$ and $L_{\rm BC}$ are the dust luminosity in the diffuse region and the birth clouds, 
respectively) 
and the AGN torus dust luminosity ($L_{\rm torus}$) of \akari-selected ULIRGs estimated from the SED decomposition. 
In order to compare with the results of the ULIRGs to other samples, here we take the dust re-emitted luminosities instead of the intrinsic primary radiation discussed in Section \ref{sec:chap2_results}. 
The AGN IR contribution can be described as $f_{\rm AGN, IR}=L_{\rm torus}/(L_{\rm torus}+L_{\rm SF, dust})$. 
The optical-IR SED fitting derives an empirical function between $f_{\rm AGN, IR}$ and the AGN bolometric contribution $f_{\rm AGN}$, i.e., 
\begin{equation}
    \frac{f_{\rm AGN, IR}^{\,\,-1}-1}{f_{\rm AGN}^{\,\,-1}-1} = 
    \frac{L_{\rm SF, dust}\,/\,L_{\rm star}}{L_{\rm torus}\,/\,L_{\rm AGN}} = 4\pm2, 
\end{equation}
for the 149 \akari-selected ULIRGs. 

In Figure \ref{fig:evolution_path} we also show the $L_{\rm SF, dust}$ and $L_{\rm torus}$ of 
LIRGs ($10^{11}$\lsun\,$\le L_{1\textup{--}1000}< 10^{12}$\lsun) and 
normal star-forming galaxies (SFG, $L_{1\textup{--}1000}< 10^{11}$\lsun), 
which are also selected from the \akari\ FIR catalog (see Section \ref{sec:chap2_sample}), 
as well as the \akari-selected ULIRGs but identified as Type-1 quasars with optical spectra (see Section \ref{subsec:spectral_fitting}). 
The $L_{\rm SF, dust}$ and $L_{\rm torus}$ are estimated using the 2-band method (Equation \ref{equ:SED_2_band_1}) 
for the LIRGs, SFGs, and quasars. 
In addition, we also plot the results of the local optically selected Seyfert 2 galaxies presented by \citet{Netzer2009}. 
The $L_{\rm torus}$ of the Seyfert 2 galaxies is estimated from the reported AGN bolometric luminosity \citep[][see their Figure 13]{Netzer2009}
with the empirical $L_{\rm AGN}$-$L_{\rm MIR}$ relationship of \citet{Ichikawa2014}. 

The time tags, i.e., $t_0$--$t_3$, in Figure \ref{fig:evolution_path} 
show a suggested evolution path for a `single event' merger \citep{Netzer2009}. 
The merger occurring at time $t_0$ leads to a catastrophic gas inflow and an intense star formation activity, 
which moves the system to SF-dominated ULIRG period at $t_1$. 
As the inflowing gas reaches into the central region of the galaxy, the luminosity of AGN increases
with a peak activity at $t_2$. 
The SFR begins to decrease due to the consuming of the gas and the feedback effect of starburst and\,/\,or AGN. 
Finally both of the star formation and AGN activities decay as a result of the diminishing supply of cold gas of the star-forming region and the central black hole ($t_3$). 

One prediction of the evolutionary scenario is that as the ULIRG evolves to AGN luminous phase, the outflow becomes intense and which provides one of the main mechanism to blow out the gas and dust, and finally terminate the star formation in the galaxy. 
The correlation between outflow and AGN activity is consistent with our results from the \oiii\ and MIR observations (Section \ref{subsec:discuss_vout_LAGN}). 
However, the ULIRGs with extreme outflows do not show the suppression of SFR compared to the ULIRGs with moderate outflows (Section \ref{subsec:discuss_vout_SFR}).
% This is consistent with the observed correlation between the outflow velocity and the luminosity of AGN. 
% %Our results show consistency with the prediction that both of the luminosity and the outflow velocity of \oiiiblong\ line is highly correlated to AGN luminosity and energy contribution to ULIRGs. 
% Only ULIRGs with $L_{\rm AGN}\ge10^{11.5}$\,\lsun\ show extremely fast outflow, e.g., $v_{\rm out}\ge 1500$\,\kms\ or $v_{\rm out}\ge v_{\rm esc, halo}$. % (purple and violet squares in Figure \ref{fig:evolution_path}). 

Another prediction of the merger scenario is that 
ULIRGs experience continuous build up of stellar mass 
as a result of the intense SFR during their lifetimes (e.g., $t_1$--$t_2$ in Figure \ref{fig:evolution_path}), 
thus the later-stage, AGN-dominated (higher $f_{\rm AGN}$) ULIRGs are expected to possess larger stellar mass than the early-stage, SF-dominated ULIRGs. 
The comparison between the total stellar mass ($M_{\rm star, tot}$) and the bolometric AGN contribution ($f_{\rm AGN}$) is shown in Figure \ref{fig:Mstar_FracAGN}. 
% An ordinary least-squares bisector fit is performed between $M_{\rm star, tot}$ and $f_{\rm AGN}$ and the best-fit function is 
% \begin{equation}
%     \log M_{\rm star, tot}= 0.38\, (\pm0.02)\,\log L_{\rm AGN} + 11.77\, (\pm0.04), 
% \end{equation}
% where the $M_{\rm star, tot}$ is in the unit of \msun. 
% However, the fitting slope is shallow with coefficient $\rho_{\rm s}$ near to zero (0.03) and a high p-value of 0.7, which indicate that there is no correalation between $M_{\rm star, tot}$ and $f_{\rm AGN}$. 
However, the coefficient $\rho_{\rm s}$ close to zero (0.03) with a high p-value of 0.7, 
which indicate that there is no correalation between $M_{\rm star, tot}$ and $f_{\rm AGN}$. 
The result of non-correlation still holds if only ULIRGs with $\beta_{\rm AGN}>3$ is considered ($\rho_{\rm s}$ of 0.07 and p-value of 0.6). 
Therefore we consider that there is no clear evidence of stellar build up from SF-dominated to AGN-dominated ULIRGs with the \akari-selected sample.

Within the evolutionary scenario the AGN-dominated ULIRGs are expected to possess more transparent star-forming regions with lower $E(B-V)$, as a result of the intense outflow which blow out the gas and dust out of the star-forming regions (e.g., \citealp{Veilleux2009, Hou2011}). 

Similar check is performed for $E(B-V)$ estimated from stellar continua, which reflect the diffuse dust extinction in the star-forming region. 
Within the evolutionary scenario the AGN-dominated ULIRGs are expected to possess more transparent star-forming regions with lower $E(B-V)$, as a result of the intense outflow which blow out the gas and dust out of the star-forming regions (e.g., \citealp{Veilleux2009, Hou2011}). 
However, the result from the \akari-selected sample still shows no correlation between $E(B-V)$ and $f_{\rm AGN}$, with a small coefficient, $\rho_{\rm s}=-0.1$, and a high p-value of 0.1. 

% \begin{figure}
%     \begin{center}
%     \includegraphics[trim=0 10pt -10pt 0, width=0.4\columnwidth]{LumAGN_FracMass.pdf}
%     \includegraphics[trim=0 10pt -10pt 0, width=0.4\columnwidth]{LumAGN_Extinction.pdf}
%     \end{center}
%     %\caption{
%     \red{just compare the average value to literature in the text}
%     Upper: Mass fraction of stars younger than 100 Myr ($f_{\rm 100 Myr}$) versus AGN bolometric luminosity ($L_{\rm AGN}$) of \akari-selected ULIRGs. 
%     Bottom: Extinction ($E(B-V)$) versus $L_{\rm AGN}$. In both of the two panels, 
%     the orange dots denote the results directly estimated from optical spectra, while the blue diamonds show the results corrected for aperture-loss and the hidden starburst (ASB). Orange dot-dashed and blue solid lines are the linear fit for orange and blue dots, respectively. 
%     The best-fits for both $f_{\rm 100 Myr}$-$L_{\rm AGN}$ and $E(B-V)$-$L_{\rm AGN}$ show small coefficients
%     ($\rho_{\rm pcc}=-0.02$, corrected $f_{\rm 100 Myr}$; 0.02, fiber $f_{\rm 100 Myr}$; -0.09, corrected $E(B-V)$; -0.06, fiber $E(B-V)$)
%     and high p-values
%     (0.83, corrected $f_{\rm 100 Myr}$; 0.88, fiber $f_{\rm 100 Myr}$; 0.39, corrected $E(B-V)$; 0.58, fiber $E(B-V)$)
%     indicating that there is no correlation for $f_{\rm 100 Myr}$-$L_{\rm AGN}$ or $E(B-V)$-$L_{\rm AGN}$. 
%     %}
%     \label{fig:LumAGN_fMass_EBV}
% \end{figure}

%The result is consistent with \citet{Rodriguez2010} and \citet{Su2013}, but against the result of \citet{Hou2011}. 

In summary, although we find the correlation between outflow and AGN activity, 
our results do not show clear evidence, e.g., stellar build up, of the transition link from SF- to AGN-dominated ULIRGs. 
%our results do not support the merger-induced evolutionary scenario of ULIRGs because no clear link between ULIRG types (SF- or AGN-dominated) and stellar properties, e.g., stellar mass, is found in our ULIRG sample. 
One possible explanation is that the timescale of the phase transition is much shorter than the lifetime of ULIRGs, 
i.e., it takes much shorter time for gas moving into the vicinity of the central black hole than the duration of intense starbursts (e.g., 100 Myr), 
thus the SF-dominated and AGN-dominated ULIRGs could possess similar stellar populations. 
Several simulation works reported that the fast-transition is possible at some conditions, e.g., the mass ratio of the progenitor galaxies of the merger is much larger than 1:1 \citep{Johansson2008}. 
It is also possible that the selected ULIRG sample is biased to massive galaxies ($10^{11}$--$10^{12}$\,\msun), 
thus the accumulated stellar mass during the ULIRG phase (e.g., $10^{10.5}$\,\msun, assuming SFR of 300\,\sfrunit and 100 Myr duration) could not significantly affect the total stellar mass of the galaxy. 

\section{Conclusions}
\label{sec:chap2_conclusions}

In order to understand the stellar population and outflow properties of ULIRGs to place observational constraints on the evolutionary path of ULIRGs, we construct a 90\,$\mu$m flux limited catalog of 1077 ULIRGs, which are selected from the \akari\ FIR all-sky survey by utilizing the SDSS optical and \wise\ MIR imaging data. 
202 out of the 1077 ULIRGs are spectroscopically identified by SDSS and Subaru/FOCAS observations, in which 149 ULIRGs possess galaxy dominated optical spectra. 
Thanks to the deeper depth and higher resolution of \akari\ compared to the previous \iras\ survey, and reliable identification from \wise\ MIR pointing, the sample is unique in identifying optically-faint (i$\sim$20) IR-bright galaxies, which could be missed in previous surveys. 
Compared to the previous works about stellar population of ULIRGs which focused on narrow-line objects (e.g., \citealp{Hou2011}), our ULIRG sample also provides the unique opportunity to study the evolution of ULIRGs with extremely fast outflows. 
The main results are as follows.

(1) A self-consistent spectrum-SED decomposition method, which constrains stellar population properties in SED modeling based on spectral fitting results, has been employed for 149 ULIRGs with host galaxy dominated spectra (Section \ref{sec:chap2_method}). 
They are identified as massive galaxies ($M_{\rm star}\sim10^{11}$-$10^{12}$\,\msun), associated with intense star formation activity (SFR\,$\sim400$-2000\,\sfrunit). 12 ULIRGs possess SFR exceeding 1000 \sfrunit\ and the ULIRG, J115458.02+111428.8, even shows SFR up to 5000 \sfrunit, indicating one of the most intense starbursts at $z\sim0.5$ (Section \ref{subsec:results_HG}). 

(2) The ULIRGs cover a large range of AGN activity, with bolometric luminosity from $10^{10}$\,\lsun\ to $10^{13}$\,\lsun, 
and the outflow velocity measured from \oiiiblong\ emission line shows a correlation with the AGN bolometric luminosity.
Several galaxies show extremely fast outflow with $v_{\rm out}$ close to 2000 \kms. The outflow velocity of five ULIRGs even exceeds the escape velocity of the host halos. 
However, the co-existence of the strong outflows and vigorous starbursts suggests that the star formation has not yet been suppressed by the outflow 
during the ULIRG phase (Section \ref{subsec:results_AGN}, \ref{subsec:results_outflow}, \ref{subsec:discuss_vout_LAGN}, and \ref{subsec:discuss_vout_SFR}). 

(3) There is no significant discrepancies of the stellar population, e.g., stellar mass and dust extinction, 
between ULIRGs with weak and powerful AGN. 
The results do not show evidence of the transition from SF-dominated to AGN dominated ULIRGs as predicted by the merger-induced evolutionary scenario. 
One possible explanation is that the timescale of the phase transition from SF- to AGN-dominated ULIRGs is much shorter than the lifetime of ULIRGs (Section \ref{subsec:discuss_evolution_path}).

%(4) Finally we discuss the possible effect of AGN-heated galactic scale dust on the estimation of the properties of the ULIRGs. A quenching pattern, i.e., SFR decreases as AGN becomes luminous, appears in the extreme case, in which the AGN is highly obscured by kpc-scale dust. Such discussion suggests the large uncertainties in the analysis of evolution of ULIRGs. The sub-millimeter interferometric observation to determine the dust distribution around AGN is required to address the question (Section \ref{subsec:Appendix_AGN_Polar}). 

\acknowledgments

We thank the anonymous referee for the constructive advice. 
We thank Marko Stalevski for providing the SKIRTor torus SED library 
and Angelos Nersesian for the discussion of the THEMIS ISM dust model. 
This work was supported by the Program for Establishing a Consortium
for the Development of Human Resources in Science and Technology, Japan Science and Technology Agency (JST) and was partially supported by the Japan Society for the Promotion of Science (JSPS) KAKENHI (18K13584; KI).
This research is based on data collected at Subaru Telescope, which is operated by the National Astronomical Observatory of Japan.
This research is based on observations with AKARI, a JAXA project with the participation of ESA. 
This publication makes use of data products from the Wide-field Infrared Survey Explorer, which is a joint project of the University of California, Los Angeles, and the Jet Propulsion Laboratory\,/\,California Institute of Technology, funded by the National Aeronautics and Space Administration.
Funding for the Sloan Digital Sky Survey IV has been provided by the Alfred P. Sloan Foundation, the U.S. Department of Energy Office of Science, and the Participating Institutions. SDSS-IV acknowledges
support and resources from the Center for High-Performance Computing at
the University of Utah. The SDSS web site is www.sdss.org.
SDSS-IV is managed by the Astrophysical Research Consortium for the 
Participating Institutions of the SDSS Collaboration including the 
Brazilian Participation Group, the Carnegie Institution for Science, 
Carnegie Mellon University, the Chilean Participation Group, the French Participation Group, Harvard-Smithsonian Center for Astrophysics, 
Instituto de Astrof\'isica de Canarias, The Johns Hopkins University, Kavli Institute for the Physics and Mathematics of the Universe (IPMU) / 
University of Tokyo, the Korean Participation Group, Lawrence Berkeley National Laboratory, 
Leibniz Institut f\"ur Astrophysik Potsdam (AIP),  
Max-Planck-Institut f\"ur Astronomie (MPIA Heidelberg), 
Max-Planck-Institut f\"ur Astrophysik (MPA Garching), 
Max-Planck-Institut f\"ur Extraterrestrische Physik (MPE), 
National Astronomical Observatories of China, New Mexico State University, 
New York University, University of Notre Dame, 
Observat\'ario Nacional / MCTI, The Ohio State University, 
Pennsylvania State University, Shanghai Astronomical Observatory, 
United Kingdom Participation Group,
Universidad Nacional Aut\'onoma de M\'exico, University of Arizona, 
University of Colorado Boulder, University of Oxford, University of Portsmouth, 
University of Utah, University of Virginia, University of Washington, University of Wisconsin, 
Vanderbilt University, and Yale University.

\bibliography{xmc_akari}

%%%%%%%%%%%%%%%%%%%%%%%%%%%%%%%%%%%%%%%%%%%%%%%%%%%%%%%%%%%%%%%%%%%%%%%%%%%%%%%%%%%%%%%%%%%%%%%%%%%%%%%%%%%%%%%%%%%%%%%%%%%%%
%%%%%%%%%%%%%%%%%%%%%%%%%%%%%%%%%%%%%%%%%%%%%%%%%%%%%%%%%%%%%%%%%%%%%%%%%%%%%%%%%%%%%%%%%%%%%%%%%%%%%%%%%%%%%%%%%%%%%%%%%%%%%
%%%%%%%%%%%%%%%%%%%%%%%%%%%%%%%%%%%%%%%%%%%%%%%%%%%%%%%%%%%%%%%%%%%%%%%%%%%%%%%%%%%%%%%%%%%%%%%%%%%%%%%%%%%%%%%%%%%%%%%%%%%%%
%%%%%%%%%%%%%%%%%%%%%%%%%%%%%%%%%%%%%%%%%%%%%%%%%%%%%%%%%%%%%%%%%%%%%%%%%%%%%%%%%%%%%%%%%%%%%%%%%%%%%%%%%%%%%%%%%%%%%%%%%%%%%
%%%%%%%%%%%%%%%%%%%%%%%%%%%%%%%%%%%%%%%%%%%%%%%%%%%%%%%%%%%%%%%%%%%%%%%%%%%%%%%%%%%%%%%%%%%%%%%%%%%%%%%%%%%%%%%%%%%%%%%%%%%%%
%%%%%%%%%%%%%%%%%%%%%%%%%%%%%%%%%%%%%%%%%%%%%%%%%%%%%%%%%%%%%%%%%%%%%%%%%%%%%%%%%%%%%%%%%%%%%%%%%%%%%%%%%%%%%%%%%%%%%%%%%%%%%

\begin{appendices}

\section{Details on the optical spectral fitting procedure} 
\label{sec:Appendix_specfit} 

The optical spectrum fitting procedure is modified from Quasar Spectral Fitting package (\texttt{QSFit}, v2.0.0, 
\citealp{Calderone2017}), which is an IDL package based on MPFIT \citep{Markwardt2009}. The original code of \texttt{QSFit} is optimized to the analyses of spectra of AGN-dominated systems, e.g., quasar spectra with broad Balmer lines. We modified the codes to support the fitting with host galaxy component with multiple stellar populations as well as the automatic separation between quasar spectra and host galaxy spectra under a given threshold. The details of the fitting procedure are explained as follows. 

\subsection{Components in spectral fitting}

The model used to fit the observed spectrum is a collection of several components, which can be classified into three categories: (1) AGN continua, e.g., power-law continuum; (2) host galaxy continua, e.g., stellar continuum; (3) emission lines, e.g, \ha\ and \oiii. The AGN continua consist of a bending power-law continuum, a Balmer emission continuum and an iron emission line pseudo-continuum. 
The power-law continuum, i.e., $L_{\lambda}\propto\lambda^{\alpha}$ represents the radiation directly from the accretion disc. 
In the spectral fitting the index $\alpha$ is a free parameter with initial value of $-1.76$ \citep{VandenBerk2001} and a range of $\pm0.5$. 
%which can be described as $L_{\lambda}\propto\lambda^{\alpha_1},\, \lambda \ll \lambda_{\rm 0}$ and $L_{\lambda}\propto\lambda^{\alpha_2},\, \lambda \gg \lambda_{\rm 0}$, where $\alpha_1$ and $\alpha_2$ are the power-law slope at blue and red end, and $\lambda_{\rm 0}$ is the bending wavelength (see Appendix B5 of \citet{Calderone2017} for more details). In the spectral fitting $\lambda_{\rm 0}$ is a free parameter which is limited within the observed wavelength range. $\alpha_1$ and $\alpha_2$ are also free parameters with initial value of $-1.76$ and $-0.44$ \citep{VandenBerk2001}, respectively, with a range of $\pm0.5$. 
The Balmer continuum is described by the electron temperature ($T_{e}$) and the optical depth ($\tau_{\rm BE}$) at Balmer edge (3645\AA, \citealp{Dietrich2012}). In order to avoid degeneracy between parameters we fix $\tau_{\rm BE}=1$, following \citet{Dietrich2012}. $T_{e}$ is set as a free parameter to obtain a variable slope of Balmer continuum, with an initial value of $T_{e}=$10000\,K and the range of 7500--30000\,K. In addition, the pseudo-continuum generated from the blending high order Balmer emission lines (H11-H50) are also considered in the fitting procedure. The line intensities are tied using the theoretical ratios of \citet{Storey1995} at a fixed electron density of $10^9$\,\ccm\ with the same temperature of Balmer continuum. 
The last AGN continuum component is the iron pseudo-continuum from blending of \feii\ emission lines. The iron templates of Narrow Line Seyfert 1 galaxy I ZW 1 from 
\citet{Vestergaard2001} (1420-3090\AA, hereafter VW2001), 
\citet{Veron2004} (3500-7200\AA, hereafter VC2004), and 
\citet{Tsuzuki2006} (2200-3500\AA, and 4200-5600\AA, hereafter TK2006) are used in the fitting procedure. In order to obtain a continuous iron template, we re-normalize the UV template of VW2001 to TK2006 using the integrated flux at 2200-2700\AA, and re-normalize the optical template of VC2004 to TK2006 using the integrated flux at 5100-5600\AA
. Finally the continuous iron template is generated by connecting the VW2001 template at 1420-2200\AA, the TK2006 template at 2200-3500\AA, and the VC2004 template at 3500-7200\AA. 
Note that the public templates of VW2001, VC2004 and TK2006 were un-reddened using different Galactic extinction law, therefore before the connection with each other we re-redden the public templates with the different extinction laws to reproduce the original observed spectra.
The generated template is then corrected for the Galactic extinction using the same CCM extinction law \citep{Cardelli1989} with $E(B-V)=0.105$ 
\citep{Vestergaard2001} and corrected for the intrinsic extinction using 
\citet{Calzetti2000} extinction law with $E(B-V)=0.10$ \citep{Tsuzuki2006} . 

\begin{table*}[ht]
\centering
\caption{Parameters AGN's continuum components. \label{tab:para_spec_AGN}}
\begin{tabular}{llll}
\hline
\hline
Component & Parameter & Range & Fixed/Tied \\
\hline
Power-law cont. & normalization  \\ 
                & extinction $E(B-V)$ & [0, 3] & Yes$^{1}$ \\
                & bending wavelength $\lambda_0$ & obs. wavelength range \\ 
                & slope at blue end $\alpha_1$ & $-1.76\pm0.5$ \\ 
                & slope at  red end $\alpha_2$ & $-0.44\pm0.5$ \\ 
                & curvature & 100 & Yes \\ 
Balmer cont.    & norm. of continuum part \\ 
                & norm. of blended high-order lines \\ 
                & velocity shift (\kms) & [-1000, 1000] & Yes \\
                & velocity FWHM (\kms)  & [ 1000, 3000] & Yes \\
                & extinction $E(B-V)$ & [0. 3] & Yes \\
                & electron temperature (K) & [7500, 30000]  \\
                & electron dentisy (\ccm) & $10^9$ & Yes \\
                & optical depth at Balmer edge & 0 & Yes \\
\feii\ cont.    & normalization  \\ 
                & velocity shift (\kms) & [-1000, 1000] \\
                & velocity FWHM (\kms)  & [ 1000, 3000] \\
                & extinction $E(B-V)$   & [0. 3] & Yes \\
\hline
\multicolumn{4}{l}{$^{1}$\footnotesize{Yes means this is not a free parameter, which is fixed to a given value or tied to }} \\
\multicolumn{4}{l}{\footnotesize{\ \ another parameter in the fitting. }} \\
\end{tabular}
\end{table*}

As for SFH in the spectral fitting, we assume two stellar populations.
One is a underlying main stellar population (MS) and the other represents the ongoing starburst population (SB). Each population is described with a composite stellar population ($S_{\rm CSP}$), which can be calculated by convolving the single stellar population ($S_{\rm SSP}(t, Z)$) and the star formation history (SFR$(t)$): 
\begin{equation}
\begin{split}
    S_{\rm CSP}(t_{\rm p}) = \int_{t_0}^{t_{\rm p}} \mathrm{SFR}(t) S_{\rm SSP}(t_{\rm p}-t, Z) dt, 
    \label{equ:CSP}
\end{split}
\end{equation}
assuming that both of the initial mass function (IMF) and metallicity are time-independent, where $t_{\rm p}$ and $t_0$ are the present time and the time when the galaxy began to form. In this work we adopt the SSP library of \citet{Bruzual2003} with a \citet{Salpeter1955} IMF and a solar metallicity ($Z=0.02$). The SFR$(t)$ of MS and SB components are modeled with $\mathrm{SFR}_{\rm MS}(t)=\mathrm{SFR}_{\rm MS}(t_0)\exp{(-(t-t_0)/\tau_{\rm SF})}$ and $\mathrm{SFR}_{\rm SB}(t)=\mathrm{constant}$, respectively. 
In order to avoid the degeneracy between the time scale of star formation ($\tau_{\rm SF}$) and the beginning time of the galaxy ($t_{\rm 0, gal}$), we fix $\tau_{\rm SF}$ to 1 Gyr, which is a typical value for local massive galaxies with stellar mass $M_{\rm star}>10^{10}$\,\msun\ in the simulation of galaxy evolution (0.5--1.5 Gyr, \citealp{Hahn2017,Wright2019}). Therefore the $S_{\rm CSP}$ of MS and SB components are only determined by the maximum stellar age ($A_{\rm max}=t_{\rm p}-t_0$) of each population. The $A_{\rm max}$ of MS and SB populations are constrained in the range of [1 Gyr, min(15 Gyr, $A_{\rm universe}$)] and [30 Myr, 300 Myr], respectively, where $A_{\rm universe}$ denotes the age of universe at the redshift of the galaxy in unit of Gyr. The range of $A_{\rm max}$ of SB population is determined according to the simulation result of 
galaxy merging \citep{Hopkins2008}, which suggests a typical duration of starburst of 100 Myr. 

\begin{table*}[ht]
\centering
\caption{Parameters of host galaxy's continuum components. \label{tab:para_spec_HG}}
\begin{tabular}{llll}
\hline
\hline
Component & Parameter & Range & Fixed/Tied \\
\hline
MS cont.        & normalization  \\ 
                & velocity shift (\kms) & [-500, 500] \\
                & velocity FWHM (\kms)  & [ 100, 500] \\
                & extinction $E(B-V)$   & [0. 3]  \\          
                & maximum stellar age (Myr)   & [1000, Age of universe] \\
                & minimum stellar age (Myr)   & 0.1 & Yes \\
                & exponential timescale ($\tau_{\rm SF}$, Myr) & 1000 & Yes \\
                & metallicity ($Z_{\odot}$) & 1 & Yes \\
TSB cont.       & normalization  \\ 
                & velocity shift (\kms) & [-500, 500] & Yes  \\
                & velocity FWHM (\kms)  & [ 100, 500] & Yes  \\
                & extinction $E(B-V)$   & [0. 3]  & Yes  \\          
                & maximum stellar age (Myr)   & [30, 300] \\
                & minimum stellar age (Myr)   & 0.1 & Yes \\
                & exponential timescale ($\tau_{\rm SF}$, Myr) & -$^{1}$ & Yes \\
                & metallicity ($Z_{\odot}$) & 1 & Yes \\
\hline
\multicolumn{4}{l}{$^{1}$\footnotesize{SFH with a constant SFR is assumed for TSB.}} \\
\end{tabular}
\end{table*}

\begin{figure}
    \begin{center}
    \includegraphics[width=\columnwidth]{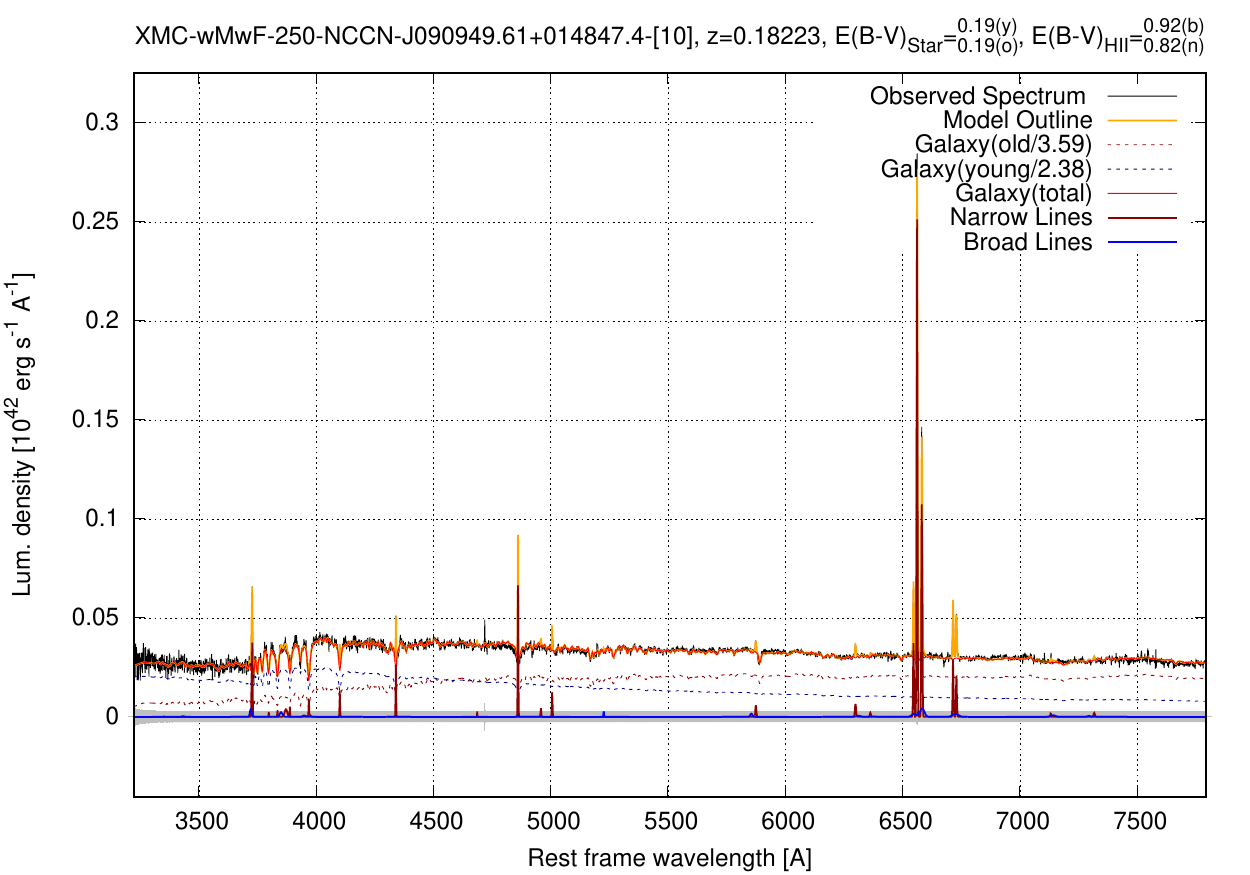}
    \includegraphics[width=\columnwidth]{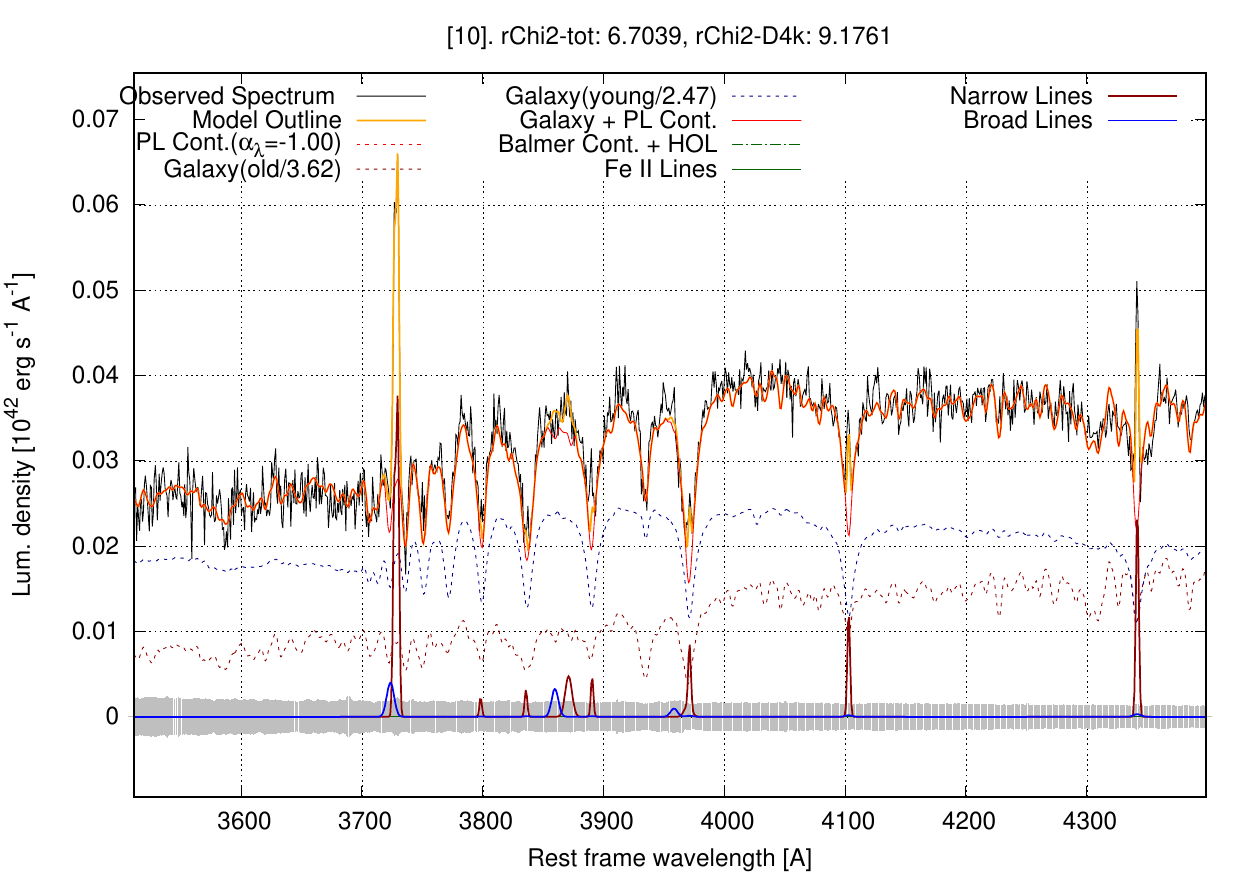}
    \end{center}
    \caption{
    Example of tied Balmer emission lines (\ha--H10, upper panel) and the fitting for Balmer absorption lines around 4000\AA\ (bottom panel) for the ULIRG, J090949.61+014748.4. 
    The observed spectrum is shown in black and the measurement error is shown in gray. 
    The red and blue dotted curves denote the best-fit continuum of MS and TSB, respectively. 
    The red and blue solid curves show the narrow and broad emission lines, respectively. }
    \label{fig:Spec_D4000}
\end{figure}

In the fitting procedure we accounts for most of the emission lines which could be covered by the observed spectrum from \hbox{Ly$\alpha$} to \hbox{[S\sc iii] 9531\AA}.
% \footnote{Permitted lines: Balmer series (\ha--H10), \hbox{Ly$\alpha$}, \hbox{N\sc v 1239\AA}, \hbox{O\sc i 1302\AA}, \hbox{C\sc ii 1335\AA}, \hbox{Si\sc iv 1394\AA}, \hbox{C\sc iv 1548\AA}, \hbox{He\sc ii 1640\AA}, \hbox{O\sc iii 1666\AA}, \hbox{Al\sc iii 1855\AA}, \hbox{C\sc iii 1909\AA}, \hbox{C\sc ii 2325\AA}, \hbox{Mg\sc ii 2796\AA}, \hbox{He\sc ii 4686\AA}, \hbox{He\sc i 5876\AA}, \hbox{He\sc ii 8237\AA}, \hbox{Pa18 8438\AA}, \hbox{O\sc i 8446\AA}}
% \footnote{Forbidden lines: \hbox{[Ne\sc v] 3346\AA}, \hbox{[Ne\sc v] 3426\AA}, \hbox{[O\sc ii] 3726\AA}, \hbox{[O\sc ii] 3729\AA}, \hbox{[Ne\sc iii] 3869\AA}, \hbox{[Ne\sc iii] 3967\AA}, \hbox{[O\sc iii] 4959\AA}, \hbox{[O\sc iii] 5007\AA}, \hbox{[O\sc i] 6300\AA}, \hbox{[O\sc i] 6364\AA}, \hbox{[N\sc ii] 6549\AA}, \hbox{[N\sc ii] 6583\AA}, \hbox{[S\sc ii] 6716\AA}, \hbox{[S\sc ii] 6731\AA}, \hbox{[Ar\sc iii] 7136\AA}, \hbox{[O\sc ii] 7320\AA}, \hbox{[O\sc ii] 7331\AA}, \hbox{[Ar\sc iii] 7751\AA}, \hbox{[S\sc iii] 9069\AA}, \hbox{[S\sc iii] 9531\AA}}. 
The majority of the emission lines are represented with a narrow and a broad Gaussian profiles\footnote{The \hbox{N\sc v 1239\AA} line only has narrow component to avoid the blending with \hbox{Ly$\alpha$}. The \hbox{Mg\sc ii 2796\AA} only has one component with wide FWHM range ([100, $10^4$] \kms) to avoid degeneracy with iron template.}. 
The velocity shift ($v_{\rm off}$) in relative to the systemic redshift for narrow and broad component is constrained in the range of [-500, 500]\,\kms\ and [-1500, 500]\,\kms, respectively. The FWHM of narrow and broad components are constrained in the range of [100, 1000]\,\kms\ and [1000, $10^4$]\,\kms, respectively. 
The line intensity of Balmer emission lines (\ha--H10) is tied with the theoretical ratios of \citet{Storey1995}. The initial ratios for both narrow and broad Balmer lines are taken with an electron temperature of $10^4$ K and an electron density of 100\,\ccm\ to represent the condition of \hii\ region or AGN NLR. If the spectrum is identified as quasar-dominated (in the following Step 2), 
broad Balmer lines from AGN BLR are considered 
and the line ratios are taken with an electron temperature of $10^4$ K and an electron density of $10^9$ \ccm. The velocity shift and FWHM of Balmer lines are tied to each other for narrow and broad components, respectively\footnote{We set a tolerance of 50 \kms for velocity shift of narrow Balmer lines for a good fitting result.}. 
The tied Balmer emission lines are not only used to estimate the extinction in the nebular gas by comparing the theoretical ratio and observed ratio, but also used to improve the fitting quality of stellar continuum by reducing the contamination of emission lines to the decomposition of absorption line features. 
The Balmer absorption line series, e.g., H7 and Ca H complex, are important features to identity the contribution of the stellar continuum and to estimate the ages of MS and SB population. 
One example is shown in Figure \ref{fig:Spec_D4000}. In addition to Balmer lines, we also fix the ratios of a series of forbidden line doublets, e.g., \oiilong\ and \oiiilong\ doublets, for which the line ratios are fixed to theoretical ratios calculated using \texttt{PyNeb} \citep{Luridiana2015} under the typical temperature ($T_{\rm e}=10^4$\,K) and electron density ($n_{\rm e}=100$ cm$^{-3}$) of \hii\ regions and AGN NLR. 
The theoretical line ratios for Balmer lines and other forbidden line doublets are summarized in Table \ref{tab:Line_ratio}. Note that although the velocity shift and FWHM of \siilong\ doublets are tied to each other, the line ratios for both narrow and broad component are not fixed to theoretical values, but are set as free parameters in the fitting procedure. 
The individual absorption component is only enabled for \hbox{Ly$\alpha$}, \hbox{Si\sc iv 1394\AA}, \hbox{C\sc iv 1548\AA}, \hbox{C\sc iii 1909\AA} to account for the absorbers in AGN BLR. 
The `unknown lines' in the original \texttt{QSFit} code are disabled to avoid invoking unknown uncertainties. 

\begin{table}[ht]
\centering
\caption{Parameters of emission line components. \label{tab:para_spec_lines}}
\begin{tabular}{llll}
\hline
\hline
Component & Parameter & Range & Fixed/Tied \\
\hline
Narrow line     & normalization & -$^{1}$ \\ 
                & velocity shift (\kms) & [ -500,  500] & -$^{1}$ \\
                & velocity FWHM (\kms)  & [\ \ 100, 1000] & -$^{1}$ \\
                & extinction $E(B-V)$   & [0. 3]  & -$^{2}$ \\          
Broad line      & normalization  & -$^{1}$ \\ 
                & velocity shift (\kms) & [-1500,  500] & -$^{1}$ \\
                & velocity FWHM (\kms)  & [ 1000, 3000]$^{3}$ & -$^{1}$ \\
                & extinction $E(B-V)$   & [0. 3]  & -$^{2}$ \\          
\hline
\multicolumn{4}{l}{$^{1}$\footnotesize{The fluxes and velocities of Balmer lines and other doublet emisison }} \\
\multicolumn{4}{l}{\footnotesize{\ \ lines are tied, respectively. See Section \ref{subsec:spectral_fitting} and Table \ref{tab:Line_ratio} for details. }} \\
\multicolumn{4}{l}{$^{2}$\footnotesize{See Table \ref{tab:Ebv_tying} for details. }} \\
\multicolumn{4}{l}{$^{3}$\footnotesize{The upper limit is 10000\,\kms\ if the spectrum is quasar identified. }} \\
\end{tabular}
\end{table}

\begin{table}[ht]
\centering
\caption{Theoretical flux ratios of emission lines \label{tab:Line_ratio}}
\begin{tabular}{lc}
\hline
\hline
Lines                           & Ratio           \\
\hline
Balmer lines$^{1}$ (\hii\ or NLR) & 0.053 : 0.073 : 0.105 : 0.159 \\
                                & : 0.259 : 0.468 : 1.000 : 2.863 \\
Balmer lines (BLR)              & 0.087 : 0.109 : 0.142 : 0.198 \\
                                & : 0.297 : 0.500 : 1.000 : 2.615 \\
\hbox{[Ne\sc v] 3346\AA\ 3426\AA} & 1.00 : 2.73 \\
\oiilong\ & 1.00 : 1.36 \\
\hbox{[Ne\sc iii] 3869\AA\ 3967\AA} & 3.32 : 1.00 \\
\oiiilong\ & 1.00 : 2.92 \\
\hbox{[O\sc i] 6300\AA\ 6364\AA} & 3.13 : 1.00 \\
\niilong\ & 1.00 : 2.94 \\
\hline
\multicolumn{2}{l}{$^{1}$\footnotesize{H10, H9, H8, H7, H$\delta$, H$\gamma$, \hb, \ha.}} \\
\end{tabular}
\end{table}

In the spectral fitting the dust reddening ($E(B-V)$) is a general parameter for all of the continuum and line components, although not all of them are free parameters. In the default case, the $E(B-V)$ of MS stellar continuum, iron pseudo-continuum, narrow and broad Balmer lines are fit independently, and the extinctions of other components are tied to one of the above four components. If the narrow (or broad) component of the second brightest Balmer line (which is usually \hb, or H$\gamma$ when \ha\ is not available) is severely weak, i.e., with S\,/\,N\,$<1$, then the extinction of narrow (or broad) lines is also tied to the value of other component, e.g., $E(B-V)$ of MS stellar continuum. A detailed tying relationship of extinction estimation is summarized in Table \ref{tab:Ebv_tying}. 

In order to build a self-consistent spectrum-SED decomposition method, we adopt the extinction law calculated by \citet{Jones2013} with \texttt{DustEM} dust extinction and emission calculator \citep{Compiegne2011} for THEMIS dust model, which is used in the following SED decomposition method to reproduce the thermal emission of interstellar dust. 
Note that the detailed extinction law depends on the size distribution of ISM dust, for the sake of simplicity, in this work we adopt a fixed extinction law from the standard core-mantle diffuse ISM dust model 
\citep{Jones2013,Kohler2014}. 
%See Section \ref{subsec:SED_decomp} for more discussion on this dust model. 

\begin{table}[ht]
\centering
\caption{Extinction fit in the spectral fitting \label{tab:Ebv_tying}}
\begin{tabular}{ll}
\hline
\hline
Independent components & Tied components           \\
\hline
\multicolumn{2}{c}{Quasar dominated spectrum} \\
Iron pseudo-continuum & Power-law continuum \\
Broad Balmer lines    & Balmer continuum \\ 
                      & Other broad permitted lines \\
Narrow Balmer lines   & All the other lines \\ 
\hline
\multicolumn{2}{c}{Host galaxy (HG) dominated spectrum} \\
MS stellar continuum  & SB stellar continuum \\
Narrow Balmer lines   & All the other lines$^{1}$ \\ 
\hline
\multicolumn{2}{c}{HG dominated spectrum, Balmer lines are weak or not available} \\
MS stellar continuum  & SB stellar continuum \\
                      & All the emission lines  \\
\hline
\multicolumn{2}{l}{$^{1}$\footnotesize{Since the broad \hb\ or H$\gamma$ in ULIRG is usually weak, we use the }} \\
\multicolumn{2}{l}{\footnotesize{$E(B-V)$ of narrow Balmer lines to correct for the extinction of narrow }} \\
\multicolumn{2}{l}{\footnotesize{and broad components of all the other emission lines.}} \\
\end{tabular}
\end{table}

In addition to the dust extinction, the other two general parameters for all continuum (except for power-law continuum) and line components are velocity shift ($v_{\rm off}$) and FWHM. The $v_{\rm off}$ and FWHM of emission lines can be directly obtained from the position and width of the Gaussian profile. The $v_{\rm off}$ of the continuum components, i.e., MS, SB, Balmer and iron (pseudo-) continua can also be estimated from the shifting distance in relative to systemic redshift. 
The $v_{\rm off}$ of all the continuum components is set as a free parameter. 
However, in order to estimate the FWHM of a given continuum component, the template should be firstly convolved with a Gaussian profile with different width (the convolution is performed in logarithmic wavelength grid), and then fit with the observed spectrum. Since the convolution is time consuming, the FWHMs of continuum components are fixed to their intrinsic value, thus no convolution is required, until the last step. The intrinsic FWHM of stellar and iron (pseudo-) continua are about 70 \kms\ \citep{Bruzual2003} and 900 \kms\ \citep{Vestergaard2001}, respectively, while for Balmer continuum and pseudo-continuum the FWHM is set as 2000 \kms, a typical dispersion velocity in AGN BLR. During the last step, the FWHM of two stellar continua and iron pseudo-continuum are set as free parameters, while the FWHM of Balmer continuum and pseudo-continuum is tied to the value of broad Balmer emission lines. 

%%%%%%%%%%%%%%%%%%%%%%%%%%%%%%%%%%%%%%%%%%%%%%%%%%%%%%%%%%%%%%%%%%%%%%%%%%%%%%%%%%%%%%%%%%%%%%%%%%%%%%%%%%%%%%%%%%%%%%%%%%%%%
%%%%%%%%%%%%%%%%%%%%%%%%%%%%%%%%%%%%%%%%%%%%%%%%%%%%%%%%%%%%%%%%%%%%%%%%%%%%%%%%%%%%%%%%%%%%%%%%%%%%%%%%%%%%%%%%%%%%%%%%%%%%%

\subsection{Spectral fitting processes}
\label{sec:Appendix_specfit_process} 

The spectral fitting for a given galaxy usually contains tens of free parameters and tens of tied parameters (see Table \ref{tab:para_spec_AGN}--\ref{tab:para_spec_lines} for details), which make it impossible to look for the global minimum $\chi^2$, where $\chi^2=\Sigma (f_{\rm obs}-f_{\rm model}) ^2 / \sigma_{\rm err}^{\,2}$, from the entire parameter space. Fast minimization method, e.g., Levenberg-Marquardt (LM) algorithm, which is the core procedure on \texttt{QSFit} and \texttt{MPFIT} package, is necessary to solve the fitting problem. The modified fitting process can be divided into the following steps. 

(1) Preliminary fit with all components enabled. 
Since the LM algorithm finds only a local minimum, which is not necessarily the global minimum. This feature indicates that the spectral fitting result could significantly depend on the initial values of the parameters. 
In order to reduce the dependency on the initial parameters, 
we follow the approach of \citet{Calderone2017} to add (or enable) the continuum and line components step by step: 
(i) only with power-law and stellar continua; 
(ii) add Balmer and iron (pseudo-) continua; 
(iii) add emission lines, 
and re-run the process at each step. 
The best-fit results from the last step are transferred into the next step, and become as proper initial (or guessing) values of existing components in the next step. Finally all of the continuum and line components discussed above are are included in the fitting to obtain the best fitting. 

(2) Spectral type identification using BLR features. 
In Step (1) all of the components are enabled in the fitting process, while it is not physical in most cases. For example, if the spectrum is host galaxy (hereafter HG) dominated, then iron pseudo-continuum is not necessary. Therefore in this step we identify the type of each spectrum, i.e., HG dominated or quasar dominated, and then update the fitting components set according to the identification results. 
The quasar dominated spectra are separated from the HG dominated spectra using two indicators of AGN BLR feature, i.e., the equivalent width of observed \feii\ and Balmer (pseudo-) continua, and the 80\% width of the line profile for permitted lines (\hb\ and \mgii). See Section \ref{subsec:spectral_fitting} for the details of the spectral identification. 

If the spectrum is identified as HG dominated (e.g., Figure \ref{fig:Spec_D4000}), the AGN related continua (power-law, Balmer and iron) are then disabled in the following steps. 
If the spectrum is quasar dominated, in order to reduce the degeneracy between power-law continuum and SB stellar continuum (which can be featureless in the observed wavelength range and similar to power-law continuum), the SB stellar component is disabled, while the MS stellar component is still kept to represents the contribution of host galaxy. In addition, in the quasar case the theoretical line ratios of broad Balmer lines are modified from NLR conditions ($T_{\rm e}=10^4$\,K, $n_{\rm e}=100$\,\ccm) to BLR conditions ($T_{\rm e}=10^4$\,K, $n_{\rm e}=10^9$\,\ccm) as discussed above. We re-run the process after updating the fitting components to obtain a more physical and reliable fitting result. 

(3) Correction for systemic redshift. 
Based on the fitting result in Step (2), we can correct systemic redshift from the SDSS archived value with the $v_{\rm off}$ of stellar continuum in HG dominated case, or the $v_{\rm off}$ of narrow permitted lines (e.g., \ha\ and \hb) in quasar dominated case. The fitting process is re-run after the correction, thus the output shift velocities of all of the components are in relative to the updated systemic velocity. 
17 galaxies show a deviation between the updated redshift and SDSS archived redshift over 300\,\kms, while the median deviation for the entire sample is estimated to be 130\,\kms. The median uncertainty of the measurement of systemic redshift is about 40 \,\kms.

\begin{figure}
    \begin{center}
    \includegraphics[trim= 0 30pt 0 0, width=0.8\columnwidth]{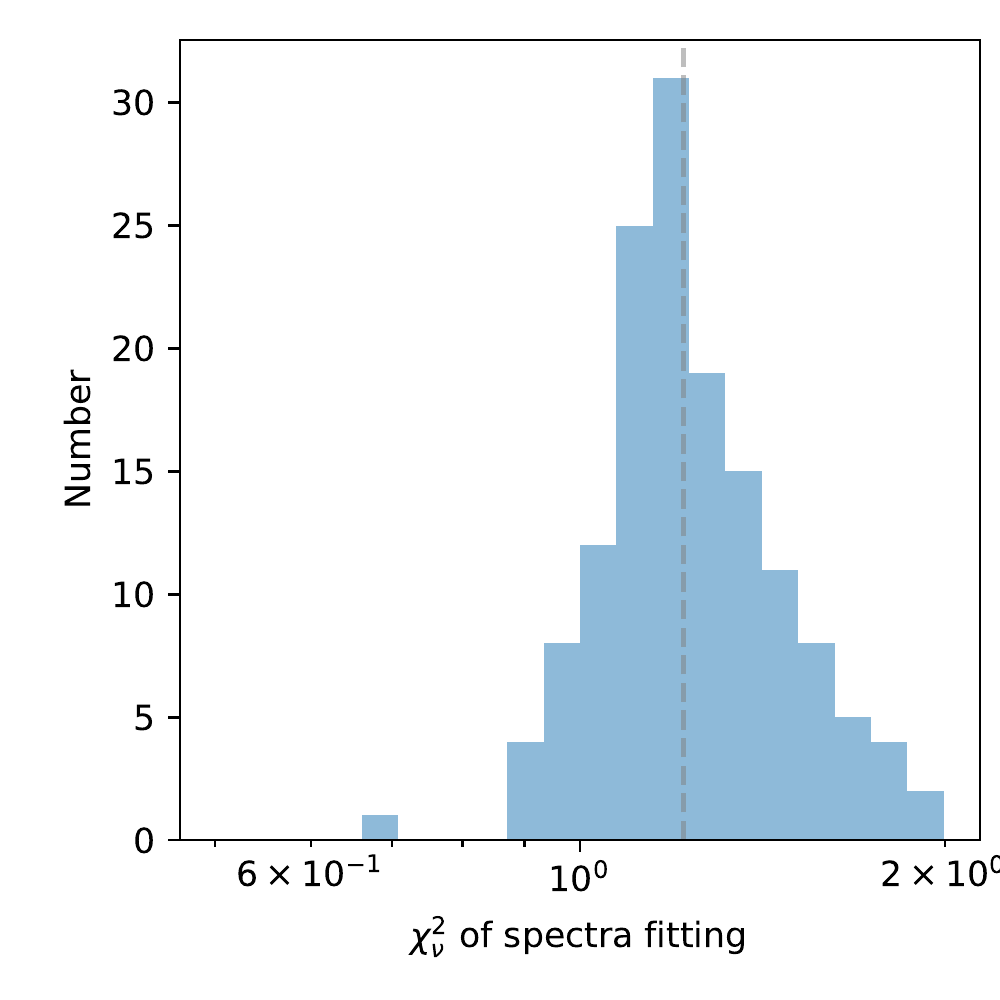}
    \end{center}
    \caption{Reduced $\chi^2$ of spectral fitting for \akari-selected ULIRGs. The vertical dashed line denotes the median $\chi^2_{\nu}$ of 1.2. }
    \label{fig:Spec_chisq}
\end{figure}

(4) Convolution of continua. 
Since the convolution of continuum is time consuming, the FWHM of (pseudo-) continuum component (MS, SB, Balmer and iron) are fixed in the previous fitting process. 
During this step, the FWHM of two stellar continua and iron pseudo-continuum are set as free parameters with a minimum limit of their intrinsic values, while the FWHM of Balmer continuum and pseudo-continuum is tied to the value of broad Balmer emission lines. 
This is the last step in the fitting process for a given spectrum. 
The distribution of reduced $\chi^2$ for the sample are shown in Figure \ref{fig:Spec_chisq}. 

(5) Uncertainty estimation with Monte Carlo simulation. 
We perform the Monte Carlo resampling method to estimate the uncertainties of output results in the fitting process. The method is based on the assumption that the uncertainties on the observed spectrum are Gaussian distributed with measurement errors as standard deviations, i.e., $f_{\rm observed}=f_{\rm intrinsic}+e_{\rm measured}$, where $f_{\rm intrinsic}$ is the ideal intrinsic spectrum without any uncertainties from the observation. A typical approach is to assume $f_{\rm intrinsic}\simeq f_{\rm model}$, where $f_{\rm model}$ is the best-fit model spectrum, and then generate mock spectra by adding random noise to the model spectrum, i.e., $f_{\rm mock}=f_{\rm model}+r_{\rm NG}e_{\rm measured}$, where $r_{\rm NG}$ is the random noise following a normal Gaussian distribution. However, the assumption of $f_{\rm intrinsic}\simeq f_{\rm model}$ can be effective only if the fitting quality is good within the entire wavelength range, otherwise the method will make the mock spectra biased to model spectrum and result in underestimated uncertainty, for example, a weak broad line but with relatively high S\,/\,N. Therefore we modified the mock function to 
\begin{equation}
\begin{split}
    f_{\rm mock}& =f_{\rm observed}+r_{\rm NG}e_{\rm measured} \\
    & = f_{\rm intrinsic}+e_{\rm measured}+r_{\rm NG}e_{\rm measured} \\
    & \simeq f_{\rm intrinsic}+\sqrt{2}e_{\rm measured}, 
    \label{equ:spec_mock}
\end{split}
\end{equation}
In this work for each observed spectrum, 30 mock spectra are generated using Equation \ref{equ:spec_mock}.
Similar fitting processes are performed for the mock spectra and finally for each of the parameters we obtain a distribution of best-fit values. The uncertainty of a given parameter ($U_{\rm p}$) can be estimated using the standard deviation ($\sigma$) of the distribution of the best-fit values, i.e., $U_{\rm p}\simeq\sigma/\sqrt{2}$, 
with the underlying assumption that the scatter of the best-fit values increases by the same times (i.e., $\sqrt{2}$) as the increasing factor of the scatters in the mock spectra. 

As discussed above the fitting procedure is a local $\chi^2$ minimizer and the output may significantly depend on the initial values of parameters. In order to include the systemic errors in the estimated uncertainties of output parameters, in the fitting process of each mock spectrum we employ a set of randomized initial values, i.e., $p_{\rm init, mock}=p_{\rm best, real}+r_{\rm NG}w_{\rm parameter}$, where $p_{\rm init, mock}$ and $p_{\rm best, real}$ are the initial value of mock spectrum and the best-fit output of the real spectrum, respectively; $w_{\rm parameter}$ is the width of the parameter range
\footnote{If the generated $p_{\rm init, mock}$ exceeds the parameter range, then $p_{\rm init, mock}$ is taken as the maximum\,/\,minimum of the parameter range.} 
\footnote{If the parameter is a normalization with the range of $p\ge0$, then we employ $w_{\rm parameter}=p_{\rm best, real}$.}
.

%%%%%%%%%%%%%%%%%%%%%%%%%%%%%%%%%%%%%%%%%%%%%%%%%%%%%%%%%%%%%%%%%%%%%%%%%%%%%%%%%%%%%%%%%%%%%%%%%%%%%%%%%%%%%%%%%%%%%%%%%%%%%
%%%%%%%%%%%%%%%%%%%%%%%%%%%%%%%%%%%%%%%%%%%%%%%%%%%%%%%%%%%%%%%%%%%%%%%%%%%%%%%%%%%%%%%%%%%%%%%%%%%%%%%%%%%%%%%%%%%%%%%%%%%%%
%%%%%%%%%%%%%%%%%%%%%%%%%%%%%%%%%%%%%%%%%%%%%%%%%%%%%%%%%%%%%%%%%%%%%%%%%%%%%%%%%%%%%%%%%%%%%%%%%%%%%%%%%%%%%%%%%%%%%%%%%%%%%
%%%%%%%%%%%%%%%%%%%%%%%%%%%%%%%%%%%%%%%%%%%%%%%%%%%%%%%%%%%%%%%%%%%%%%%%%%%%%%%%%%%%%%%%%%%%%%%%%%%%%%%%%%%%%%%%%%%%%%%%%%%%%

\section{Details on the multi-band SED decomposition} 
\label{sec:Appendix_sedfit} 

\subsection{Dust absorbers and emitters in host galaxies}
\label{subsec:Appendix_ISM_dust}

A widely adopted strategy of multi-band SED decomposition is so-called `energy balance' which requires the conservation between the attenuated primary radiation from stars and\,/\,or AGN, and the re-emitted emission by the dust surrounding the primary emitters. 

We firstly consider the dust absorption and re-emission in host galaxies. The real conditions and properties of dust in the galaxies can be very complicated (e.g., \citealp{Galliano2008, Galliano2018}). In the fitting procedure we employ a modified two-components model presented by 
\citet{Charlot2000} (hereafter CF2000). The CF2000 model provides a simple and brief scenario that 
the young stars are embedded in their birth clouds during a finite time scale (e.g, $t_{\rm BC}$), which are dispersed by strong stellar self-winds or the winds due to nearby supernovae explosions after $t_{\rm BC}$, and then the stars are exposed into the diffuse ISM dust. 
Therefore the emission of young stars is absorbed by their natal clouds at $t<t_{\rm BC}$ and by the diffuse ISM dust at $t>t_{\rm BC}$, while the old stars are only attenuated by diffuse ISM. 

In the spectral fitting (Appendix \ref{sec:Appendix_specfit}) we tie the extinction of starburst (SB) component to the value of main stellar (MS) component. Within the scenario of CF2000 model, the optically observed starburst components represents the young stars for which the natal clouds have been destroyed and migrate to the diffuse ISM dust. For the sake of simplicity, hereafter we name the young stellar population embedded in optically-thin diffuse ISM as `transparent starburst' (TSB) component, and the young stars which are almost fully absorbed by optically-thick natal clouds as `attenuated starburst' (ASB). Under the framework of CF2000 model, ASB and TSB components are separated by a certain timescale $t_{\rm BC}$. 
The timescale $t_{\rm BC}$ are considered to be from 10 Myr for normal star-forming galaxies to 100 Myr for ULIRGs 
\citep{daCunha2008, daCunha2010}, which is similar to the range of age of starburst components in our spectral fitting procedure. 
%Observations of local ULIRGs show that these galaxies are forming stars at high rates in a very concentrated central region with typical scale of at most a few kpc (e.g. CO interferometer observations by Bryant & Scoville 1996; Downes & Solomon 1998; also observations of Arp220 by Sakamoto et al. 2008). In this scenario, the central region of the ULIRG would be similar to a huge, optically-thick molecular cloud with a lifetime typically larger than $10^7$ yr. Further evidence by Lahuis et al. (2007) suggests the presence of high abundances of warm, dense gas, associated with deeply embedded star formation in ULIRGs. In this environment, H ii regions are prevented from expanding by large pressure gradients of gravity, thus increasing the lifetime of the star formation process. Therefore, in the framework of the attenuation model of Charlot & Fall (2000), we adopt tBC = 108 yr for ULIRGs. 
In this work we employ another partly different scenario, that the ASB and TSB components are not separated by $t_{\rm BC}$, but with a similar age. This can be explained by a variable $t_{\rm BC}$ for young stars, or the natal clouds are not fully isotropic and the the UV light is allowed to intersect the clouds in some directions to freely stream away into the diffuse ISM regions \citep{Popescu2011}. 
Within this scenario, we can assume the ASB component has the same stellar population as the TSB component, except for extinctions. The extinction estimated from optical spectral fitting can be used for MS and TSB components that are embedded in optically-thin diffuse ISM. The birth clouds in star-forming regions (e.g., giant molecular clouds) can have heavy extinction ($A_{\rm v}\sim$\,50–-150, \citealp{Reipurth2008}). Using the 9.7 $\mu$m silicate features, \citet{daCunha2010} estimated the optical depths for 16 local ULIRGs, with $\tau_{\rm V}\sim$30--40. In this work we assume $A_{\rm V}=100$ for ASB component for all the galaxies, which corresponds to a typical dense collapsing cloud with spatial scale of 0.2 pc, $n_{\rm H2}=2\times10^5$\ccm, and V-band opacity of $3\times10^3$\,m$^2$/Kg \citep{Shu1987}. 

The distribution of starlight intensities need to be assumed to model the dust radiation. The diffuse ISM dust is usually considered to be exposed in ambient starlight with a constant intensity. 
% Suggesting the stars and dust are uniformly distributed in a given region, the flux intensity $\mathrm{d}U$ (usually in unit of Inter Stellar Radiation Field (ISRF) intensity) received by a dust particle from the stars with a distance $[r, r+\mathrm{d}r]$ can be described as
% \begin{equation}
% \begin{split}
%     \mathrm{d}U &= 4\pi r^2 \, \mathrm{d}r \, n_{\star} \frac{L_{\star}}{4\pi r^2} 10^{- K_{\lambda} \rho r}  = n_{\star}L_{\star} 10^{-K_{\lambda} \rho r} \, \mathrm{d}r , 
% \end{split}
% \end{equation}
% and the total received flux in a spherical volume with radius $R$ is 
% \begin{equation}
% \begin{split}
%     U = \int_{0}^{R} n_{\star}L_{\star} 10^{-K_{\lambda} \rho r}\, \mathrm{d}r  
%       = n_{\star}L_{\star}\frac{10^{-K_{\lambda} \rho R}}{K_{\lambda}\rho \ln{10}}  = \mathrm{constant}, 
%     \label{equ:Dust_diffuse}
% \end{split}
% \end{equation}
% where $n_{\star}$ and $L_{\star}$ are the number density and average luminosity of stars; $K_{\lambda} = 0.4\times1.086 \kappa_{\lambda}$ and $\kappa_{\lambda}$ is the dust opacity; $\rho$ is the mass density of mass. 
The heating for dust in the birth cloud is considered to be dominated by its stellar progeny.
% , then heating intensity received by a dust at a distance $R$ from the central star is:
% %
% \begin{equation}
% \begin{split}
%     U(R) = \frac{L_{\star}}{4\pi r^2} 10^{-K_{\lambda} \int_{0}^{R} \rho(r) dr}. 
%     \label{equ:Dust_PDR}
% \end{split}
% \end{equation}
In the literature the dust in the region with intense radiation field (so-called `PDR' components) is usually described using a power-law distributed starlight intensity, i.e., $\mathrm{d}M/\mathrm{d}U \propto U^{-\alpha}$ 
\citep{Dale2002, Dale2014, Draine2007}. 
Suggesting the dust surrounding the star follows $\rho \propto r^{-\beta}$, 
%$\rho=K_{\rho}r^{-\beta}$
then the two parameters $\alpha$ and $\beta$ are correlated as $\alpha\simeq(5-\beta)/2$ in moderate extinction condition, indicating that the dust with a more concentrating distribution is exposed in more intense heating intensity, and heated to higher temperature when the dust is in thermal equilibrium with the radiation field. 

A detailed ISM dust model is required to reproduce the observed IR SED, which depends on a series of dust properties, e.g, the abundance of carbon, silicate, and other elements; the compositions and structures of dust particle (e.g., crystalline, amorphous, etc.); the distribution of size of dust grains, and so on. In this work we employ the THEMIS model, which is based on the optical properties of amorphous hydrocarbon and amorphous silicate materials measured in the laboratory \citep{Jones2013,Kohler2014}. 
% The dust in THEMIS framework consists of 
% a power-law distribution of small H-poor hydrocarbon grains (a-C) and log-normal distributions of large H-rich hydrocarbon grains (a-C(:H)), as well as an amorphous forsterite-type silicate with iron nano-particle inclusions (a-Sil$_{\rm Fe}$). THEMIS was developed to study the dust evolution in the interstellar medium, which has also been used to explained the diffuse ISM dust extinction and emission in the MW \citep{Jones2013, Ysard2015, Fanciullo2015}. 
%Recently, with THEMIS model Nersesian et al. (2019) reported the relationship between dust radiation and the galaxy morphology parameterized by the Hubble stage for 814 galaxies from DustPedia sample (Davies et al. 2017; Clark et al. 2018). 
%\red{\texttt{DustEM} optically thin limit (no radiative transfer) ???}
The dust extinction curve and radiation SED in THEMIS model are calculated with \texttt{DustEM} codes using the inter stellar radiation field (ISRF, \citealp{Mathis1983}). We take the updated version presented by \citet{Nersesian2019}, in which the model is mainly described with five parameters: 
(1) the mass fraction of small aromatic feature emitting grains, $q_{\rm HAC}$; 
(2) the average heating intensity for diffuse ISM dust, $<U>$; 
(3) the minimum and (4) maximum heating intensity, i.e., $U_{\rm min}$ and $U_{\rm max}$, as well as (5) the power-law slope for dust in birth clouds, $\alpha$. 
$q_{\rm HAC}$ controls the intensity of line features in MIR range (3--20 $\mu$m). 
In the fitting process, we take the range of $q_{\rm HAC}$ from minimum value in THEMIS model, i.e., 0.02, to the value of MW, i.e., 0.17, with an additional extended lower limit of 0.005 to represent the extreme region where the small dust grains are destroyed by strong radiation field, e.g., close to a O/B star \citep{Galliano2018}. 
With THEMIS model, the dust temperature can be approximately estimated from the intensity of radiation field \citep{Nersesian2019}, i.e., 
$T_{\rm dust} = 18.3\,U^{1/5.79}$, where $T_{\rm dust}$ and $U$ are in units of Kelvin and the intensity of ISRF, respectively. 
% \begin{equation}
%     T_{\rm dust}\, (\mathrm{K}) = 18.3\,U^{1/5.79}. 
% \end{equation}
Following \citet{Draine2007} and \citet{Nersesian2019} we assume $U_{\rm min}=<U>$, which represents a smooth temperature transition from the external layer of birth clouds to diffuse ISM dust. 
We also assume the diffuse ISM dust heated by MS and TSB components has the same $U_{\rm min}$, because in practice they are hard to be distinguished. 
The value of $U_{\rm min}$ is constrained in the range of [2, 80], which corresponds to the temperature range of [20, 40]\,K. 
For the galaxies with only one or two photometric detections in FIR range ($> 50 \mu$m), we only test four temperatures: the typical temperature of elliptical and star-forming galaxies, 30 K ($U=17$) and 25 K ($U=7$), respectively \citep{Nersesian2019}; two typical temperature of ULIRGs, 35 K ($U=40$) and 40 K ($U=80$, \citealp{daCunha2010}). 
The $U_{\rm max}$ in the model is fixed to $10^7$, corresponding to temperature about 300\,K, and a distance of 7 AU to a massive star of 10 \lsun. 
The heating intensity power-law slope $\alpha$ is taken from three values, i.e., 2, 2.5, and 3, which corresponds to the mass density index of 1, 0, and -1 with the approximate relationship $\alpha=(5-\beta)/2$ assuming mass density follows $\rho=K_{\rho}r^{-\beta}$. 
$\beta=1$ indicates that the dust concentrated to the center of the birth clouds, representing the early phase of star-formation when the star just formed in the center of the collapsing natal cloud; whereas $\beta=-1$ suggests the majority of the dust locates at the outer region, representing the late phase when the natal cloud is blown out by the stellar wind. 
In summary, in our SED fitting process the dust surrounding stars can be describes by three parameters, $q_{\rm HAC}$, $U_{\rm min}$ (or $<U>$), and $\alpha$. 

%%%%%%%%%%%%%%%%%%%%%%%%%%%%%%%%%%%%%%%%%%%%%%%%%%%%%%%%%%%%%%%%%%%%%%%%%%%%%%%%%%%%%%%%%%%%%%%%%%%%%%%%%%%%%%%%%%%%%%%%%%%%%
%%%%%%%%%%%%%%%%%%%%%%%%%%%%%%%%%%%%%%%%%%%%%%%%%%%%%%%%%%%%%%%%%%%%%%%%%%%%%%%%%%%%%%%%%%%%%%%%%%%%%%%%%%%%%%%%%%%%%%%%%%%%%

\subsection{MIR radiation from AGN torus}

In addition to the dust surrounding stars, the AGN can also contribute to the IR SED of a galaxy by the thermal emission from a thick layer of dust surrounding an accretion disc. The UV-optical radiation from accretion disk is absorbed by the torus and then re-emitted as torus thermal radiation. Since the inner radius of the dusty torus can reach 0.5--1 pc scale, the dust can be heated up to the sublimation temperature, e.g., 1000--1500\,K, which is much higher than the typical temperature of star heated dust, e.g., $\sim100$\,K in PDR, or 20--40\,K in star-forming regions. The torus with much higher temperature shows a significant MIR excess, peaking at $\sim3\mu$m, in the SED of the galaxy. Therefore the MIR torus feature is widely used to identify the AGN activity in galaxies (e.g., \citealp{Ciesla2015,Malek2017}). 

In this work, we employ the SKIRTor torus model developed by 
\citet{Stalevski2012} and updated in \citet{Stalevski2016} with 3D Monte Carlo radiative transfer code \texttt{SKIRT} 
\citep{Baes2003, Baes2011}. The SKIRTor model consist of two-phase medium, i.e., a large number of high-density clumps embedded in a smooth dusty component of low density grains. Compared to the pure smooth or clumpy torus model, the SKIRTor with two-phase medium can produce attenuated silicate features and a pronounced NIR emission at the same time. 
In the current SED library, the fraction of total dust mass inside clumps is fixed to 0.97, corresponding to a volume filling factor of 0.25. The clumpy and smooth dust grains are spatially distributed with density following $\rho(r, \theta)\propto r^{-p}e^{-q|\cos{\theta}|}$, where the $r$ and $\theta$ are the radius and angle in the polar coordinate system. The value of $r$ is limited by the inner and outer radius, i.e.,  $R_{\rm in}$ and $R_{\rm out}$. 
The model is assumed to be scaled with the AGN bolometric luminosity 
% with the function \citep{Barvainis1987}:
% \begin{equation}
%     R_{\rm in} = 1.3\, \mathrm{pc}\times\left( \frac{L_{\rm AGN}}{10^{46}\, \rm erg\,s^{-1}} \right)^{0.5} \left( \frac{T_{\rm sub}}{1500\, \rm K} \right)^{-2.8}, 
% \end{equation}
% assuming an average dust grain size of 0.05 $\mu$m, 
% where $T_{\rm sub}\simeq 1180$\,K is the sublimation temperature of the dust grains in SKIRTor model. Therefore 
and the radial scale of the torus is described using the radius ratio $R_{\rm out}/R_{\rm in}$. 
A larger $R_{\rm out}/R_{\rm in}$ corresponds to 
cooler SED extended SED to FIR range. 
The polar angle $\theta$ of torus is limited by the half opening angle $\Theta$, which is related to the covering factor of the torus, $\mathrm{CF}=\sin{\Theta}$. 
The amount of dust of the torus is described using the 9.7 $\mu$m optical depth in the equatorial direction, i.e., $\tau_{9.7}$. 

Within the framework of SKIRTor model, not only the dusty torus but also the primary central source is anisotropic, with the flux follows $F(\theta)\propto \cos{\theta}(1+2\cos{\theta})\sim \cos{\theta}$. 
It is a more reasonable assumption to describe a disc-like heating source compared to other torus model with isotropic point-like central source. 
The primary source follows a commonly adopted multi-slope bending power-law SED \citep{Schartmann2005}. 
The anisotropic torus is the fundamental scenario for the AGN unified scheme. If the inclination $\varphi$ follows $\varphi+\Theta\le\pi /2$, the primary source is blocked in the line of sight and the AGN is in type-2 geometry; otherwise the central radiation freely streams away and the AGN is in type-1 geometry. 

Except for the fixed clumpy fraction, there are six parameters in SKIRTor model. 
As explained in Section \ref{subsec:spectral_fitting}, in this work we focus on type-2 AGN (for which the optical spectrum is dominated by galaxy component), therefore we only take $\varphi+\Theta\le\pi /2$. 
Recently, the study on X-ray selected AGN reported that the typical covering factor is about 0.6 \citep{Stalevski2016, Mateos2017, Ichikawa2019}, which corresponds to $\Theta\sim30\arcdeg$. 
Due to the degeneracy between $\varphi$ and $\Theta$ and the limited number of data points, it is a proper choice to fix it to the typical value from the literature, i.e., $30\arcdeg$. 
Thus the inclination $\varphi$ is constrained in the range of $[60\arcdeg, \, 70\arcdeg, \, 80\arcdeg, \, 90\arcdeg]$. 
We also fix the spatial distribution to one case with $p=1$ and $q=0.5$, and then take the full available range of $R_{\rm out}/R_{\rm in}$, i.e., [10,\,20,\,30], and the extinction $\tau_{9.7}$, i.e., [3, 5, 7, 9, 11]. 
%Examples of the SED templates in SKIRTor torus model are shown in Figure \ref{fig:SKIRTor_example}. 

% \begin{figure}
%     \begin{center}
%     \includegraphics[trim=0 20pt 0 20pt, width=0.4\columnwidth]{skirtor_sed.pdf}
%     \end{center}
%     %\caption{
%     \red{REMOVE}
%     Examples of the SED templates in the SKIRTor torus model. 
%     The colored curves show the SED of accretion disk (solid) and torus (dashed) with different amounts of extinction ($\tau_{9.7}$) and inclinations ($\varphi$). In the plot the radius ratio ($R_{\rm out}/R_{\rm in}$) is fixed to 30 and the half opening angle ($\Theta$) is fixed to $30\arcdeg$. 
%     The gray solid line denote the intrinsic radiation of the accretion disk in face-on direction ($\varphi=0\arcdeg$). 
%     %}
%     \label{fig:SKIRTor_example}
% \end{figure}

Note that even though we only focus on type-2 AGN with $\varphi+\Theta\le\pi /2$, the primary radiation of the accretion disk is not fully extinct. A part of the radiation of the accretion disk could penetrate the dusty torus, and the fraction of the penetrating radiation depends on the inclination, the amount of the extinction, and the spatial distribution of the dust. 
Since the penetrating component extends from MIR to NIR and optical bands, it is necessary to evaluate whether it significantly affect the fitting of the optical spectrum or not. 
In the SEDs (torus + accretion disk) used in this work, the one with $\tau_{9.7}=11$ and $\varphi=60\arcdeg$ shows the bluest optical-IR color 
%(blue solid + dashed curves in Figure \ref{fig:SKIRTor_example})
, i.e., the largest contribution to optical bands, and the flux density ratio of $S_{3.4\mu \rm m}/S_{0.8\mu \rm m}\sim100$, in which the wavelengths are taken as the central wavelength of \wise\ w1 band and SDSS \textit{i} band, respectively. 
On the other hand, for the 149 ULIRGs with galaxy dominated optical spectra, the average flux ratio $F_{\rm w1}/F_{i\rm -band}$ is $3.8\pm4.7$. The result suggests that even the bluest SED in the template is still much redder than the observed SED in optical-NIR bands, therefore the contamination of the penetrating component on the optical spectrum could be ignored. 

%%%%%%%%%%%%%%%%%%%%%%%%%%%%%%%%%%%%%%%%%%%%%%%%%%%%%%%%%%%%%%%%%%%%%%%%%%%%%%%%%%%%%%%%%%%%%%%%%%%%%%%%%%%%%%%%%%%%%%%%%%%%%
%%%%%%%%%%%%%%%%%%%%%%%%%%%%%%%%%%%%%%%%%%%%%%%%%%%%%%%%%%%%%%%%%%%%%%%%%%%%%%%%%%%%%%%%%%%%%%%%%%%%%%%%%%%%%%%%%%%%%%%%%%%%%

\subsection{Connection from optical spectral fitting to SED decomposition}
\label{sec:Appendix_sedfit_connection} 

In order to connect the spectral fitting results to the SED fitting process, we consider the energy conservation in each emitter-absorber system and the entire galaxy. The equations which represents the energy balance in the dust surrounding stars are listed in Equation \ref{equ:EB_HG}: 
\begin{equation}
\begin{split}
    \frac{
    \int_{912\rm \textup{\AA}}^{1\textup{mm}} S_{\rm MS}^{\rm\,sp} \left[ 1 - 10^{ -0.4(\tau_{\lambda}/\tau_{\rm V})A_{\rm V, MS}^{\rm\,sp} } \right]\,
    \mathrm{d}\lambda 
    }{
    \int_{100\rm \textup{\AA}}^{1\textup{mm}} S_{\rm MS}^{\rm\,sp}\,\mathrm{d}\lambda 
    }
    & = \frac{L_{\rm ISM, MS}}{L_{\rm MS}}  = \frac{L_{\rm ISM, MS}}{C_{\rm ape, MS}L_{\rm MS}^{\rm\,sp}},  \\
    \frac{
    \int_{912\rm \textup{\AA}}^{1\textup{mm}} S_{\rm TSB}^{\rm\,sp} \left[ 1 - 10^{ -0.4(\tau_{\lambda}/\tau_{\rm V})A_{\rm V, TSB}^{\rm\,sp} } \right]\,
    \mathrm{d}\lambda 
    }{
    \int_{100\rm \textup{\AA}}^{1\textup{mm}} S_{\rm TSB}^{\rm\,sp}\,\mathrm{d}\lambda 
    } 
    & = \frac{L_{\rm ISM, TSB}}{L_{\rm TSB}}  = \frac{L_{\rm ISM, TSB}}{C_{\rm ape, TSB}L_{\rm TSB}^{\rm\,sp}},  \\
    \frac{
    \int_{912\rm \textup{\AA}}^{1\textup{mm}} S_{\rm ASB}^{\rm\,sp} \left[ 1 - 10^{ -0.4(\tau_{\lambda}/\tau_{\rm V})A_{\rm V, ASB} } \right]\,
    \mathrm{d}\lambda 
    }{
    \int_{100\rm \textup{\AA}}^{1\textup{mm}} S_{\rm ASB}^{\rm\,sp}\,\mathrm{d}\lambda 
    } 
    & = \frac{L_{\rm ISM, ASB}}{L_{\rm ASB}}, 
    \label{equ:EB_HG}
\end{split}
\end{equation}
where the $S_{\rm MS}^{\rm\,sp}$ means the intrinsic stellar spectrum of MS stars, $A_{\rm V, MS}^{\rm\,sp}$ means the extinction of diffuse ISM dust around MS stars; $L_{\rm MS}$ and $L_{\rm ISM, MS}$ denote the total luminosity for the MS population and the dust heated by MS stars. Similar variables are defined for TSB and ASB components and dust around them. 
All variables with `sp' suffix suggest that the spectra and values are taken from the spectral fitting results, e.g., $S_{\rm MS}^{\rm\,sp}$ is the best-fit continuum for MS components in spectral fitting results. Note that two factors, i.e., $C_{\rm ape, MS}$ and $C_{\rm ape, TSB}$ are employed to correct for the aperture-loss of the SDSS fibers. We allow the MS and TSB components have different correctors. $\tau_{\lambda}$ denotes the extinction (opacity) curve used in the SED fitting, which is the same one used in the spectral fitting. In order to build a self-consistent method, we adopt the extinction curve derived from THEMIS dust model, which is used to reproduce the dust emission in the SED fitting process. 
Note that in the integration of absorbed luminosity by dust, we set a lower limit of wavelength of 912\AA, with an underlying assumption that most of the photons with energy higher than 13.6 eV are absorbed by 
the photoionization process of Hydrogen and Helium gas in \hii\ region, rather than attenuated by the dust \citep{Draine2007}. 
As explained in Section \ref{subsec:Appendix_ISM_dust}, within the variables in Equation \ref{equ:EB_HG}, we assume $A_{\rm V, TSB}^{\rm\,sp}=A_{\rm V, MS}^{\rm\,sp}$ and $A_{\rm V, ASB}=100$; $S_{\rm ASB}^{\rm\,sp} = S_{\rm TSB}^{\rm\,sp}$, thus the ASB component has the same intrinsic spectrum as TSB components. Only three variable, i.e., $C_{\rm ape, MS}$, $C_{\rm ape, TSB}$, and $L_{\rm ASB}$, are the independent parameters which will be estimated in the fitting process. 
 
As for the energy conservation of AGN and torus radiation, we directly adopt the result in SKIRTor model, which is calculated by a radiation transfer program. A similar equation can be explained as Equation \ref{equ:EB_AGN}. Note that since both of the primary and dust SED are anisotropic, they are shown in flux unit in the equation. 

\begin{equation}
\begin{split}
    \frac{
    \int_{\Omega_{\rm torus}}  \int_{912\textup{\AA}}^{1\textup{mm}} F_{\rm AD}  \left[ 1 - 10^{ -0.4(\hat{\tau}_{\lambda}/\hat{\tau}_{\rm V})A_{\rm V, Tor} } \right] 
     \mathrm{d}\lambda \mathrm{d}\omega
    }{
    \int_{4\pi} \int_{100\rm \textup{\AA}}^{1\textup{mm}} F_{\rm AD} \, \mathrm{d}\lambda \mathrm{d}\omega 
    } \\
    = \frac{
    \int_{4\pi} \int_{100\rm \textup{\AA}}^{1\textup{mm}} F_{\rm torus}\, \mathrm{d}\lambda \mathrm{d}\omega 
    }
    {L_{\rm AGN} / D_{\rm L}(z)^2}
    \label{equ:EB_AGN}
\end{split}
\end{equation}

\begin{figure}
    \begin{center}
    \includegraphics[trim=0 30pt 0 0, width=0.8\columnwidth]{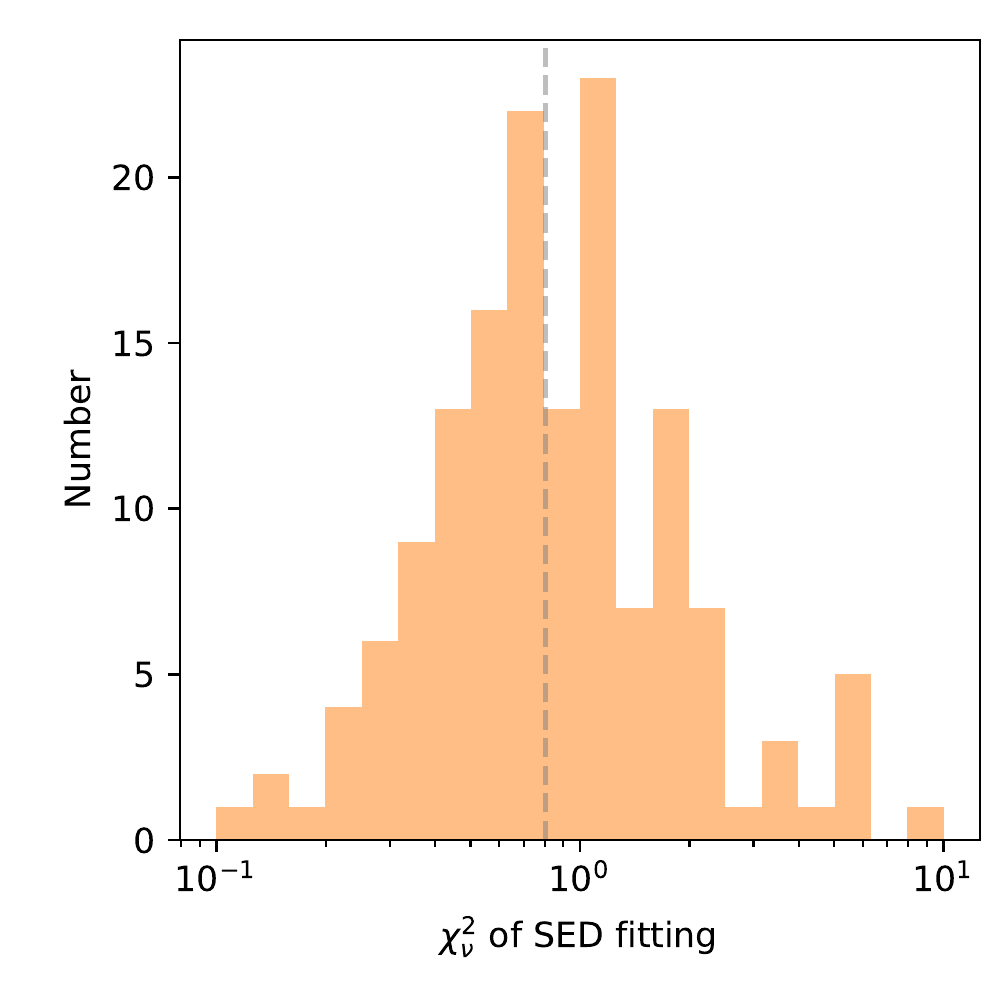}
    \end{center}
    \caption{Reduced $\chi^2$ of the SED decomposition. The vertical dashed line denotes the median $\chi^2_{\nu}$ of 0.8. }
    \label{fig:SSFIT_chisq_1}
\end{figure}

For the SED model with parameters of a given values, the expected fluxes in each band can be calculated by convolving the model SED $S_{\rm comp}$ with the transmission curve of each band $T_{\rm band}$: 
\begin{equation}
\begin{split}
    F_{\rm band}^{\rm mod} = \Sigma_{\rm comp}\, \frac{L_{\rm comp}}{4\pi D_{\rm L}^2(z)} 
    \frac{\int S_{\rm comp}(z) T_{\rm band}\,\mathrm{d}\lambda}{\int S_{\rm comp}(z)\,\mathrm{d}\lambda \, \int T_{\rm band}\, c/\lambda^2\,\mathrm{d}\lambda}\,,\\
    \label{equ:SED_LA}
\end{split}
\end{equation}
and the best-fit SED parameters (e.g., $q_{\rm HAC}$) and the normalizations, $L_{\rm comp}$ (e.g., $L_{\rm AGN}$) can be obtained by minimizing $\chi^2=(F_{\rm band}^{\rm obs}-F_{\rm band}^{\rm mod})^2/\sigma_{\rm band}^{\rm obs\, 2}$, where $F_{\rm band}^{\rm obs}$ and $\sigma_{\rm band}^{\rm obs}$ are the observed fluxes and their measurement errors. 
All of the primary radiation components, i.e., MS, TSB, ASB, and AGN (accretion disk), and the dust re-emitted components. i.e., ISM heated by MS and TSB, BC heated by ASB, and torus heated by AGN, are included in the calculation. 
In order to reproduce the contribution of emission lines to the broad band photometry in optical bands, we also include the optical emission line spectrum from the decomposition of observed spectrum (Section \ref{subsec:spectral_fitting}) in Equation \ref{equ:SED_LA}. The emission line spectrum is corrected for aperture-loss using the average of $C_{\rm ape, MS}$ and $C_{\rm ape, TSB}$. 
The four outputs directly from Equation \ref{equ:SED_LA} are: 
two aperture correction factors ($C_{\rm ape, MS}$ and $C_{\rm ape, TSB}$), 
the total luminosity of primary radiation of ASB component ($L_{\rm ASB}$), 
and AGN accretion disk ($L_{\rm AGN}$). 
The inputs and outputs of the SED decomposition are summarized in Table \ref{tab:SED_var}. 
The distribution of the $\chi^2_{\nu}$ is shown in Figure \ref{fig:SSFIT_chisq_1}, with the median reduced $\chi^2$ of 0.8. 

\begin{table}[ht]
\centering
\caption{Summary of the variables in SED decomposition equation. \label{tab:SED_var}}
\begin{tabular}{cc}
\hline
\hline
\multicolumn{2}{c}{Input (from spectral fitting)} \\
Host galaxy & $S_{\rm MS}^{\rm\,sp}$, $S_{\rm TSB}^{\rm\,sp}$$^{1}$, 
            $A_{\rm V, MS}^{\rm\,sp}$$^{2}$, 
            $L_{\rm MS}^{\rm\,sp}$, $L_{\rm TSB}^{\rm\,sp}$ \\
\hline
\multicolumn{2}{c}{Input (parameters of undetermined SED)} \\
Host galaxy & $q_{\rm HAC}$, $U_{\rm min}$, $\alpha$ \\
AGN & $\varphi$, $R_{\rm out}/R_{\rm in}$, $\tau_{7.9}$ \\
\hline
\multicolumn{2}{c}{Output (directly from Equation \ref{equ:SED_LA})} \\
Host galaxy & $C_{\rm ape, MS}$, $C_{\rm ape, TSB}$, $L_{\rm ASB}$ \\
AGN & $L_{\rm AGN}$\\
\hline
\multicolumn{2}{l}{$^{1}$\footnotesize{the same for $S_{\rm ASB}^{\rm\,sp}$.}} \\
\multicolumn{2}{l}{$^{2}$\footnotesize{the same for $A_{\rm V, TSB}^{\rm\,sp}$; we assume $A_{\rm V, ASB}$=100.}} \\
\end{tabular}
\end{table}

%We also show the $\chi^2_{\nu}$ and relative residuals ($F_{\rm band}^{\rm obs}-F_{\rm band}^{\rm mod}/ F_{\rm band}^{\rm obs}$) in each band, and we can find that the quite large $\chi^2_{\nu}$ is dominated by the $\chi^2_{\nu}$ in \textit{r}- and w1 bands, which could be due to the small photometric errors in these two bands. The relative residuals in \textit{r}- and w1 bands are $-0.021$ and $0.003$, respectively, which are similar to the relative residuals in other SDSS and \wise\ bands. Note that in this work the $\chi^2_{\nu}$ is only used to select the best-fit SED model, the uncertainties of the results are also estimated with Monte Carlo resampling method. 
For each ULIRG, 100 mock photometric observations are generated by adding the random noise to the observed fluxes. 
The random noise is Gaussian distributed with the amplitude from the measurement error in each band. 
The same fitting processes are employed for the mock photometric data and the scatters of the best-fit values are taken as the uncertainties of the parameters.

\section{Possible effect of the galactic-scale dust heated by AGN}
\label{subsec:Appendix_AGN_Polar}

An underlying assumption of the typical SED decomposition method for galaxies is that the dust heated by stellar light dominates the bulk of the observed FIR radiation, and the AGN dusty torus mainly contributes to the MIR excess due to a higher dust temperature (e.g., Section \ref{subsec:SED_decomp}, see also \citealp{Ciesla2015, Malek2017}). 
However, an equatorial optically-thick torus is only a first-order approximation of the circumnuclear dust environment around AGN, 
and the dust can also exist in the polar direction, i.e., the direction perpendicular to the equatorial plane of the torus \citep[e.g.,][]{Lyu2018}. 
The dust emission in polar direction is supported by recent high spatial resolution MIR observations \citep{Asmus2016, Asmus2019, Fuller2019}. 
The polar dust can also contribute to the FIR emission if it extends to a large scale ($\sim10^2$--$10^3$ pc). 
Dust model{}s combing both of the torus dust and polar dust with sophisticated geometry has been developed using radiation transfer codes to explain the observed interferometric image (e.g., \citealp{Honig2017, Stalevski2017}). The best-fit geometry is usually depends on the particular observational constraints for individual AGNs. 
Here we introduce a simpler method to model the AGN-heated dust at a large scale to quickly test the possible effect of the additional dust component on the analyses of the evolution of ULIRGs. 
{}
Since at a large distance the dust can be assumed to be heated by a point-like radiation source, we consider the polar dust component follows a scaled PDR-like structure (Appendix \ref{subsec:Appendix_ISM_dust}), and the energy conservation between the absorption and the re-emitted thermal radiation can be calculated as:
\begin{equation}
\begin{split}
    \frac{
    \int_{912\textup{\AA}}^{1\textup{mm}} F_{\rm AD}(i=0)  \left[ 1 - 10^{ -0.4(\tau_{\lambda}/\tau_{\rm V})A_{\rm V, Pol} } \right] 
     \mathrm{d}\lambda 
    }{
    \int_{100\textup{\AA}}^{1\textup{mm}} F_{\rm AD}(i=0) \, \mathrm{d}\lambda 
    } \\
    = \frac{L_{\rm polar}}{L_{\rm AGN}} \left( 1-\left( \frac{\Omega_{\rm torus}}{4\pi} \right)^2 \right)^{-1}, 
    \label{equ:EB_Polar}
\end{split}
\end{equation}
where $F_{\rm AD}(i=0)$ denotes the assumed SED of the AGN accretion disk in polar direction; $\tau_{\lambda}/\tau_{\rm V}$ and $A_{\rm V, Pol}$ are the dust opacity and extinction in the polar direction; $L_{\rm polar}$ and $L_{\rm AGN}$ denote the luminosity of polar dust and the bolometric luminosity of AGN; $\Omega_{\rm torus}/4\pi$ is covering factor of the torus, which is related to the half opening angle ($\Theta$) by $\Omega_{\rm torus}/4\pi=\sin{\Theta}$. 
The polar dust is assumed to be isotropic, but only the dust directly facing the AGN, i.e., in the direction that is not covered by the torus, is included in the energy conservation equation. We neglect the contribution from scattered light. 
The component is considered to be optically-thin and the extinction measured from the optical stellar continuum is assumed for the polar dust, i.e., $A_{\rm V, Pol}=A_{\rm V, MS}$. Following the diffuse interstellar dust, we still employ the THEMIS dust model and its opacity in the calculation. 
The inner edge of the polar dust is fixed using $U=10^7$, which is similar to the sublimation radius of the torus. 
We assume the dust at the outer edge is with the same temperature as the ISM dust, thus the outer radius of the polar dust can be parameterized using the average heating intensity in ISM region. For an AGN with $L_{\rm AGN}=10^{11}$\,\lsun\ and the outer dust temperature of 30 K, the inner and outer radius of the polar dust is approximately 3 pc and 2 kpc, respectively. 
For simplicity, we only consider two cases with fixed torus opening angle $\Theta=30\arcdeg$ (hereafter Case 1) and $\Theta=10\arcdeg$ (hereafter Case 2). 
The half opening angle in Case 1 is the same as that in the previously discussed default case without polar dust (hereafter Case 0). 
A smaller $\Theta$ in Case 2 means that AGN can eject more energy into the polar dust, i.e., contributes more FIR radiation. 

\begin{figure*}
    \begin{center}
    \includegraphics[trim=0 10pt -10pt 0, width=\columnwidth]{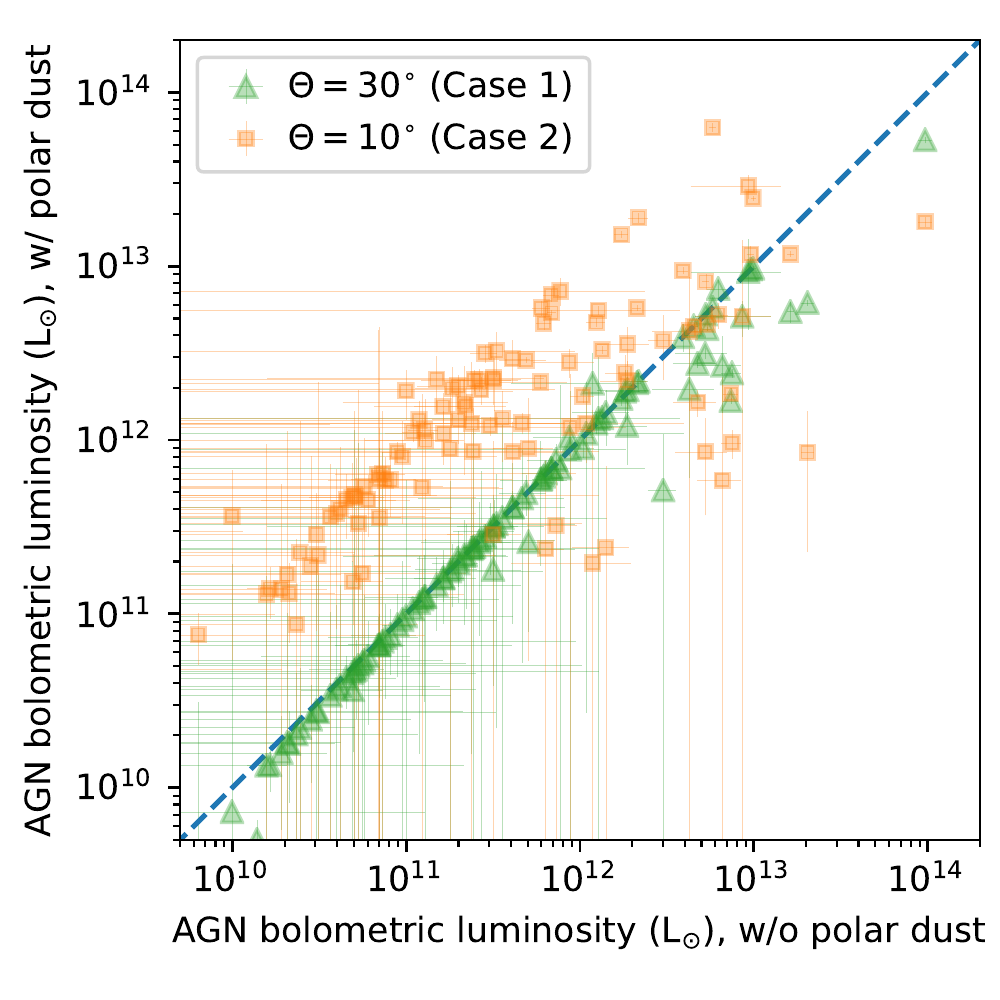}
    \includegraphics[trim=0 10pt -10pt 0, width=\columnwidth]{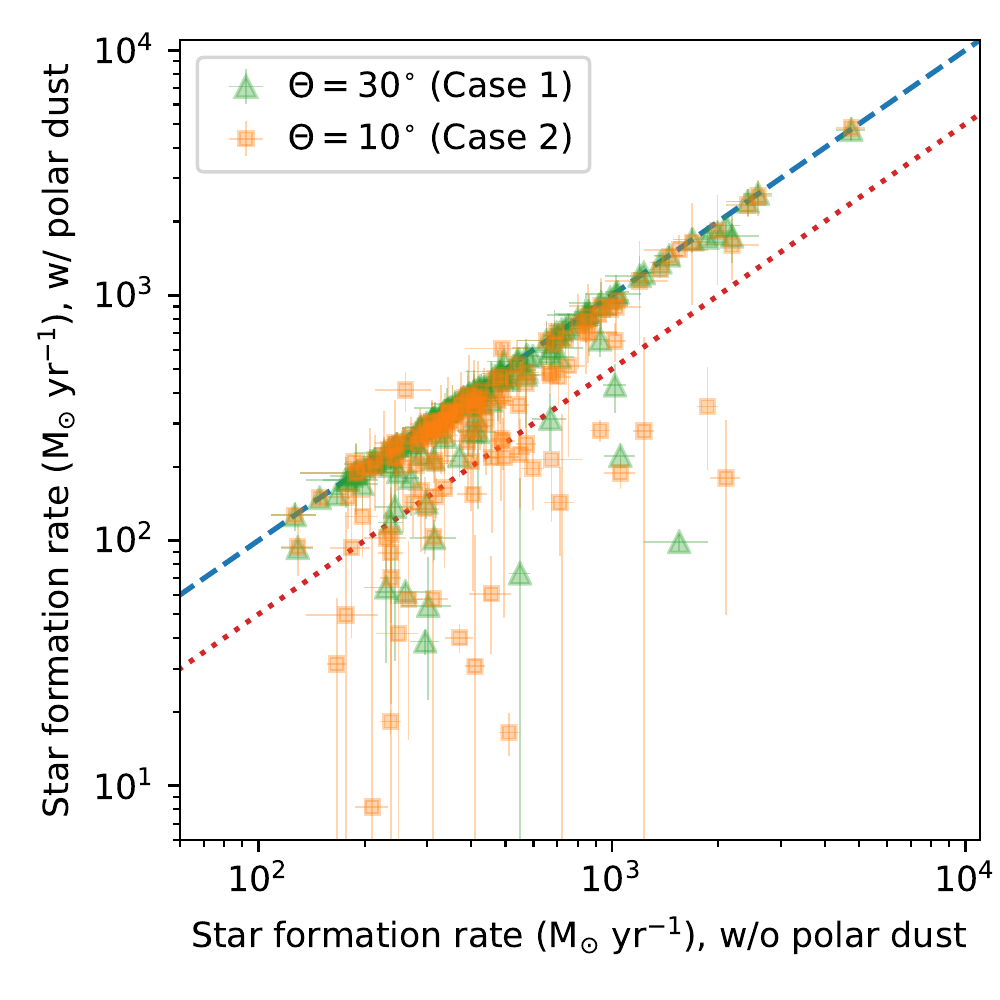}
    \end{center}
    \caption{
    Left: Comparison between AGN bolometric luminosity ($L_{\rm AGN}$) estimated with (\textit{y}-axis) and without (\textit{x}-axis) polar dust component. 
    The green triangles and orange squares denote the results with $\Theta=30\arcdeg$ and $\Theta=10\arcdeg$, respectively, where $\Theta$ is the half opening angle of the torus. Blue dashed line shows 1:1 ratio position. 
    Right: Comparison between SFR estimated with (\textit{y}-axis) and without (\textit{x}-axis) polar dust component. The legends are the same as those in the left panel. Red dotted line denotes 2:1 ratio position, e.g., the estimated SFR decreases by 50\% with polar dust component. 
    }
    \label{fig:LumAGN_SFR_polar_1}
\end{figure*}

The two panels in Figure \ref{fig:LumAGN_SFR_polar_1} show the $L_{\rm AGN}$ and SFR estimated from the SED decompositions with the polar dust component, and the comparison to the results in the default case without polar dust. 
The $L_{\rm AGN}$ from Case 1 are nearly the same as the $L_{\rm AGN}$ from Case 0, because the $L_{\rm AGN}$ is limited by the MIR torus emission, and the same $\Theta=30\arcdeg$ corresponds to (almost) the same $L_{\rm AGN}$-$L_{\rm torus}$ correlation ($L_{\rm AGN}$ and $L_{\rm torus}$ are the bolometric luminosity of AGN (accretion disc) and the luminosity of torus). 
$\Theta=10\arcdeg$ in Case 2 suggests a smaller energy conversion fraction from AGN to torus, and yields out a ten times higher $L_{\rm AGN}$. 
%The $L_{\rm AGN}$-$L_{\rm torus}$ correlation at $\Theta=30\arcdeg$ and $\Theta=10\arcdeg$ are shown in Figure \ref{fig:AGN_Torus_OA}. 
The average energy conversion fractions of the polar dust ($f_{\rm polar}=L_{\rm polar}/L_{\rm AGN}$) for Case 1 and Case 2 are $0.23\pm0.05$ and $0.28\pm0.05$, respectively. 
If we compare the two cases, the average conversion fractions of the torus dust ($f_{\rm torus}=L_{\rm torus}/L_{\rm AGN}$) for Case 1 and Case 2 are $0.21\pm0.01$ and $0.03\pm0.01$, respectively. 

Since the radiation of polar dust is assumed extend to FIR wavelength range, the SFR from the SED decomposition with polar dust can be reduced. 
For simplicity we define the reduced ratio as $f_{\rm SFR, red}=1-\mathrm{SFR}_{\rm with\, polar}/\mathrm{SFR}_{\rm no\, polar}$. 
In Case 1, 11 out of the 149 ULIRGs show $f_{\rm SFR, red}>50\%$, while 114 ULIRGs show $f_{\rm SFR, red}<10\%$, 
indicating that with the torus of $\Theta=30\arcdeg$, 
the effect of polar dust on the estimation of SFR is small for most of the galaxies. %, unless the AGN is very luminous. 
However, in Case 2, 33 ULIRGs show $f_{\rm SFR, red}>50\%$, in which 6 galaxies even show $f_{\rm SFR, red}>90\%$, 
which is due to the much cooler and redder AGN dust SED ($L_{\rm polar}/L_{\rm torus}\sim9$, to be compared, the ratio for Case 1 is $\sim1$). 
The high $f_{\rm SFR, red}$ in Case 2 could suggest a quenching pattern, i.e., the SFR decreases as AGN becomes luminous, which we do not find in the default analysis. 
The quenching correlation should be treated with caution since the result is highly dependent on the assumption of the model, e.g., the distribution and extinction of the polar dust component. 
Since the merger can trigger massive inflows of gas and dust into the nuclear region, the highly obscured AGN embedded in galactic scale dusty environment could exist in some cases, e.g., hot DOGs \citep{Lyu2018}. 
The possible effect of AGN on large scale dust indicates that considerable uncertainties remain in the 
study of ULIRG evolution. 
The submm observation, e.g., ALMA, to directly determine the dust emission distribution around the AGN is required to address the question. 

%%%%%%%%%%%%%%%%%%%%%%%%%%%%%%%%%%%%%%%%%%%%%%%%%%%%%%%%%%%%%%%%%%%%%%%%%%%%%%%%%%%%%%%%%%%%%%%%%%%%%%%%%%%%%%%%%%%%%%%%%%%%%
%%%%%%%%%%%%%%%%%%%%%%%%%%%%%%%%%%%%%%%%%%%%%%%%%%%%%%%%%%%%%%%%%%%%%%%%%%%%%%%%%%%%%%%%%%%%%%%%%%%%%%%%%%%%%%%%%%%%%%%%%%%%%

\end{appendices}

\end{document}